\documentclass[a4paper,11pt]{article}
\usepackage[x11names,dvipsnames]{xcolor}
\usepackage{mathtools, siunitx, physics, array, bm, graphicx}
\usepackage{booktabs}
\usepackage{tikz}
\usepackage{wrapfig}
\usepackage[compat=1.1.0]{tikz-feynman}
\usepackage{subcaption, caption}
\usepackage{adjustbox}
\usepackage{enumitem, comment}
\usepackage{slashed}
\usepackage{float}
\usepackage{pdfpages}
\usepackage{jheppub}
\usepackage{academicons}
\usepackage{hyperref}
\usepackage{mathrsfs}
\usepackage{orcidlink}

\tikzfeynmanset{
every crossed dot@@/.style={
    /tikz/fill=none,
    /tikz/draw=black,
    /tikz/shape=crossed circle,
    /tikz/minimum size=0.25cm,
    /tikz/inner sep=0pt,
},
wblob/.style={
    /tikz/fill=none,
    /tikz/draw=red,
    /tikz/shape=crossed circle,
    /tikz/minimum size=0.25cm,
    /tikz/inner sep=0pt,
},
gblob/.style={
    /tikz/fill=none,
    /tikz/draw=blue,
    /tikz/shape=crossed circle,
    /tikz/minimum size=0.25cm,
    /tikz/inner sep=0pt,
},
wgblob/.style={
    /tikz/fill=none,
    /tikz/draw=green!50!black,
    /tikz/shape=crossed circle,
    /tikz/minimum size=0.3cm,
    /tikz/inner sep=0pt,
},
}

\def\sl#1{\slashed{#1}}

\def\cM{{\cal M}}
\def\cL{{\cal L}}
\def\cO{{\cal O}}
\def\cC{{\cal C}}
\def\cR{{\cal R}}

\usetikzlibrary{positioning,arrows.meta}
\tikzfeynmanset{ with arrow/.style = {
   decoration={
     markings,
     mark=at position 0.5
          with {\arrow[xshift=1mm]{Triangle[black,width=0.8mm,length=1.2mm]}}
     },
   postaction=decorate}
}

\title{A comprehensive study of ALPs from $B$-decays}

\author{Deepanshu Bisht\orcidlink{0009-0009-7047-773X},}
\author{Sabyasachi Chakraborty\orcidlink{0000-0001-5356-7607},}
\author{and Atanu Samanta\orcidlink{0009-0006-2782-2911}.}

\affiliation{Department of Physics, Indian Institute of Technology Kanpur, Kanpur-208016, India}

\emailAdd{dbisht22@iitk.ac.in}
\emailAdd{sabyac@iitk.ac.in}
\emailAdd{asamanta23@iitk.ac.in}

\abstract{We present a comprehensive study of axion-like particles (ALPs) through flavor changing neutral current processes, such as $B\to K a$ followed by $a\to\text{hadronic}, \gamma\gamma,\mu^+\mu^-$ channels. Our generic framework encompasses different ultraviolet scenarios similar to KSVZ, DFSZ and flavorful axions etc. Starting from the effective Lagrangian written at the high scale, we compute the anomalous dimension matrix, taking into account all one-loop and relevant two-loop contributions. The latter is most important for the KSVZ and heavy QCD axion scenarios. We recognized that such two-loop diagrams can have both ultraviolet (UV) and infrared (IR) divergences. We show explicitly that UV divergences cancel by inserting appropriate counterterms, which are new operators involving the axion field and required to be present at the UV itself, to renormalize the theory. On the other hand, the cancellation of IR divergences is subtle and demonstrated through matching with the effective theory at the electroweak scale. We also utilize chiral perturbation theory and vector meson dominance framework to compute the decay and branching fractions of the ALP pertaining to our framework. We find that for KSVZ-like scenario, axion decay constant, $f_a \lesssim 1$ TeV can be ruled out. The bound becomes stronger for the DFSZ and Flaxion-like models, reaching upto $10^4$ TeV and $10^6$ TeV, respectively. We also provide projections on the parameter space based on 3 ab$^{-1}$ data from Belle II and 300 fb$^{-1}$ data from LHCb.}

\begin{document}

\hypersetup{
pdfauthor={Deepanshu Bisht, Sabyasachi Chakraborty, Atanu Samanta},
pdftitle={A comprehensive study of ALPs from B-decays},
pdfsubject={},
pdfkeywords={}
}

\maketitle
\flushbottom
%%%%%%%%%%%%%%%%%%%%%

\section{Introduction}
\label{sec:intro}
Searches for new physics significantly lighter than the TeV scale have garnered considerable attention in the recent past. The reason is primarily twofold: Firstly, the lack of any signatures of new physics at the Large Hadron Collider (LHC) has prompted us to pursue a broader perspective on where new physics could hide. Secondly, any new light physics particles offer the potential to be tested at multiple frontiers, such as intensity, cosmic frontier, etc. The axion~\cite{Weinberg:1977ma,Wilczek:1977pj} is one of the most well-motivated and well-studied examples for such light new physics. They usually emerge as pseudo-Nambu-Goldstone bosons resulting from the spontaneous breaking of a $U(1)$ symmetry, for example, the Peccei-Quinn (PQ) symmetry~\cite{Peccei:1977hh,Peccei:1977ur} and hence are naturally light. The QCD-induced potential of the axion is naturally minimized at the zero expectation value of the axion field, providing a compelling solution to the strong-$CP$ problem~\cite{tHooft:1976rip}. Over the years, the axion has been demonstrated to be an excellent cold dark matter candidate~\cite{Preskill:1982cy,Dine:1982ah,Abbott:1982af} or to facilitate SIMP (Strongly interacting massive particle) cold dark matter by providing a portal to SM~\cite{Hochberg:2018rjs}, resolve matter-antimatter asymmetry~\cite{Co:2019wyp,Chakraborty:2021fkp} and, in some incarnations, address the long-standing hierarchy problem of the Standard Model (SM)~\cite{Graham:2015cka,Hook:2016mqo,Trifinopoulos:2022tfx}. As a result, the axion provides a natural testing ground for physics beyond the SM.

For a prototypical axion, QCD solely determines its mass, predicting the relation $m_a f_a \simeq m_\pi f_\pi$, where, like pions, $m_a$ and $f_a$ are the mass and decay constants for the axion. It is important to note that $f_a$ also indicates the scale at which the PQ symmetry 
\begin{wrapfigure}{l}{0.51\textwidth}
\includegraphics[width=0.51\textwidth]{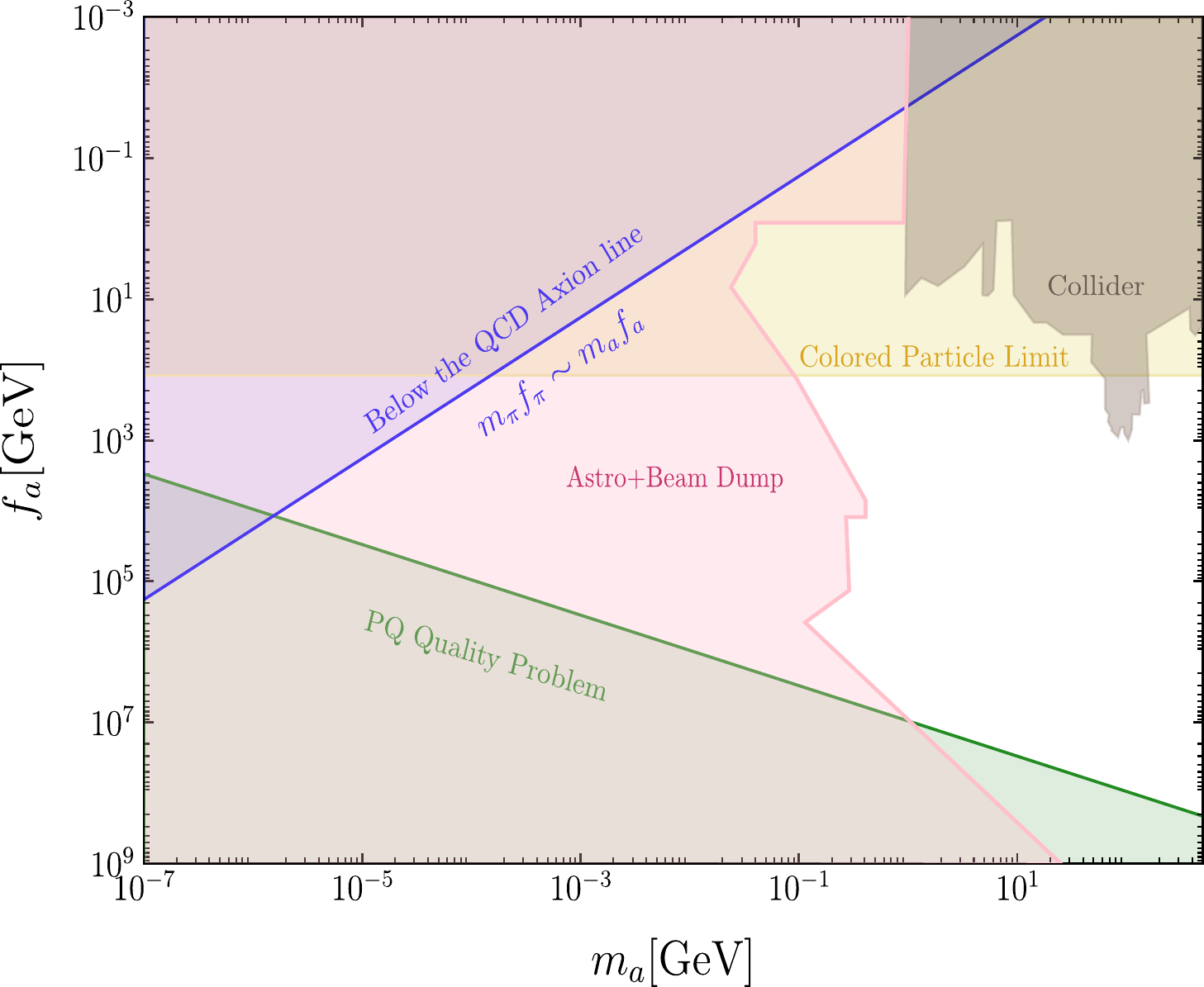}
\caption{A schematic representation of axion parameter space inspired from~\cite{Hook:2019qoh}. Colored regions are ruled out by different frontier experiments.}
\label{fig:schematic}
\end{wrapfigure}
is spontaneously broken. Although $f_a$ could theoretically take any value, it is rather well motivated from both theoretical and experimental perspectives to choose it around the TeV scale. However, such a choice automatically places the axion mass around the keV scale, which is already excluded by astrophysical observations~\cite{CAST:2017uph,Raffelt:2006cw,Friedland:2012hj,Caputo:2024oqc} and beam dump experiments~\cite{CHARM:1985anb,Bjorken:1988as,Blumlein:1990ay,Dobrich:2015jyk}.  Furthermore, the straightforward and elegant bottom-up solution to the strong-$CP$ problem suffers from a major top-down flaw. The QCD-induced potential is extremely shallow. As a result, any explicit PQ symmetry-breaking effects, even those suppressed by the Planck scale, such as $V\sim f_a^5/M_{Pl}\lesssim\Lambda_{\text{QCD}}^4$ can shift the minima of the potential away from zero, thus jeopardizing the solution of the strong-$CP$ problem. This particular issue is known as the axion quality problem in the literature~\cite{Kamionkowski:1992mf,Holman:1992us,Barr:1992qq,Ghigna:1992iv}. It therefore becomes imperative to choose frameworks that deviate away from the aforementioned mass relation. This can be realized by considering an additional contribution to the axion potential. Several model building efforts have been made in this direction in the recent past~\cite{Dimopoulos:1979pp,Holdom:1982ex,Dine:1986bg,Flynn:1987rs,Choi:1988sy,Rubakov:1997vp,Choi:1998ep,Berezhiani:2000gh,Choi:2003wr,Hook:2014cda,Fukuda:2015ana,Dimopoulos:2016lvn,Agrawal:2017ksf,Agrawal:2017eqm,Gaillard:2018xgk,Hook:2019qoh,Gherghetta:2020keg,Gupta:2020vxb,Delgado:2024pcv}, giving rise to axion-like particles (ALPs). Similar construction can also be realized in DFSZ-like scenarios~\cite{Berezhiani:1999qh} and its implications for astrophysical observations. It is evident that higher mass axions or lower values of the $f_a\sim\mathcal O$(TeV) scale not only reintroduce the advantages of the electroweak scale axion but also improve the quality problem. Indeed, such a region (see Fig.~\ref{fig:schematic}) with $m_a\gtrsim \mathcal O$(MeV) and $f_a\gtrsim\mathcal O$ (TeV) has yet to be probed by experiments. Moreover, slightly heavier axions around the GeV scale also present considerable theoretical challenges, which is the primary focus of our work.  

Such regions are also ideally suited for intensity frontier experiments, which utilize intense sources of beam to investigate rare processes in the SM. The main advantage is that, since SM processes are rare, any potential new physics might not be masked by a large SM background. Motivated by this, we will focus on axion-like particles with $f_a \gtrsim \mathcal{O}(\text{TeV})$ and which couple to the SM gluon and fermion fields. As a result, the production is purely hadronic, whereas the decay rates of the ALPs would depend on particular UV models under consideration.
For ALP masses $m_a\lesssim \mathcal O$(GeV), it is convenient and customary to rotate away the axion-gluon $a-G- \widetilde{G}$ term in the Lagrangian by performing a chiral transformation on the SM quark fields. Under chiral rotation, the measure of the action is not invariant and can be chosen to completely rotate away the axion-gluon term. The subsequent Lagrangian generates the axion-quark mass and kinetic mixing terms, which can be matched with the effective theory of the chiral Lagrangian (ChPT)~\cite{GrillidiCortona:2015jxo,Bauer:2021wjo,Bandyopadhyay:2021wbb}. This effective potential derived from the chiral Lagrangian is well suited to study non-perturbative effects such as $a-\pi^0$ and $a-\eta (\eta^\prime)$ mixing. This framework is frequently used to probe the axion parameter space by studying rare decays of pions~\cite{PIENU:2019usb,Pocanic:2003pf,Altmannshofer:2019yji,Bandyopadhyay:2021wbb} and kaons~\cite{Georgi:1986df,Bardeen:1986yb,Alves:2017avw,Gori:2020xvq,E949:2005qiy,NA62:2014ybm,KTeV:2008nqz,KOTO:2018dsc,Bauer:2021wjo, Girmohanta:2024nyf}. Similarly, fixed-target experiments~\cite{GlueX:2017zoo,Aloni:2019ruo} and colliders~\cite{OPAL:2002vhf,Knapen:2016moh,Bauer:2017ris,CMS:2017dcz,Mariotti:2017vtv,Cacciapaglia:2021agf,Biswas:2023ksj,Bao:2024pyt} also provide competitive constraints in the $m_a-f_a$ plane, albeit for slightly heavier masses. Noticeably, for $m_a\gtrsim\mathcal O$ (GeV), the power counting of the ChPT breaks down, necessitating perturbative calculations.

The region with $\mathcal O(100\;\text{MeV})\lesssim m_a\lesssim \mathcal O(\text{GeV})$ is also sensitive to flavor-changing neutral current (FCNC) processes, mostly from decays of the $B$ meson, and can leave telltale signatures.  Before discussing the primary objective of our work, let us briefly chart various attempts that have been made to probe such heavy axions in this particular ballpark using FCNC processes.

\begin{itemize}[leftmargin=*]
    \item {\bf Generic ALP framework:} Ref.~\cite{Bauer:2020jbp} presented a relevant anomalous dimension matrix for the most general dimension-5 interactions of axion and SM fields. Partial two-loop matching coefficients were also worked out for flavor-changing effects. This framework was extended to comprehensively study quark and lepton flavor violations~\cite{Bauer:2021mvw} in different low-energy experiments.
    
    \item {\bf Heavy QCD axion:} Starting from a heavy QCD axion that only couples to gluon fields, Ref.~\cite{Chakraborty:2021wda} studied for the first time, $b\to s$ transitions generated at the two-loop order. To renormalize the theory, one requires additional counterterms, essentially operators involving axion and quark fields. After the Renormalization-group (RG) evolution of the Wilson coefficients to the weak scale, branching ratios for $B\to K a$ were computed and augmented with the decay of axions. The associated branching fractions of axions were taken from~\cite{Aloni:2018vki}, which prescribed a data-driven approach. The bounds obtained both from prompt searches~\cite{Chakraborty:2021wda} and displaced vertex searches~\cite{Bertholet:2021hjl} were demonstrated to be robust and conservative, as flavor-changing couplings were not considered in UV. 

    \item {\bf Flavorful axions:} Flavor-dependent axions are also well motivated, as they can address the strong $CP$ problem, as well as the flavor hierarchy of the SM in the quark and lepton sector~\cite{Berezhiani:1983hm, Berezhiani:1985in, Ema:2016ops}. This is usually achieved by augmenting the SM with a complex scalar field. Such axions can also explain the observed dark matter abundance~\cite{Planck:2015fie} and baryon asymmetry, while the radial mode can play the role of an inflaton. Similar frameworks where axions possess non-universal PQ charges~\cite{MartinCamalich:2020dfe,Li:2024thq,Greljo:2024evt} as well as Froggatt-Nielson~\cite{Froggatt:1978nt} motivated frameworks where ALPs only interact with up-type quarks~\cite{Carmona:2021seb} can also provide interesting discovery potential at the $B$-factories and other low energy experimental facilities. Originally, Refs.~\cite{Berezhiani:1989fs, Berezhiani:1989fp} presented a scenario in which the diagonal flavor and the off-diagonal couplings of a Nambu Goldstone boson are generated from the spontaneous breaking of a global $U(1)_H$ symmetry. The decay widths of various lepton and meson production channels and the subtleties in the form factor values were emphasized. 
\end{itemize}

    Other studies of heavier ALP searches using FCNC processes involve ALP coupling to weak gauge SM bosons~\cite{Izaguirre:2016dfi}, exploring the interplay of ALP-Higgs and bosonic couplings~\cite{Gavela:2019wzg} etc. Studies of a dilaton in the MeV-GeV mass range~\cite{Girmohanta:2023tdr} are also notable, showing similarities with the axion phenomenology, with the important distinction of being CP-even. In this paper, we take a similar approach as presented in~\cite{Chakraborty:2021wda} with the following differences. The minimal heavy QCD axion generates a flavor-changing neutral current process at the two-loop order. Such diagrams can give rise to ultraviolet (UV), infrared (IR), and UV-IR mixed divergences. Working in the dimensional regularization scheme with the $\overline{\text{MS}}$ scheme, we show that the UV and UV-IR mixed divergences are canceled by incorporating appropriate counterterms. Meanwhile, the purely IR divergences cancel through the process of matching with the effective theory written at the electroweak scale.  In addition to the full two-loop and one-loop matching coefficients, we also improved the previous analysis by computing additional pieces in the anomalous dimension matrix. This helps us to consider different UV complete scenarios motivated by KSVZ (Kim-Shifman-Vainshtein-Zakharov)~\cite{Kim:1979if,Shifman:1979if}, DFSZ (Dine-Fischler-Srednicki-Zhitnitsky)~\cite{Dine:1981rt,Zhitnitsky:1980tq} and flavorful axions ~\cite{Ema:2016ops, Berezhiani:1989fp}. In contrast,~\cite{Chakraborty:2021wda} only considered the most minimal case of heavy QCD axions. We emphasize that we only consider the UV values of the Wilson coefficients from these frameworks while deviating away from the axion-pion mass relation. From the perspective of phenomenology, we present limits and projections for these frameworks by including only prompt signatures of ALPs. In addition, displaced vertex signatures of light new physics particles such as heavy sterile neutrinos~\cite{Dib:2019tuj}, light neutralinos~\cite{Dey:2020juy}, dark photons~\cite{Duerr:2019dmv,Ferber:2022ewf,Galon:2022xcl,Bandyopadhyay:2022klg} as well as ALPs~\cite{Bertholet:2021hjl} have gained a lot of attention in the recent past. A detailed study on such signatures for different UV-complete scenarios will be treated separately.

   We organize the paper in the following way. Section~\ref{sec:eff_lag} presents the effective Lagrangian pertinent to the study of the $b \to s$ transition in the context of an ALP. This is followed by a discussion of the renormalization group equations in Section~\ref{sec:RGEs}. We provide a detailed analysis of the dominant contributions to the anomalous dimension matrix. We emphasize that heavy QCD axions contribute to the $b \to s$ transition at the two-loop level. The cancellation of UV divergences is explicitly demonstrated through counterterms. These counterterms are nothing but operators defined at the UV scale itself, which renormalize the theory. Once identified, we calculate upto the leading two-loop contribution to the anomalous dimension matrix and compute the renormalization group evolution of the Wilson coefficients of such operators. This is most relevant for the heavy QCD axion and KSVZ-like scenarios, where at low energy the dominant interaction is the axion-gluon interaction. The finite terms of these two loop diagrams are presented in Section~\ref{sec:finite}, which also contains purely IR divergent pieces.  In Section~\ref{sec:Matching}, we explicitly show the cancelation of IR divergences by matching the theory to the effective theory at the electroweak scale $M_W$, as the IR divergence is the same between the two frameworks.  After computing the matching coefficients up to two-loop order, in Section~\ref{sec:AxionDecays} we provide results of the axion decay widths and branching fractions. The detailed discussion can be found in the Appendix~\ref{Axion_DW}. Next, we discuss the experimental methodologies for the relevant decay channels considered in our analysis in Section~\ref{sec:Exp}. This is followed by the presentation of the current and projected exclusion limits in the $m_a-f_a$ plane for several well-motivated UV frameworks such as KSVZ, DFSZ, etc. in Section~\ref{sec:results}. We consider different scenarios in which the UV scale is independent of $f_a$, as well as $\Lambda_{\text{UV}}=f_a$ or $4\pi f_a$. Finally, we conclude our findings in Section~\ref{sec:Conclusion}. In this work, we focus mainly on the hadronic production and decay modes of ALPs. We have also studied $a\to \gamma\gamma$ and $a\to\mu\mu$ prompt decay channels using the data from Belle and LHCb experiments. Such channels are less relevant for KSVZ, but more promising for DFSZ and Flavorful axion-like models due to non-zero tree-level couplings of the axion with fermions at the UV.
   
\section{The Effective Lagrangian}
\label{sec:eff_lag}

As mentioned in the Introduction, we start with an ALP, where $m_a\gg m_\pi f_\pi/f_a$. The effective Lagrangian takes the form 
\begin{equation}\label{eq:lag}
\mathcal{L} = \mathcal{L}_{\text{\tiny SM}} + \frac{1}{2} \left(\partial_\mu a\right)^2 -\frac{m_a^2}{2} a^2 + \sum_{n} \mathcal{C}_n \mathcal{O}_n\;,
\end{equation}
where $\cL_{\text{\tiny SM}}$ denotes the SM Lagrangian, and $\mathcal{C}_n$'s are the Wilson Coefficients corresponding to the operators $\mathcal{O}_n$. Notice that the bare-mass term for the axions breaks the axion-pion mass relation. At a fundamental level, this also breaks the shift symmetry, which might jeopardize the solution to the strong-$CP$ problem, requiring an appropriate shift in the axion field itself~\cite{Lella:2024gqc}.

For our $b\to sa$ study, we first take into account all the relevant operators pertaining to flavor changing neutral current processes, such as 
\begin{equation}
\begin{aligned}
    \mathcal{O}_1  = \frac{a}{f_a}\; G_{\mu\nu}^b \widetilde{G}^{\mu\nu b} \;, \quad
    \mathcal{O}_{2L/R}^{ij}  = \frac{\partial_\mu a}{f_a}\ \bar{u}_i \gamma^\mu P_{L/R}\, u_j \;, \quad \mathcal{O}_{3L/R}^{ij}  = \frac{\partial_\mu a}{f_a}\ \bar{d}_i \gamma^\mu P_{L/R}\, d_j \;,
\end{aligned}
 \label{eq:operators}
\end{equation}
where $\widetilde{G}^{\mu\nu b}\equiv (1/2)\epsilon^{\mu\nu\rho\sigma}G_{\rho\sigma}^b$, with $G_{\mu\nu}^b$ being the field strength tensor for the gluon fields, while $i, j$ are generation indices running over up- or down-type quark flavors ($\epsilon^{0123} = +1$). Here, $P_L = (1 - \gamma_5)/2$ and $P_R = (1 + \gamma_5)/2$. In this work, $L$ and $R$ will denote left- and right-handed states, respectively. For convenience in incorporating electroweak effects, we choose the chiral basis involving $L$ and $R$ fermions, rather than working with vector and axial vector currents. The operators of Eq.~\eqref{eq:operators} are generated at the new physics scale $\Lambda_{\text{UV}}$~\footnote{Apart from the operators mentioned in Eq.~\eqref{eq:operators}, axions can also have Wess-Zumino-Witten (WZW) interactions~\cite{Chakraborty:2024tyx,Bai:2024lpq} (also see~\cite{Harvey:2007rd,Harvey:2007ca,Chakraborty:2023wgl} for SM WZW interactions). We ignore them as they do not play a significant role in the construction or phenomenology under consideration.}. Moreover, axion interactions with leptons are also possible and well motivated in many UV complete frameworks. We consider such contributions as well. 

In addition, axions can also interact with the Higgs field via the dimension 5 operator $\mathcal{O}_H \equiv \partial^\mu a\; H^\dagger i\overleftrightarrow D_\mu H $. However, this particular operator is redundant and can be rotated away using field redefinition~\cite{Georgi:1986df}. 
Still, the same operator can be generated via RG evolution~\cite{Bauer:2020jbp}, and we provide a detailed discussion of this in Appendix~\ref{app:RedOp}.

The Lagrangian Eq.~\eqref{eq:lag} is written in the mass basis. We assume that it has the SM gauge symmetries manifest when expressed in the flavor basis. The $SU(2)_L$ symmetry in particular imposes a relationship between the mass basis Wilson coefficients corresponding to left-handed up and down-type quark operators 
% changed normal C --> \mathcal{C}
\begin{equation}\label{eq:udrelation}
    \mathcal{C}^{ij}_{2L} = \sum_{k,l} V_{ik}\mathcal{C}^{kl}_{3L} V^*_{jl} \;,
\end{equation}
where $V_{ij}$ are the elements of the CKM matrix in the Standard Model. Thus, not all Wilson coefficients appearing in the mass basis Lagrangian are independent~\cite{Bauer:2020jbp}.  The relation Eq.~\eqref{eq:udrelation} is an imprint of the manifest $SU(2)_L$ symmetry in the flavor basis Lagrangian. This point will be made more explicit in Section~\ref{sec:results} when we discuss particular UV models. The discussion above will have important consequences for the RG equations.

\section{Renormalization Group Evolutions}
\label{sec:RGEs}
The Wilson coefficients, defined at the UV scale, must evolve to the electroweak scale using the renormalization group equations (RGEs). Since the observations are at the $B$-physics scale, these coefficients, in principle, would run again after matching, from the electroweak scale $M_W$ to the bottom mass scale $m_b$. Following the notation of~\cite{Buras:1992zv,Buras:2020xsm}, in this section, we first establish the general framework for the RGEs applicable to composite operators and then specifically address the RGEs relevant to the process under consideration.

\subsection{Generic Framework}
 Renormalization of the fields is required to remove any divergences associated with the computation of a matrix element, such as
 \begin{equation}
\begin{aligned}
     {G_{\mu}^{b}}^{(0)}  &= \cR_{gg}^{1/2} G_{\mu}^{b} \approx \qty(1 + \frac{1}{2} \delta \cR_{gg})G_{\mu}^{b} \,, \\
     q_{_L,_i}^{(0)}  &= \qty[\cR_{q,L}^{1/2}]_{ij} q_{_L,j} \approx \qty[I + \frac{1}{2}\delta \cR_{q,L}]_{ij}q_{_L,_j}\,, \\
     q_{_R,_i}^{(0)}  &= \qty[\cR_{q,R}^{1/2}]_{ij} q_{_R,j} \approx \qty[I + \frac{1}{2}\delta \cR_{q,R}]_{ij}q_{_R,_j}\,,
\end{aligned}
\label{eq:wave_function_renormalization}
\end{equation}
with $\cR = (I + \delta\cR)$, where $I$ is an Identity matrix, and $q$'s are the up or down type quark fields with a generation index $i$. Repeated indices imply summation in this section. This wavefunction renormalization relates bare fields to renormalized fields. We do not consider any axion field renormalization, as it is suppressed by additional powers of $f_a$. 

In addition, composite operators such as defined in Eq.~\eqref{eq:operators} often give rise to other divergent contributions and hence require operator renormalization described by~\cite{Buras:2020xsm}
\begin{equation}
    \mathcal{O}_m^{(0)} = \mathcal{Z}_{mn} \mathcal{O}_n\;. 
\end{equation}
Here, the superscript $(0)$ denotes operators composed of bare fields, while the absence of a superscript indicates renormalized operators. Consequently, the bare and renormalized amputated Green functions, defined by $\langle \hdots \rangle$, are connected by the following relationship:
\begin{align}
    \langle \mathcal{O}_m \rangle = \mathcal{Z}^{-1}_{mn}\; \mathcal{R}_{np} \langle \mathcal{O}_p \rangle^{(0)} \Rightarrow \langle \mathcal{O}_m\rangle^{(0)} = \mathcal{R}^{-1}_{mn}\mathcal{Z}_{np} \langle \mathcal{O}_p\rangle = \chi_{mp} \langle \mathcal{O}_p\rangle\;.
\end{align}
Here $\mathcal{R}$ is the wavefunction renormalization matrix of the composite operators. The complete renormalization matrix $\chi$, introduced above, is defined as $\chi \equiv  \mathcal{R}^{-1}\mathcal{Z}$ and renormalizes the matrix elements (amputated Green functions with insertion of operators). Finally, the effective UV amplitude can be written in terms of the renormalized matrix elements as
\begin{equation}
\label{eq:aeff1}
\mathcal{M}_{\text{UV}} = \mathcal{C}_n\langle \mathcal{O}_n\rangle = \mathcal{C}_n \mathcal{Z}^{-1}_{np}\mathcal{R}_{pm} \langle \mathcal{O}_m\rangle^{(0)}\;.
\end{equation}
On the other hand, the running of the Wilson coefficients $\mathcal{C}_m = (\mathcal{C}_1, \mathcal{C}_{2},\hdots)$ can be deduced by renormalizing in a distinct but equivalent procedure. The relationship between the bare and the renormalized Wilson coefficients is governed by the renormalization matrix $\mathcal{Z}^c_{ij}$, which is defined by 
\begin{equation}
\label{eq:barewilson}
    \mathcal{C}_m^{(0)} = \mathcal{Z}^c_{mn}\mathcal{C}_n\;.
\end{equation}
In this approach, instead of operator renormalization, one uses the {\it counterterm} method, leading to a finite renormalized effective amplitude
\begin{equation}
    \mathcal{M}_{\text{UV}} = \mathcal{C}_n \mathcal{Z}^c_{pn}\;\mathcal{R}_{pm} \langle \mathcal{O}_m\rangle^{(0)} \;.
\end{equation}
Compared with Eq.~\eqref{eq:aeff1}, we see $\mathcal{Z}^c_{pn} = \mathcal{Z}^{-1}_{np}$. As $\mathcal{C}_m^{(0)}$ are independent of the renormalization scale $\mu$, Eq.~\eqref{eq:barewilson} leads to the running of the renormalized Wilson coefficients in terms of the anomalous dimension matrix (ADM)~\cite{Pich:1998xt,Buras:2020xsm}, defined to be $\gamma$, as
\begin{equation}
\label{eq:RGEequation}
\mu\frac{d\mathcal{C}_m(\mu)}{d\mu} = (\gamma^T)_{mn}\; \mathcal{C}_n(\mu)\;, \quad     \gamma \equiv \mathcal{Z}^{-1}\mu \frac{d \mathcal{Z}}{d\mu}\;,
\end{equation}
where the elements in $m$th column of $\gamma$ gives the RGE for $\mathcal{C}_m$.
 
In $\overline{\text{MS}}$, one can now expand these matrices in orders of $1/\epsilon$ poles as
\begin{equation}
\begin{aligned}
    \mathcal{Z}  &= I + \frac{1}{\epsilon}\mathcal{Z}_{(1)}(\alpha_s, \alpha_w, \alpha_t) + \frac{1}{\epsilon^2}\mathcal{Z}_{(2)}(\alpha_s, \alpha_w, \alpha_t) + \hdots \;, \\
     \mathcal{R} &= I + \frac{1}{\epsilon}\mathcal{R}_{(1)}(\alpha_s, \alpha_w, \alpha_t) + \frac{1}{\epsilon^2}\mathcal{R}_{(2)}(\alpha_s, \alpha_w, \alpha_t) + \hdots \;,  \\
     \chi & = I + \frac{1}{\epsilon}\chi_{(1)}(\alpha_s, \alpha_w, \alpha_t) + \frac{1}{\epsilon^2}\chi_{(2)}(\alpha_s, \alpha_w, \alpha_t) + \hdots \;,
\end{aligned}
\label{eq:Renormalization_matrices_expansion}
\end{equation}
where, $\mathcal{Z}_{(1)}$ represents the co-efficient of $1/\epsilon$ poles and so on. It is now straightforward to show that $\gamma(\alpha_s,\alpha_w, \alpha_t)$ only depends on the $1/\epsilon$ pole pieces
\begin{equation}\label{eq:ADMfirst}
    \gamma(\alpha_s, \alpha_w, \alpha_t) = -2\alpha_s \frac{\partial \mathcal{Z}_{(1)}}{\partial \alpha_s} - 2\alpha_w \frac{\partial \mathcal{Z}_{(1)}}{\partial \alpha_w} - 2\alpha_t\frac{\partial \mathcal{Z}_{(1)}}{\partial \alpha_t} \;,
\end{equation}
where $\alpha_s, \alpha_w$ are the coupling constants for the strong and weak force while $\alpha_t$ is related to top Yukawa coupling: $\alpha_t \equiv y_t^2/4\pi\;.$ These couplings will be relevant to this work.

As $\mathcal{Z} = \mathcal{R}\chi$, this easily implies $\mathcal{Z}_{(1)} = \mathcal{R}_{(1)} + \chi_{(1)}$. Therefore, we only require the $1/\epsilon$ poles of the wavefunction renormalization and the full renormalization diagrams to compute $\gamma$. Expanding the relevant matrices, i.e., $R_{(1)}, \chi_{(1)}$ and $\gamma$ order-by-order in the strong and electroweak gauge couplings and the Yukawa coupling, we obtain 
\begin{equation}
\begin{aligned}
    \mathcal{R}_{(1)} &= \frac{\alpha_s}{4\pi} \mathcal{A}_s + \frac{\alpha_w}{16\pi }\mathcal{A}_w + \frac{\alpha_t}{16\pi }\mathcal{A}_t + \frac{\alpha_s\alpha_w}{64\pi^2}\mathcal{A}_{sw} + \hdots \;, \\
    \chi_{(1)} &= \frac{\alpha_s}{4\pi}\mathcal{B}_s + \frac{\alpha_w}{16\pi }\mathcal{B}_w + \frac{\alpha_t}{16\pi }\mathcal{B}_t  + \frac{\alpha_s\alpha_w}{64\pi^2}\mathcal{B}_{sw} + \hdots \;, \\
    \gamma &= \frac{\alpha_s}{4\pi}\gamma_s + \frac{\alpha_w}{16\pi }\gamma_w + \frac{\alpha_t}{16\pi }\gamma_t + \frac{\alpha_s\alpha_w}{64\pi^2}\gamma_{sw} + \hdots \;,
    \end{aligned}
    \label{eq:totalADM}
\end{equation}
where, evidently
\begin{equation}
\begin{aligned}
    \gamma_s &= -2\left(\mathcal{A}_s + \mathcal{B}_s\right) \;,  \quad~ \gamma_w = -2\left(\mathcal{A}_w + \mathcal{B}_w\right) \;,\\
    \gamma_t &= -2\left(\mathcal{A}_t + \mathcal{B}_t\right) \;, \quad  \gamma_{sw} = -4\left(\mathcal{A}_{sw} + \mathcal{B}_{sw}\right)\;.
\end{aligned}
\label{eq:ADMdecomposition}
\end{equation}
 The elements $\mathcal{A}(m,n)$ at a given coupling order are derived from the self-energy diagrams at that order which renormalize the SM fields of $\mathcal{O}_m$, changing it to $\mathcal{O}_n$ ; for axion, $a^{(0)} \approx a$ upto leading order in $1/f_a$. Indices $m, n$ are generic and uniquely identify an operator. In contrast, the elements of $\mathcal{B}(m,n)$ are obtained from diagrams generating operator $\mathcal{O}_n$ by insertion of operator $\mathcal{O}_m$, where $m$ can also be the same as $n$. Finally, Eq.~\eqref{eq:ADMdecomposition} gives the anomalous dimension matrix of that particular order. We provide the general structure of the field strength renormalization and full renormalization matrices obtained in this work in a compact form using flavor indices $i,j$ and $k,l$.

\begin{table}[H]
\centering
\begin{adjustbox}{max width=\textwidth}
\begin{tabular}{c|ccccc}
 &  $a\,G\,G$ & $ a\,\bar u_{_L,_k}\,u_{_L,_l}$ & $ a\,\bar u_{_R,_k}\,u_{_R,_l}$ & $ a\,\bar d_{_L,_k}\,d_{_L,_l}$  & $ a\,\bar d_{_R,_k}\,d_{_R,_l}$\\[0.1cm]\hline 
 \\ [- 2ex]
$a^{_{(0)}}\,G^{^{(0)}}G^{^{(0)}} $ & $\mathcal{A}_s(1,1)$  & 0  & 0  & 0  &  0 \\[0.2cm]
$a^{_{(0)}}\,\bar u_{_L,_i}^{_{(0)}}u_{_L,_j}^{_{(0)}}$ & 0  & $\mathcal{A}_{(s+w+t)}(2L, 2L)_{ij;kl} $  & 0  & 0 & 0\\[0.2cm]
$a^{_{(0)}}\,\bar u_{_R,_i}^{_{(0)}}u_{_R,_j}^{_{(0)}}$ & 0 & 0  & $\mathcal{A}_{(s+w+t)}(2R, 2R)_{ij;kl} $ & 0 & 0 \\[0.2cm]
$a^{_{(0)}}\,\bar d_{_L,_i}^{_{(0)}}d_{_L,_j}^{_{(0)}}$ & 0 & 0  & 0 & $\mathcal{A}_{(s+w)}(3L, 3L)_{ij;kl}$ & 0  \\[0.2cm]
$a^{_{(0)}}\,\bar d_{_R,_i}^{_{(0)}}d_{_R,_j}^{_{(0)}}$ & 0 & 0 & 0 &  0 &  $\mathcal{A}_{(s+w)}(3R, 3R)_{ij;kl} $ \\[0.2cm]
\hline

\end{tabular}
\end{adjustbox}
\caption{Structure of the wavefunction renormalization matrix. We denote different up/down quark flavors by indices $i,j; k,l$ throughout. Subscripts $s$ and $w$ denote renormalization at the $\mathcal{O}(\alpha_s)$ and $\mathcal{O}(\alpha_w)$ via the gluon and electroweak gauge boson loop, respectively. While subscript $t$ denotes the top-Yukawa contribution to renormalization. The matrix is completely block diagonal. We ignore electromagnetic contributions in our analysis.} 
\label{tab:I}
\end{table}

\begin{table}[H]
\centering
\begin{adjustbox}{max width=\textwidth}
\begin{tabular}{c|ccccc}
 & $\langle \mathcal{O}_1 \rangle^{tree}$ & $\langle \mathcal{O}_{2L}^{kl}\rangle^{tree}$ & $\langle \mathcal{O}_{2R}^{kl}\rangle^{tree}$ & $\langle \mathcal{O}_{3L}^{kl}\rangle^{tree}$  & $\langle \mathcal{O}_{3R}^{kl}\rangle^{tree}$ \\[0.1cm]\hline
  \\ [- 2ex]
$\langle\mathcal{O}_1\rangle^{(0)}$ & $\mathcal{B}_s(1,1)$  & $\mathcal{B}_s(1,2L)_{;kl}$   & $\mathcal{B}_s(1,2R)_{;kl}$   & $\boxed{\mathcal{B}_{(s + sw)}(1,3L)_{;kl}}$  & $\mathcal{B}_s(1,3R)_{;kl}$   \\[0.2cm]
$\langle \mathcal{O}_{2L}^{ij}\rangle^{(0)}$ & 0   & $\mathcal{B}_{(s+w)}(2L,2L)_{ij;kl}$ & $\mathcal{B}_{(w + t)} (2L, 2R)_{ij;kl}$ & $\mathcal{B}_w(2L, 3L)_{ij;kl}$ &  $\mathcal{B}_w(2L, 3R)_{ij;kl}$  \\[0.2cm]
$\langle \mathcal{O}_{2R}^{ij}\rangle^{(0)}$ & 0 & $\mathcal{B}_{ (w + t) } (2R, 2L)_{ij;kl}$  & $\mathcal{B}_{(s+w)}(2R, 2R)_{ij;kl}$  & $\mathcal{B}_w(2R, 3L)_{ij;kl}$ & 0 \\[0.2cm]
$\langle \mathcal{O}_{3L}^{ij}\rangle^{(0)}$ & 0 & $\mathcal{B}_w(3L, 2L)_{ij;kl}$  & $\mathcal{B}_w(3L, 2R)_{ij;kl}$ & $\mathcal{B}_{(s+w)}(3L, 3L)_{ij;kl}$ & $\mathcal{B}_w(3L, 3R)_{ij;kl}$ \\[0.2cm]
$\langle \mathcal{O}_{3R}^{ij}\rangle^{(0)}$ & 0  & $\mathcal{B}_w(3R, 2L)_{ij;kl}$  & 0  &$\mathcal{B}_w(3R, 3L)_{ij;kl}$ & $\mathcal{B}_{(s+w)}(3R, 3R)_{ij;kl}$ \\[0.2cm]
\hline
\end{tabular}
\end{adjustbox}
\caption{Structure of the full renormalization matrix. Subscripts and notations are the same as discussed in Table~\ref{tab:I} and the text. The two-loop RG contribution to $b\to sa$ is contained inside the boxed entry (indices $k=2, l=3$). The reason for some entries being identically zero will be explained in later sections. We will also show that some entries appearing here are proportional to light quark masses; hence, their contributions to RGEs can be neglected.}
\label{tab:II}
\end{table}

In Table~\ref{tab:I}, we write down the wavefunction renormalization of our composite operators Eq.\eqref{eq:operators}, whereas in Table~\ref {tab:II}, we provide the complete structure of the renormalization matrix. The subscripts $s, w$ and $t$ represent renormalization at the $\alpha_s, \alpha_w$ and $\alpha_t$ order, respectively. For example, $\mathcal{B}_{sw}(1,3L)_{;23}$ represents full renormalization at $\mathcal{O}(\alpha_s\alpha_w)$. This particular element encapsulates the contribution to the $b-s-a$ operator from the prototypical $a-G-\widetilde{G}$ operator. The leading order effect arises at the two-loop level, explaining the order of $\alpha$. In addition, depending on the particular operator structures, the indices $i,j;k, l$ represent the flavor of the up-/down-type quarks. The entries without any flavor index represent a single element of the matrix. Those with indices $;kl$ are row matrices containing nine elements each, while those with both sets $ij; kl$ of indices are $9\times 9$ matrices containing $81$ elements each.

\subsection{Anomalous dimension matrix}
To determine the dominant contributions to the anomalous dimension matrix, we adhere to the following schemes:

\begin{enumerate}[leftmargin=*]
\item \textbf{Naive Dimensional Regularization:} Throughout the paper, we work in the Feynman gauge and employ the widely used naive dimensional regularization~\cite{Buras:1989xd} in $d = 4-2\epsilon$ dimensions to regulate the UV divergences. We also use $d$-dimensional Dirac algebra for all $\gamma^\mu$ matrices, i.e.,
\begin{equation}
\label{eq:gamma_algebra}
    \left\{\gamma_\mu,\gamma_\nu\right\} = 2 g_{\mu\nu}\;, \; \; \text{and} \; \; \left\{\gamma_\mu,\gamma_5\right\} = 0\;.
\end{equation}
Furthermore, explicit relations between the $\epsilon$ tensor and $\gamma_5$ are used only when all UV divergences are evaluated in terms of $1/\epsilon$ poles and we return to $d=4$. We note in passing that Eq.~\eqref{eq:gamma_algebra} is valid as long as we are not dealing with any closed-parity fermion loop. Such diagrams are usually finite and do not affect the determination of the anomalous dimension matrix. A more rigorous treatment, involving the 't Hooft-Veltman scheme~\cite{tHooft:1972tcz}, will be addressed in future work~\cite{HV2loop}.

\item We consider the limit in which all external momenta and light-quark masses are set to zero. This approximation induces infrared divergences, which we regulate separately. We also expand about the axion momenta $l = 0$ and take only the leading non-vanishing terms in all diagrams. We have checked that the inclusion of non-zero quark masses introduces a correction of order $(m_b/M_W)^2$. Additionally, our approximation simplifies the analysis by avoiding complexities associated with two-loop master integrals in the presence of non-vanishing external momentum.

\item {\bf NDR with mass regularization:} To regularize the IR divergences in scaleless integrals or otherwise, we use a fictitious IR mass $m_R$ for the gluon propagators  within the loop diagrams. A naive power counting of the light-quark momenta indicates that an IR mass for the light quarks is not required. This is also helpful as any IR mass for light quarks might generate chirality flipping evanescent operators. As a result, we make the following substitutions to the gluon propagator:
\begin{equation}\label{eq:IRprop}
\frac{-ig_{\mu\nu}}{\widetilde{q}^2} \to \frac{-ig_{\mu\nu}}{\widetilde{q}^2 - m_R^2}\;, 
\end{equation}
where $\widetilde{q}$ can be a generic combination of the loop momenta $q$ with some external momenta. Choices for other infrared (IR) regulators, such as the assignment of nonzero off-shell momenta to the external legs, may also be considered. However, these approaches can result in non-trivial master integrals.

\item Renormalization of composite operators deals with matrix elements which are amputated Green's functions~\cite{Buras:2020xsm, Buras:1992zv, Collins:1984xc}. As a result, our analysis requires only one-part irreducible (1PI) diagrams to be considered. 
\end{enumerate}

We will now proceed with the calculation of different elements in the ADM, i.e. $\mathcal{A}$ and $\mathcal{B}$ elements at each order.

\subsubsection{$\mathcal{O}(\alpha_s)$}

{\underline{\bf $\mathcal{A}_s$ elements:}}  At $\mathcal{O}(\alpha_s)$ order, the wave function renormalization for the SM gluon and quark fields are well established and given by
\begin{equation}
    \mathcal{R}_{gg} = 1 + \frac{\alpha_s}{4\pi \epsilon} (\beta_0 - 2C_A) \;, \quad  \mathcal{R}_{qq} = 1 - \frac{\alpha_s C_F}{4\pi\epsilon}\;,
    \label{R_s}
\end{equation}
where $\beta_0 = 11C_A/3 - 2N_f/3$ and $N_f = 6$ are the number of quark flavors and $C_A = 3, C_F = 4/3$, are the quadratic Casimirs. The renormalization of the gluon wavefunction $\cR_{gg}$ gives the entry as $\mathcal{A}_s(1,1) = \beta_0 - 2C_A$ since the $a-G-\widetilde{G}$ operator is quadratic in the gluon field. We note that the gluon and quark wave functions do not mix under field renormalization, resulting in all other entries in the first row and column of Table~\ref{tab:I} being zero at any order in the couplings. The renormalization of the quark field $\mathcal{R}_{qq}$ in $\mathcal{O}(\alpha_s)$ contributes uniformly to all other diagonal entries in $\mathcal{A}_s$, since it is independent of the handedness, flavor, or charge of the quark.  Moreover, each diagonal entry itself is a diagonal matrix as there is no mixing between different flavors of the same handedness or charge. Therefore, all non-diagonal entries of $\mathcal{A}_s$ are zero and the diagonal entries are: 
\begin{equation}
\begin{aligned}
& \mathcal{A}_s(1,1)  = \beta_0 - 2C_A\;,  \\
& \mathcal{A}_s (2L, 2L) = \mathcal{A}_s(2R, 2R) = \mathcal{A}_s(3L, 3L) = \mathcal{A}_s (3R, 3R) = -C_F \delta_{ik}\delta_{jl}\;.
\end{aligned}
\label{eq:As}
\end{equation}
Note that the operator with bare fields has flavor indices $i,j$ and the one with renormalized fields has $k,l$, so all the $\mathcal{A}_s (m, n)$ have the same indices as in Table~\ref{tab:I}.

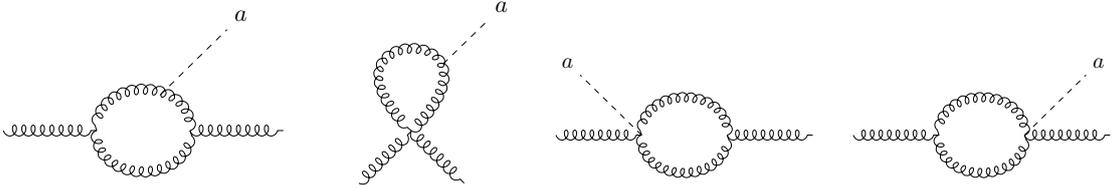
\begin{figure}[H]
\begin{center}
\resizebox{3.8cm}{!}{%
\begin{tikzpicture}
    \begin{feynman}
    \vertex (a) ;
    \vertex [right=of a] (b);
    \vertex [right=of b] (c);
    \vertex [right=of c] (d) ;
    \vertex at ($(b)!1.0!(c)!0.5!-50:(b)$) (e) ;
    \vertex [above right=of e] (f) {\(a\)};
    \diagram* {
      (a) -- [gluon] (b) -- [gluon, half right] (c) -- [gluon] (d), (c) -- [gluon, half right] (b), (e) -- [scalar] (f)
    };
    \end{feynman}
\end{tikzpicture}
}
\resizebox{3.0cm}{!}{%
\begin{tikzpicture}
    \begin{feynman}
    \vertex (b) ;
    \vertex [below left= 1.2cmof b] (a);
    \vertex [below right=1.2cm of b] (c);
    \vertex[above=0cm of b] (d);
    \vertex at ($(b)!0!(c)!1!110:(c)$) (e);
    \vertex [above right=1cm of e] (f) {\(a\)}; ;
    \diagram* {
      (a) -- [gluon ] (b) -- [gluon ] (c), (b) -- [gluon, out = 45, in = 135, distance = 2.5cm] (d), (e) -- [scalar] (f)
    };
    \end{feynman}
\end{tikzpicture}
}\hspace{0.2cm}
\resizebox{3.5cm}{!}{%
\begin{tikzpicture}
    \begin{feynman}
    \vertex (a) ;
    \vertex [right=of a] (b);
    \vertex [right=of b] (c);
    \vertex [right=of c] (d) ;
    \vertex [above left=of b] (e) {\(a\)};
    \diagram* {
      (a) -- [gluon ] (b) -- [gluon, half right  ] (c) -- [gluon] (d), (c) -- [gluon, half right ] (b), (b) -- [scalar] (e)
    };
    \end{feynman}
\end{tikzpicture}
}\hspace{0.3cm}
\resizebox{3.5cm}{!}{%
\begin{tikzpicture}
    \begin{feynman}
    \vertex (a) ;
    \vertex [right=of a] (b);
    \vertex [right=of b] (c);
    \vertex [right=of c] (d) ;
    \vertex [above right=of c] (e) {\(a\)};
    \diagram* {
      (a) -- [gluon ] (b) -- [gluon, half right  ] (c) -- [gluon] (d), (c) -- [gluon, half right] (b), (c) -- [scalar] (e)
    };
    \end{feynman}
\end{tikzpicture}
}
\end{center}
\caption{The diagrams contributing to $\mathcal{B}_s(1,1)$. The dashed line represents the axion. The Lorentz and color indices of gluons are not shown explicitly since the operator structure does not enter in the elements of the ADM.}
\label{fig:2}
\end{figure}

\noindent{\underline{\bf $\mathcal{B}_s$ elements:}} To determine the $\mathcal{B}_s$ entries, we evaluate diagrams where the operator $\mathcal{O}_n$ is generated by inserting the operator $\mathcal{O}_m$ with a gluon loop ($m$ can be the same as $n$). For example, $\mathcal{B}_s(1,1)$ represents the pole structure associated with the generation of the operator $\mathcal{O}_1$ due to the insertion of the same operator, as illustrated in Fig.~\ref{fig:2}. 

At $\mathcal{O}(\alpha_s)$, axion-fermion insertions of $\mathcal{O}_{2L/R}^{ij}$, $\mathcal{O}_{3L/R}^{ij}$ can also generate $\mathcal{O}_1$ when $i=j$. However, such diagrams involve triangle fermionic loops and are finite, not contributing to the entries of $\mathcal{B}_s$. When $i\neq j$ there are no diagrams in this order to generate $\mathcal{O}_1$. Therefore, the other entries in the first column of $\mathcal{B}_s$ are zero. 
\color{black}
\begin{figure}[H]
\centering
\resizebox{5.5cm}{!}{%
\begin{tikzpicture}
    \begin{feynman}
    \vertex (a) {\(q_{_{l}}\)};
    \vertex [right=of a] (b);
    \vertex [right=of b] (c);
    \vertex [right=1cm of c] (d) {\(q_{_{k}}\)};
    \vertex at ($(b)!1.0!(c)!0.5!-50:(b)$) (e) ;
    \vertex [above right=of e] (f) {\(a\)};
    \vertex [below=of e] (h) {};
    \diagram* {
      (a) -- [fermion] (b) -- [fermion] (c) -- [fermion] (d), (c) -- [gluon, half right, looseness = 1.5] (b), (e) -- [scalar] (f)
    };
    \end{feynman}
\end{tikzpicture}
}
\hspace{3ex}
\resizebox{5.5cm}{!}{%
\begin{tikzpicture}
    \begin{feynman}
    \vertex (a) {\(q_{_{l}}\)};
    \vertex [right=of a] (b) ;
    \vertex [right=0.9cm of b] (c);
    \vertex [right=0.9cm of c] (d) ;
    \vertex [right= of d] (e) {\(q_{_{k}}\)};
    \vertex [below right = 1.1cm of c] (f) {\(a\)} ;
    \diagram* {
        (a) -- [fermion] (b) -- [fermion, edge label' = \(q_{_{j}}\)] (c) -- [fermion, edge label' = \(\quad q_{_{i}}\)] (d) -- [fermion] (e), (d) -- [gluon, half right] (b), (c) -- [scalar] (f)
    };
    \end{feynman}
\end{tikzpicture}
}
\caption{Left panel: Diagram for the first row entries: $\mathcal{B}_s(1,2L/2R)$ and $\mathcal{B}_s(1, 3L/3R)$. Right panel: Diagram for evaluating diagonal entries of $\mathcal{B}_s$.}
\label{fig:3}
\end{figure}
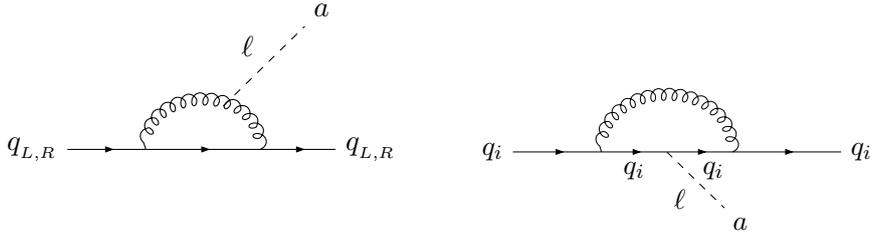
On the other hand, $\mathcal{O}_1$ insertion generates all axion-quark diagonal operators at $\mathcal{O}(\alpha_s)$ via the type of diagram shown to the left of Fig.~\ref{fig:3}, thereby giving rise to non-zero first row entries except those requiring a flavor change. We note that the Dirac structure of the left- and right-handed fermion-axion interaction operators of our Lagrangian is $(-\slashed l P_L)$ and $(-\slashed l P_R)$, and we take the axion's momentum $l$ to be outgoing for all diagrams (see Appendix~\ref{app:FD}). This leads to a relative sign difference in these elements when a diagram result shows in terms of a $\slashed l \gamma_5$ Dirac structure. Finally, axion-fermion operators arise from the insertion of the same operators as depicted in the right panel of Fig.~\ref{fig:3}.

\color{black}
These exhaust all non-zero contributions to $\mathcal{B}_s$. We find
\begin{equation}
\begin{aligned}
&\mathcal{B}_s(1,1) = 2C_A \;,\\
&\mathcal{B}_s(1,2L) = \mathcal{B}_s(1, 3L) = -\mathcal{B}_s(1,2R) = -\mathcal{B}_s(1, 3R) = 6C_F \delta_{kl} \;, \\
& \mathcal{B}_s(2L, 2L) = \mathcal{B}_s(2R, 2R) = \mathcal{B}_s(3L, 3L) = \mathcal{B}_s(3R, 3R) = C_F \delta_{ik}\delta_{jl} \;.
\end{aligned}
\label{eq:Bs}
\end{equation}

All $\mathcal{B}_w(m,n)$ have the same flavor indices as in Table~\ref{tab:II}. Recall that, at the $\mathcal{O}(\alpha_s)$ order, the ADM is defined as $\gamma_s = -2\left(\mathcal{A}_s+\mathcal{B}_s\right)$, which we obtain using Eq.~\eqref{eq:As},~\eqref{eq:Bs}.

We find that $\mathcal{C}_1$ and all diagonal quark Wilson coefficients $\mathcal{C}_{2L/R}^{ii}, \mathcal{C}_{3L/R}^{ii}$ receive their leading-order RGE contribution at $\mathcal{O}(\alpha_s)$. We will show that the leading-order contribution for the running of the off-diagonal coefficients, in particular the $b \to sa$ Wilson coefficient $\mathcal{C}_{3L}^{23}$ (or $\mathcal{C}_{3L}^{sb}$) appears at $\mathcal{O}(\alpha_w)$ and $\mathcal{O}(\alpha_s\alpha_w)$. As is standard, the flavor indices $i,j,k,l = \{1,2,3\}$ correspond to $\{d,s,b\}$ for down-type and $(u,c,t)$ for up-type quarks, respectively.

\subsubsection{$\mathcal{O}(\alpha_w)$}

\begin{figure}[h]
\begin{center}
\resizebox{4.6cm}{!}{%
    \begin{tikzpicture}
        \begin{feynman}
            \vertex (a) {\(u_i\)};
            \vertex [right=1cm of a] (b);
            \vertex [right=of b] (c);
            \vertex [right=1cm of c] (d) {\(u_j\)};
            \vertex [above right=of b] (e);

            \diagram* {
            (a) -- [fermion] (b) -- [fermion, edge label' = {$d,s,b$}] (c) -- [fermion] (d), (b) -- [boson, half left, edge label = \(W\) ] (c)
            };
        \end{feynman}
    \end{tikzpicture}
    }
    \hspace{2ex}
    \resizebox{4.6cm}{!}{%
\begin{tikzpicture}
        \begin{feynman}
            \vertex (a) {\(d_i\)};
            \vertex [right=1cm of a] (b);
            \vertex [right=of b] (c);
            \vertex [right=1cm of c] (d) {\(d_j\)};
            \vertex [above right=of b] (e);

            \diagram* {
            (a) -- [fermion] (b) -- [fermion, edge label' = {$u,c,t$}] (c) -- [fermion] (d), (b) -- [boson, half left, edge label = \(W\) ] (c)
            };
        \end{feynman}
    \end{tikzpicture}
    }
    \hspace{2ex}
    \resizebox{4.6cm}{!}{%
    \begin{tikzpicture}
        \begin{feynman}
            \vertex (a) {\(q_i\)};
            \vertex [right=1cm of a] (b);
            \vertex [right=of b] (c);
            \vertex [right=1cm of c] (d) {\(q_i\)};
            \vertex [above right=of b] (e);

            \diagram* {
            (a) -- [fermion] (b) -- [fermion, edge label' = \(q_i\)] (c) -- [fermion] (d), (b) -- [boson, half left, edge label = \(Z\)] (c)
            };
        \end{feynman}
    \end{tikzpicture}
    }
\end{center}
\caption{Diagrams contributing to fermionic diagonal elements of $\mathcal{A}_w$. We work in the Feynman gauge, where Goldstone loops of $W$ and $Z$ denoted by $W_G$ and $Z_G$ are also considered. We do not label them explicitly in the diagrams. The left and middle diagrams can change quark flavor.}
    \label{fig:5}
\end{figure}
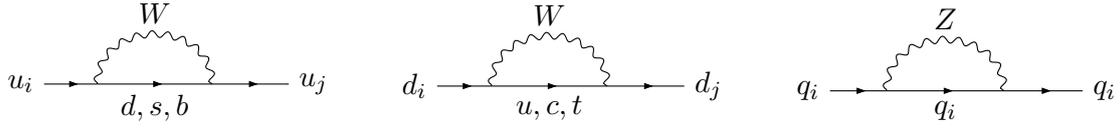

{\underline{\bf $\mathcal{A}_w$ elements:}} The gluon field doesn't mix with itself and the quark field at this order (i.e. via weak interaction), so the first row and column of $\mathcal{A}_w$ have all elements equal to zero. However, the renormalization of the wave function of quark operators receives contributions from the $W$ and $Z$ quark self-energy graphs, as shown in Fig.~\ref{fig:5}.
\color{black}
The resulting renormalizations from the diagrams of Fig.~\ref{fig:5} are~\footnote{In these equations and the following sections, summations on flavor indices are explicitly shown and repeated indices alone \textit{do not} imply summation.}
\begin{equation}
\begin{aligned}
    [\cR_{u,L}]_{ij} &= \delta_{ij} - \frac{\alpha_w}{16\pi \epsilon}\qty[\sum_{r = d,s,b} V_{ir}^\ast V_{jr}  (2 + \xi_r) +  \qty{\qty(\frac{c_{L}^{u}}{c_w})^2 + \frac{\xi_{i}}{2}} \delta_{ij}] \;, \\
    [\cR_{u,R}]_{ij} &= \delta_{ij} - \frac{\alpha_w}{16\pi \epsilon} \qty[\sum_{r=d,s,b} V_{ir}^\ast V_{jr} \sqrt{\xi_i \xi_j} +  \qty{\qty(\frac{c_{R}^{u}}{c_w})^2 + \frac{\xi_{i}}{2}} \delta_{ij}]\;, \\ 
    [\cR_{d,L}]_{ij} &= \delta_{ij} - \frac{\alpha_w}{16\pi \epsilon}\qty[ \sum_{r=u,c,t} V^\ast_{rj} V_{ri} (2 + \xi_r) +  \qty{\qty(\frac{c_{L}^{d}}{c_w})^2 + \frac{\xi_{i}}{2}} \delta_{ij}]\;, \\
    [\cR_{d,R}]_{ij} &= \delta_{ij} - \frac{\alpha_w}{16\pi \epsilon} \qty[\sum_{r=u,c,t} V_{rj}^\ast V_{ri} \sqrt{\xi_i \xi_j} +  \qty{\qty(\frac{c_{R}^{d}}{c_w})^2 + \frac{\xi_{i}}{2}} \delta_{ij}]\;.
\end{aligned}
\label{eq:R_w}
\end{equation}
Here, $V_{ri}$' s are the elements of the CKM matrix~\footnote{Note: For CKM matrix elements, we always put up-type quarks at the row indices and down-type quarks at the column indices}, $\xi_i \equiv m_i^2/M_W^2$, and $c_w=\cos\theta_w$ with $\theta_w$ the weak mixing angle. $c^{u,d}_{L/R} $ are the usual $Z$ boson couplings with the left- and right-handed quarks. However, they do not appear in the RGEs, as they cancel out after adding the full renormalization while calculating the ADM.

The field renormalization matrix at $\mathcal{O}(\alpha_w)$ thus contains the following non-zero entries.
\begin{equation}
\begin{aligned}
\mathcal{A}_w(2L,2L) &= - \frac{1}{2}\qty[4\,\delta_{ik}\delta_{jl} + \sum_{r } (   V_{ir}^* V_{kr} \delta_{jl} + V_{lr}^* V_{jr} \delta_{ik}) \xi_r   +  \qty{2\qty(\frac{c_{L}^{u}}{c_w})^2 + \frac{\xi_{i} + \xi_j}{2}} \delta_{ik}\delta_{jl} ] \;, \\
\mathcal{A}_w(2R, 2R) &=  - \frac{1}{2}\qty[(\xi_i + \xi_j)\delta_{ik}\delta_{jl} + \qty{2\qty(\frac{c_{R}^{u}}{c_w})^2 + \frac{\xi_{i} + \xi_j}{2}} \delta_{ik}\delta_{jl}] \;, \\
\mathcal{A}_w(3L,3L) &= - \frac{1}{2}\qty[4\,\delta_{ik}\delta_{jl} + \sum_{r} (V_{rk}^* V_{ri} \delta_{jl} + V_{rj}^* V_{rl} \delta_{ik}) \xi_r   +  \qty{2\qty(\frac{c_{L}^{d}}{c_w})^2 + \frac{\xi_{i} + \xi_j}{2}} \delta_{ik}\delta_{jl}]\;,  \\
\mathcal{A}_w(3R, 3R) &=  - \frac{1}{2}\qty[(\xi_i + \xi_j)\delta_{ik}\delta_{jl} + \qty{2\qty(\frac{c_{R}^{d}}{c_w})^2 + \frac{\xi_{i} + \xi_j}{2}} \delta_{ik}\delta_{jl}] \;.
\end{aligned}
\label{eq:Aw}
\end{equation}
Firstly, we notice that wavefunction renormalization at this order is completely block diagonal; a left-handed quark cannot renormalize to right-handed, and vice versa, and a down-type quark cannot renormalize to up-type and vice versa. We also notice that the operators $\mathcal{O}^{ij}_{2R}$ only renormalize to $\mathcal{O}^{ij}_{2R}$ (flavor remains unchanged) as the Kronecker deltas imply $k=i$ and $l=j$. The same happens for $\mathcal{O}^{ij}_{3R}$. The reason is that $W$ bosons do not talk to right-handed quarks, so flavor can not change. However, $\mathcal{O}^{ij}_{2L}$ can mix with operators like $\mathcal{O}^{kj}_{2L}$ or $\mathcal{O}^{il}_{2R}$, resulting in a flavor change of one of the quark fields in the bare operator. The same flavor-changing effect is observed for the $\mathcal{O}^{ij}_{3L}$ operators. Notice that the flavors of both quarks coupling to the axions would not change, as that effect is of $\mathcal{O}(\alpha_w^2)$.\\

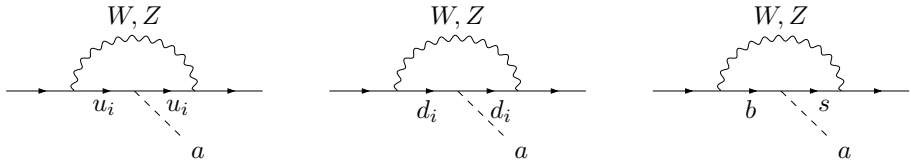
\begin{figure}[H]
\centering
\resizebox{3.5cm}{!}{%
\begin{tikzpicture}
    \begin{feynman}
    \vertex (a) ;
    \vertex [right=1cm of a] (b);
    \vertex [right=0.9cm of b] (c);
    \vertex [right=0.9cm of c] (d);
    \vertex [right= 1cm of d] (e) ;
    \vertex [below right = 1cm of c] (f) {\(a\)};
    \diagram* {
      (a) -- [fermion] (b) -- [fermion, edge label' = {$u_j$} ] (c) -- [fermion,  edge label' = {$\quad u_i$}] (d) -- [fermion] (e), (b) -- [boson, half left, edge label = {$W, Z$}] (d), (c) -- [scalar] (f)
    };
    \end{feynman}
\end{tikzpicture}
}
\hspace{3ex}
\resizebox{3.5cm}{!}{%
\begin{tikzpicture}
    \begin{feynman}
    \vertex (a) ;
    \vertex [right=1cm of a] (b);
    \vertex [right=0.9cm of b] (c);
    \vertex [right=0.9cm of c] (d);
    \vertex [right=1cm of d] (e) ;
    \vertex [below right = 1cm of c] (f) {\(a\)};
    \diagram* {
        (a) -- [fermion] (b) -- [fermion, edge label' = \(d_j\)] (c) -- [fermion, edge label' = \(\quad d_i\)] (d) -- [fermion,] (e), (b) -- [boson, half left, edge label = {$W,Z$}] (d), (c) -- [scalar] (f)
    };
    \end{feynman}
\end{tikzpicture}
}
\caption{Operator insertion diagrams for $\mathcal{B}_w$ entries. The left diagram is the insertion of $\mathcal{O}_{2L/R}$ in $W$ and $Z$ loops (along with their Goldstones). Explicit final quark states decide which operator is being generated. They depend on the insertion and the boson. 
The right diagram shows the insertion of $\mathcal{O}_{3L/R}$, i.e., the down-type axion quark flavor diagonal operators.}
\label{fig:7}
\end{figure}

\noindent{\underline{\bf $\mathcal{B}_w$ elements:}} At the $\mathcal{O}(\alpha_w)$ order, the complete renormalization introduces diagonal and off-diagonal entries. In particular, there is a generation of non-diagonal operators $\mathcal{O}_{3L}^{kl}$ from the insertion of $\mathcal{O}_{2L}^{ij}, \mathcal{O}_{2R}^{ij}$ which will lead to running of the $b\to sa$ coefficient at $\mathcal{O}(\alpha_w)$. The possible diagrams with operator insertions are shown in Fig.~\ref{fig:7}.

\color{black}
\noindent We thus obtain the full renormalization matrix $\mathcal{B}_w$ with the following non-zero entries

\footnotesize
\begin{equation}
\begin{aligned}
    \mathcal{B}_w(2L, 2L) &= \bigg(\frac{c_L^u}{c_w}\bigg)^2 \delta_{ik}\delta_{jl}\;,\qquad \mathcal{B}_w(2L, 2R) = \frac{\sqrt{\xi_i\xi_j}}{2}\delta_{ik}\delta_{jl} \;, \\
    \mathcal{B}_w(2L, 3L) & = 2 V_{ik}^*V_{jl} \;, \qquad \mathcal{B}_w(2L, 3R) =  V_{ik}^*V_{jl} \sqrt{\xi_k\xi_l} \;,  \\
    \mathcal{B}_w(2R, 2L) &= \frac{\sqrt{\xi_i\xi_j}}{2}\delta_{ik}\delta_{jl}\;, \quad \mathcal{B}_w(2R, 2R) =  \bigg(\frac{c_R^u}{c_w}\bigg)^2\delta_{ik}\delta_{jl}\;, \quad \mathcal{B}_w(2R, 3L) = V_{ik}^*V_{jl}\sqrt{\xi_i\xi_j} \;, \\
   \mathcal{B}_w(3L, 2L) &= 2V_{lj}^*V_{ki}\;, \qquad \mathcal{B}_w (3L, 2R) = V_{lj}^*V_{ki} \sqrt{\xi_k\xi_l} \;,\\
    \mathcal{B}_w(3L, 3L) &= \bigg(\frac{c_L^d}{c_w}\bigg)^2\delta_{ik}\delta_{jl}\;, \quad \mathcal{B}_w(3L, 3R) = \frac{\sqrt{\xi_i \xi_j}}{2}\delta_{ik}\delta_{jl} \;, \\
    \mathcal{B}_w(3R, 2L) &= V_{lj}^*V_{ki} \sqrt{\xi_i\xi_j}\;, \quad \mathcal{B}_w(3R, 3L) = \frac{\sqrt{\xi_i\xi_j}}{2}\delta_{ik}\delta_{jl}\;, \quad  \mathcal{B}_w(3R, 3R) = \bigg(\frac{c_R^d}{c_w}\bigg)^2\delta_{ik}\delta_{jl}\;.
\end{aligned}
\label{eq:Bw}
\end{equation}
\normalsize

\noindent The matrix $\gamma_w$ can now be obtained from Eq.~\eqref{eq:Aw},\eqref{eq:Bw}, and Eq.~\eqref{eq:ADMdecomposition}.

\subsubsection{$\order{\alpha_t}$}

Here, the top Yukawa contribution is the only relevant contribution compared to the other flavors because $y_t = \sqrt{2}m_t/v \approx 1$, where $v$ is the electroweak symmetry breaking scale. \\

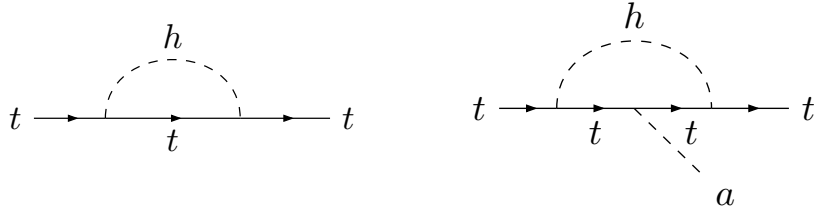
\begin{figure}[t]
\centering
\resizebox{5cm}{!}{%
\begin{tikzpicture}
    \begin{feynman}
        \vertex (a) {\(q_{_i}\)};
        \vertex [right=1cm of a] (b);
        \vertex [right=1.6cm of b] (c);
        \vertex [right=1cm of c] (d) {\(q_{_j}\)};
        \vertex [above right=of b] (e);
        \vertex [below right=1.2cm of c] (h) {};

        \diagram* {
        (a) -- [fermion] (b) -- [fermion, edge label' = \(q_{_r}\)] (c) -- [fermion] (d), (b) -- [scalar, half left, edge label = \(h\)] (c)
        };
    \end{feynman}
\end{tikzpicture}
}
\hspace{5ex}
\resizebox{4.5cm}{!}{%
\begin{tikzpicture}
    \begin{feynman}
    \vertex (a) {\(q_{_{l}}\)};;
    \vertex [right=0.8cm of a] (b);
    \vertex [right=0.8cm of b] (c);
 %   \vertex [right=1cm of c] (d);
    \vertex [right=0.8cm of c] (e);
    \vertex [right=0.8cm of e] (f) {\(q_{_{k}}\)};;
%    \vertex at ($(c)!0.5!(d)!0.5!135:(c)$) (g);
    \vertex [below right=1cm of c] (h) {\(a\)};
%    \vertex at ($(a)!0.5!(f)!0.9!90:(a)$) (i) {$(a)$};
        
    \diagram* {
            (a) -- [fermion] (b) -- [fermion, edge label' = \(q_{_{j}}\)] (c) -- [fermion, edge label' = \(\quad q_{_{i}}\)] (e) -- [fermion] (f), (b) -- [scalar, half left, edge label = \(h\)] (e), (c) -- [scalar] (h)
        };
    \end{feynman}
\end{tikzpicture}
}
\caption{The self-energy diagram for $A_t(2L,2L)$ and $A_t(2R,2R)$, and $\cO_{2L,2R}^{ij}$ operator inserted diagram for $\mathcal{B}_t(2L, 2R)$ and $\mathcal{B}_t(2R, 2L)$.}
\label{fig:Yukawa}
\end{figure}

\noindent{\underline{\bf $\mathcal{A}_t$ elements:}} The Yukawa contribution to the wave-function renormalization of top-quark is (see Fig.~\ref{fig:Yukawa} left-panel)  
\begin{equation}\label{eq:topwfn}
    [\cR_{q}]_{ij} = \qty(1 - \frac{y_i^2}{64 \pi^2}\frac{1}{\epsilon})\delta_{ij} \approx \qty(1 - \frac{\alpha_t}{16\pi \epsilon}\delta_{i3})\delta_{ij}\;.
\end{equation}
This gives the following non-zero $\mathcal{A}_t$ entries:

\begin{equation}\label{eq:At}
    \mathcal{A}_{t}(2L, 2L) = \mathcal{A}_t(2R, 2R) = - \frac{1}{2}(\delta_{i3} + \delta_{j3})\delta_{ik}\delta_{jl} \;.
\end{equation}

\noindent{\underline{\bf $\mathcal{B}_t$ elements:}} The insertion of $\cO_{2L}$ and $\cO_{2R}$ (Fig.~\ref{fig:Yukawa} right-panel) gives the full renormalization $\mathcal{B}_t$ entries: 
\begin{equation}\label{eq:Bt}
    \mathcal{B}_t(2L, 2R) = \mathcal{B}_t(2R, 2L) = \delta_{i3}\delta_{j3}\delta_{k3}\delta_{l3}\;.
\end{equation}

\noindent $\gamma_t$ is thus obtained by Eq.~\eqref{eq:ADMdecomposition}, \eqref{eq:At} and Eq.~\eqref{eq:Bt}.

\subsubsection{$\mathcal{O}(\alpha_s\alpha_w)$}\label{sec:ADMsw}
\vskip 0.2cm
For clarity, we use the following symbols interchangeably for the Wilson coefficients and operators relevant to $b\to sa$ transition
\begin{equation}
    \mathcal{C}_{4L} \equiv \mathcal{C}^{23}_{3L} \;, \mathcal{O}_{4L} \equiv \mathcal{O}^{23}_{3L}\;,  \quad \mathcal{C}_{4R} \equiv \mathcal{C}^{23}_{3R} \;, \mathcal{O}_{4R} \equiv \mathcal{O}^{23}_{3R} \;.
\end{equation}

A significant contribution to $\mathcal{C}_{4L}$ from $\mathcal{C}_1$ arises in the two-loop order, specifically $\mathcal{O}(\alpha_s\alpha_w)$. This is crucial since the leading-order effect pertaining to the $b\to s$ transition is generated only at the two-loop level for heavy QCD axion and KSVZ-like scenarios. Given this, we focus our analysis on computing the $\mathcal{A}_{sw}(1, 4L)$ and $\mathcal{B}_{sw}(1, 4L)$ elements. Other coefficients have already accounted for the contributions from $\mathcal{C}_1$ to their running. \\

\noindent{\underline{\bf $\mathcal{A}_{sw}$ elements:}} Self-energy corrections by definition do not mix quark and gluon fields; therefore $\mathcal{A}_{sw}(1, 4L)$ is zero.  \\

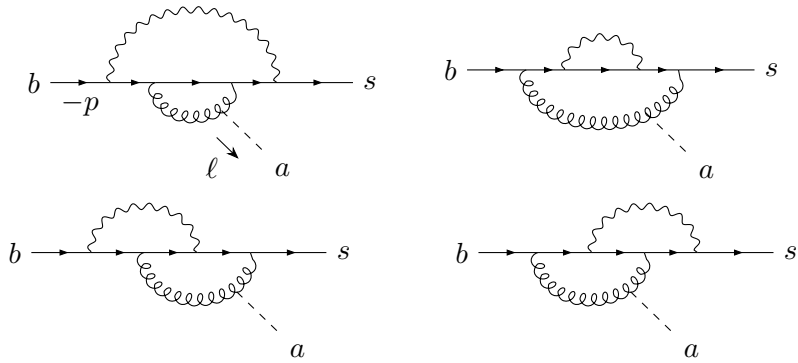
\begin{figure}[t]
\centering
\begin{tikzpicture}
    \begin{feynman}
    \vertex (a) {\(b\)};
    \vertex [right=1cm of a] (b);
    \vertex [right=0.6cm of b] (c);
    \vertex [right=1cm of c] (d);
    \vertex [right=0.6cm of d] (e);
    \vertex [right=1cm of e] (f) {\(s\)};
    \vertex at ($(c)!0.5!(d)!1.08!135:(c)$) (g);
    \vertex [below right=0.8cm of g] (h) {\(a\)};
    \vertex [below = 1.55cm of a] {};
    \diagram* {
            (a) -- [fermion, edge label' = \(-p\)] (b) -- [fermion] (c) -- [fermion] (d) -- [fermion] (e) -- [fermion] (f), (b) -- [boson, half left] (e), (c) -- [gluon, half right] (d), (g) -- [scalar, momentum' = {[arrow shorten=0.25] \(l\)}] (h)
        };
        \end{feynman}
\end{tikzpicture}\hspace{1cm}
\begin{tikzpicture}
    \begin{feynman}
    \vertex (a) {\(b\)};
    \vertex [right=1cm of a] (b);
    \vertex [right=0.6cm of b] (c);
    \vertex [right=1cm of c] (d);
    \vertex [right=0.6cm of d] (e);
    \vertex [right=1cm of e] (f) {\(s\)};
    \vertex at ($(b)!0.5!(e)!0.9!135:(b)$) (g);
    \vertex [below right=0.8cm of g] (h) {\(a\)};
%    \vertex []
    \diagram* {
            (a) -- [fermion] (b) -- [fermion] (c) -- [fermion] (d) -- [fermion] (e) -- [fermion] (f), (c) -- [boson, half left] (d), (b) -- [gluon, half right, looseness = 1.2] (e), (g) -- [scalar] (h)
        };
        \end{feynman}
\end{tikzpicture}
\\
\begin{tikzpicture}
    \begin{feynman}
    \vertex (a) {\(b\)};
    \vertex [right=1cm of a] (b);
    \vertex [right=0.7cm of b] (c);
    \vertex [right=0.8cm of c] (d);
    \vertex [right=0.7cm of d] (e);
    \vertex [right=1cm of e] (f) {\(s\)};
    \vertex at ($(c)!0.5!(e)!1.05!135:(c)$) (g);
    \vertex [below right=0.8cm of g] (h) {\(a\)};
%    \vertex []
    \diagram* {
            (a) -- [fermion] (b) -- [fermion] (c) -- [fermion] (d) -- [fermion] (e) -- [fermion] (f), (b) -- [boson, half left] (d), (c) -- [gluon, half right] (e), (g) -- [scalar] (h)
        };
        \end{feynman}
\end{tikzpicture}\hspace{1cm}
\begin{tikzpicture}
    \begin{feynman}
    \vertex (a) {\(b\)};
    \vertex [right=1cm of a] (b);
    \vertex [right=0.7cm of b] (c);
    \vertex [right=0.8cm of c] (d);
    \vertex [right=0.7cm of d] (e);
    \vertex [right=1cm of e] (f) {\(s\)};
    \vertex at ($(b)!0.5!(d)!1.05!135:(b)$) (g);
    \vertex [below right=0.8cm of g] (h) {\(a\)};
%    \vertex []
    \diagram* {
            (a) -- [fermion] (b) -- [fermion] (c) -- [fermion] (d) -- [fermion] (e) -- [fermion] (f), (c) -- [boson, half left] (e), (b) -- [gluon, half right] (d), (g) -- [scalar] (h)
        };
        \end{feynman}
\end{tikzpicture}

\caption{All possible two-loop 1PI diagrams relevant for the process $b\to s a$.}
\label{fig:8}
\end{figure}

\noindent{\underline{\bf $\mathcal{B}_{sw}$ elements:}}
On the other hand, the operator renormalization gets a non-zero contribution from the following diagrams, Fig.~\ref{fig:8} where $\mathcal{O}_1$ generates $\mathcal{O}_{4L}$. We reiterate that for the computation of amputated Green's functions, loops on the external legs have not been considered.

To evaluate the two-loop diagrams shown in Fig.~\ref{fig:8}, we use the following procedures.
\begin{enumerate}[leftmargin=*]
\item In principle, the diagrams shown in Fig.~\ref{fig:8} exhibit both ultraviolet (UV) and infrared (IR) divergences. To correctly determine the anomalous dimension, it is essential to distinguish between these divergent structures. As a result, we have considered a fictitious mass for the gluon fields only (see paragraph above~\eqref{eq:IRprop} for more details).

\item We expand the axion momentum about $l=0$ in the denominator. However, we keep the terms proportional only to $\slashed l$ in the numerator after the Dirac algebra. This is reasonable as we ignore $m_a$ compared to other scales in the theory.

\item As mentioned earlier, we work in the limit where all the external momenta, i.e., $p=0$. This helps us to simplify Feynman integrals as well as the corresponding master integrals. For the tensor reduction of the numerator, we have used FORM~\cite{Ruijl:2017dtg} whereas integration by parts was performed using KIRA~\cite{Maierhofer:2017gsa}. We provide a detailed sample of a two-loop computation in Appendix~\ref{app:two_loop}. 
% Somewhere before ref [115]
\nocite{Davydychev:1992mt, Adams:2015gva}
\end{enumerate}

The divergent parts of the two-loop amplitudes for individual diagrams shown in Fig.~\ref{fig:8} are given by

\begin{equation}
\begin{aligned}
i\Sigma_1  & = \kappa \qty[-\frac{\xi_i }{2 \epsilon ^2} + \frac{1}{\epsilon}\left\{-\xi_i  \log \left(\frac{\mu ^2}{M_W^2}\right)+\frac{ (5 \xi_i -17)\xi_i}{4 (\xi_i -1)} + \frac{(\xi_i -4) (\xi_i -2) \xi_i  \log \xi_i}{(\xi_i -1)^2}\right\}] (-\slashed l P_L)\;,\\
i\Sigma_2 & =\kappa \left[-\frac{\xi_i }{2 \epsilon ^2} + \frac{1}{\epsilon}\qty{ - \xi_i\log \qty(\frac{\mu^2}{m_R^2}) - \frac{3\xi_i}{4}}\right](-\slashed l P_L)\;, \\
i\Sigma_{3,4} & =\kappa  \left[\frac{\xi_i }{2 \epsilon ^2} + \frac{1}{\epsilon}\qty{\xi_i \log \qty(\frac{\mu^2}{m_R^2})-\frac{\xi_i}{4}}\right](-\slashed l P_L)\;,
\end{aligned}
\label{eq:double_pole}
\end{equation}

where the prefactor $\kappa$ is defined as
\begin{equation}\label{eq:kappa}
    \kappa\equiv 6 \left(\frac{\alpha_s}{4\pi}\right) \left(\frac{\alpha_w}{16\pi}\right)\qty(\frac{\mathcal{C}_1}{f_a}) C_F \sum_i V_{is}^\ast V_{ib}\;.
\end{equation}

\noindent
\subsection{Counterterms} 
\begin{figure}[h]
\centering
\begin{tikzpicture}
    \begin{feynman}
    \vertex (a);
    \vertex [right=1.2 of a, gblob] (b) {};
    \vertex [right= of b] (c);
    \vertex [below right= of b] (d) {\(a\)};
    \vertex [right= of c](a1) {};
    \vertex [right=of a1] (b1);
    \vertex [right=of b1] (c1);
    \vertex [right=of c1] (d1) {}; 
    \vertex at ($(b1)!0.5!(c1)!1.05!315:(c1)$) (e1);
    \vertex [below right=1cm of e1] (f1) {\(a\)};
    \vertex at ($(c)!0.5!(a1)!0.8!90:(c)$)  (e) {\(\equiv\)};
    \diagram* {
        (a) -- [fermion] (b) -- [fermion] (c), (b) -- [scalar] (d)
    };
    \diagram* {
        (a1) -- [fermion ] (b1) -- [fermion] (c1) -- [fermion] (d1), (b1) -- [gluon, half right] (c1), (e1) -- [scalar,] (f1)
    };
    \end{feynman}
\end{tikzpicture}
\begin{tikzpicture}
    \begin{feynman}
    \vertex (a);
    \vertex [right=1.2 of a, wblob] (b) {};
    \vertex [right= of b] (c);
    \vertex [below= of b] (d) {\(g\)};
    \vertex [right= of c](a1) {};
    \vertex [right=of a1] (b1);
    \vertex [right=0.7cm of b1] (c1);
    \vertex [right=0.7cm of c1] (d1);
    \vertex [right=of d1] (e1);
    \vertex [below=of c1] (f1) {\(g\)};
    \vertex at ($(c)!0.5!(a1)!0.8!90:(c)$)  (e) {\(\equiv\)};
    \diagram* {
        (a) -- [fermion] (b) -- [fermion] (c), (b) -- [gluon] (d)
    };
    \diagram* {
        (a1) -- [fermion ] (b1) -- [fermion] (c1) -- [fermion] (d1) -- [fermion] (e1), (d1) -- [boson, half right] (b1), (c1) -- [gluon] (f1)
    };
    \end{feynman}
\end{tikzpicture}
\begin{tikzpicture}
    \begin{feynman}
    \vertex (a);
    \vertex [right=1.2 of a, wblob] (b) {};
    \vertex [right= of b] (c);
    \vertex [right= of c](a1) {};
    \vertex [right=of a1] (b1);
    \vertex [right=0.7cm of b1] (c1);
    \vertex [right=0.7cm of c1] (d1);
    \vertex [right=of d1] (e1);
    \vertex at ($(c)!0.5!(a1)!0!90:(c)$)  (e) {\(\equiv\)};
    \diagram* {
        (a) -- [fermion] (b) -- [fermion] (c)
    };
    \diagram* {
        (a1) -- [fermion ] (b1) -- [fermion] (d1) -- [fermion] (e1), (d1) -- [boson, half right] (b1)
    };
    \end{feynman}
\end{tikzpicture}
\caption{The colored crosses indicate $1/\epsilon$ pieces from the diagrams on the right.}
\label{fig:9}
\end{figure}
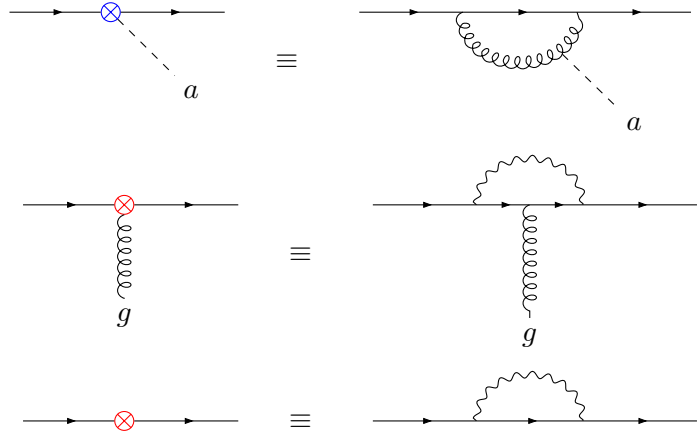

\begin{figure}[h]
\centering
%1
\begin{tikzpicture}
    \begin{feynman}
    \vertex (a) {\(b\)};
    \vertex [right=1cm of a] (b);
    \vertex [right=0.6cm of b, gblob] (c) {};
    \vertex [right=0.8cm of c] (e);
    \vertex [right=0.8cm of e] (f) {\(s\)};
    \vertex [below right=of c] (h) {\(a\)};
    \diagram* {
            (a) -- [fermion] (b) -- [fermion] (c) -- [fermion] (e) -- [fermion] (f), (b) -- [boson, half left] (e), (c) -- [scalar] (h)
        };
    \end{feynman}
\end{tikzpicture}\hspace{1cm}
\begin{tikzpicture}
    \begin{feynman}
    \vertex (a) {\(b\)};
    \vertex [right=0.9cm of a] (b);
    \vertex [right=0.6cm of b, wblob] (c) {};
    \vertex [right=0.8cm of c] (e);
    \vertex [right=1cm of e] (f) {\(s\)};
    \vertex at ($(b)!0.8!(e)!0.40!120:(b)$) (g);
    \vertex [below right=0.6cm of g] (h) {\(a\)};
    \diagram* {
            (a) -- [fermion] (b) -- [fermion] (c) -- [fermion] (e) -- [fermion] (f), (b) -- [gluon, half right, looseness = 1.5] (e), (g) -- [scalar] (h)
        };
        \end{feynman}
\end{tikzpicture}  \\ \vspace*{0.6cm}
%2
\begin{tikzpicture}
    \begin{feynman}
    \vertex (a) {\(b\)};
    \vertex [right=1cm of a, wblob] (b) {};
    \vertex [right= of b] (e);
    \vertex [right=1cm of e] (f) {\(s\)};
    \vertex at ($(b)!0.5!(e)!1.05!135:(b)$) (g);
    \vertex [below right=0.8cm of g] (h) {\(a\)};
    \diagram* {
            (a) -- [fermion] (b) -- [fermion] (e) -- [fermion] (f), (b) -- [gluon, half right] (e), (g) -- [scalar] (h)
        };
        \end{feynman}
\end{tikzpicture}\hspace{1cm}
%3
\begin{tikzpicture}
    \begin{feynman}
    \vertex (a) {\(b\)};
    \vertex [right=1cm of a] (b);
    \vertex [right= of b, wblob] (e) {};
    \vertex [right=1cm of e] (f) {\(s\)};
    \vertex at ($(b)!0.5!(e)!1.05!135:(b)$) (g);
    \vertex [below right=0.8cm of g] (h) {\(a\)};
    \diagram* {
            (a) -- [fermion] (b) -- [fermion] (e) -- [fermion] (f), (b) -- [gluon, half right] (e), (g) -- [scalar] (h)
        };
        \end{feynman}
\end{tikzpicture}
%4
\caption{1PI counter diagrams after incorporating the one-loop diagrams from Fig.~\ref{fig:9}}
\label{fig:10}
\end{figure}
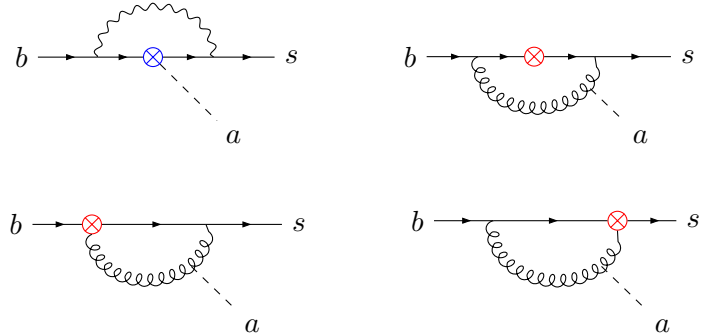

Note that the first diagram does not exhibit any IR poles. This is because the subdiagram with the gluon loop has a scale from the top-quark propagator. In the rest of the two-loop diagrams, the presence of light quarks associated with the gluon loop generates this IR divergence. In addition, we used CKM unitarity relations to evaluate these amplitudes. The divergences need to be canceled to make the theory renormalized at the UV scale itself. This requires the introduction of counterterms that can be identified from the divergent subdiagrams in Fig.~\ref{fig:8}. Some of these counterterms are already present in the SM, for example, the $b-s$ mixing and the $b-s-g$ operator generated via the $W$-boson loop. In addition, we require an additional operator that involves the interaction between the axion and diagonal quark fields. This has already been taken into account in our effective Lagrangian, given in Eq.~\eqref{eq:lag}. We illustrate these operators in Fig.~\ref{fig:9},
where the colored crosses indicate $1/\epsilon$ pieces with appropriate coefficients, which are generated from the one-loop diagrams on the right. Once the counterterms are identified, they are plugged into the counterdiagrams shown in Fig.~\ref{fig:10}. We emphasize that in the $\overline{\text{MS}}$ scheme, only $1/\epsilon$ pieces of these diagrams are considered for the counterterm insertion.
The amplitudes from the counterdiagrams, Fig.~\ref{fig:9}, are obtained as 
\small
\begin{equation}
\begin{aligned}
i\Pi_1 & =\kappa \qty[\frac{\xi_i }{\epsilon ^2} + \frac{1}{\epsilon }\left\{\xi_i  \log \left(\frac{\mu ^2}{M_W^2}\right)-\frac{(\xi_i -7) \xi_i }{2(\xi_i -1)} - \frac{ (\xi_i -4) (\xi_i -2) \xi_i  \log \xi_i}{(\xi_i -1)^2} \right\}](-\sl l P_L)\;, \\
i\Pi_2 & =\kappa \bigg[\frac{\xi_i}{\epsilon^2} + \frac{1}{\epsilon}\left\{\xi_i\log\bigg(\frac{\mu^2}{m_R^2} \bigg) + \frac{\xi_i}{2} \right\} \bigg](-\sl l P_L)\;, \\
i\Pi_{3,4} & =\kappa \bigg[-\frac{\xi_i}{\epsilon^2} + \frac{1}{\epsilon}\qty{ - \xi_i \log\bigg(\frac{\mu^2}{m_R^2}\bigg) - \frac{\xi_i}{2} } \bigg](-\sl l P_L)\;. 
\end{aligned}
\label{eq:cts}
\end{equation}
\normalsize

\begin{figure}[h]
\centering
\begin{tikzpicture}
    \begin{feynman}
    \vertex (a);
    \vertex [right=1.5cm of a, wgblob] (b) {};
    \vertex [right= 1.7cm of b] (f);
    \vertex [below right=1.8cm of b] (h) {\(a\)};
    \diagram* {
            (a) -- [fermion] (b) -- [fermion] (f), (b) -- [scalar] (h)
        };
        \end{feynman}
\end{tikzpicture}
\caption{Overall counter term diagram, i.e., $b-s-a$ operator, which absorbs the remaining $1/\epsilon$ UV pole as presented in Eq.~\eqref{eq:Coverall}.}
\label{fig:11}
\end{figure}
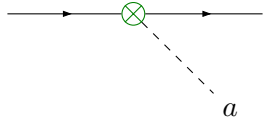

Notice that after including the amplitudes from the counterdiagrams shown in Eq.~\eqref{eq:cts}, all the double UV poles, as well as the UV-IR mixed poles, cancel from Eq.~\eqref{eq:double_pole}. However, we are still left with an overall pole $1/\epsilon$, which necessitates the introduction of a different counterterm or operator of the form axion-quark off-diagonal interaction, i.e., $b-s-a$ operator, also included in Eq.~\eqref{eq:lag} and shown in Fig.~\ref{fig:11}. \\

The corresponding amplitude for the overall counterterm for the operator $b-s-a$ is
\begin{align}
i\Pi_{\text{Overall}} & = - \qty(\sum_{j=1}^{4} i\Sigma_j + \sum_{j=1}^{4} i\Pi_j) = \kappa \qty[\frac{\xi_i}{\epsilon}](-\sl l P_L)\;.
\label{eq:Coverall}
\end{align}
This is also expected from the bottom-up approach of constructing an effective field theory since all possible operators are required to be written down in the Lagrangian to make the theory renormalizable order by order. Moreover, this remaining UV pole also contributes to $\mathcal{B}_{sw}$ and eventually to the anomalous dimension matrix. We therefore get
\begin{equation}
    \mathcal{B}_{sw}(1,4L) = -6C_F\sum_i V_{is}^*V_{ib}\xi_i\;,
\end{equation}
which implies that at $\mathcal{O}(\alpha_s\alpha_w)$ there is only one relevant non-zero entry in the ADM. We note that in this work we have not computed any other two-loop contribution to the anomalous dimensions. There can be non-zero entries pertaining to other Wilson coefficients, but they are subdominant for studying the $b\to sa$ transition.

This concludes the discussion about the anomalous dimension matrix $\gamma$, which can be formed by adding $\gamma_s, \gamma_w, \gamma_t$ and $\gamma_{sw}$ as in Eq.~\eqref{eq:totalADM}. There are subtleties that involve the finite terms that act as a matching coefficient for the overall Wilson coefficient $\mathcal{C}_{4L}$ on the $M_W$ scale. We will discuss this in detail in Section~\ref{sec:finite}.

\subsection{Complete set of RGEs} 
The two-loop anomalous dimension matrix for the analysis $b\to sa$ can now be translated to the running of Wilson coefficients, and we obtain 
 using Eq.~\eqref{eq:RGEequation}

{\small
\begin{equation}
\begin{aligned}
    \mu \frac{d\mathcal{C}_1}{d\mu} &= -\frac{\alpha_s \beta_0}{2\pi}\, \mathcal{C}_1\;, \\
    \mu \frac{d\mathcal{C}^{ij}_{2L}}{d\mu} &= -\frac{3\alpha_s C_F}{\pi}\, \tilde{\mathcal{C}}_1 \,\delta_{ij}  +  \frac{\alpha_t}{16\pi}\qty{\qty( \delta_{i3} + \delta_{j3})\,\cC^{ij}_{2L}  - 2 \delta_{i3}\,\delta_{j3}\, \cC^{ij}_{2R}} + \frac{\alpha_w}{16\pi}\xi_t \bigg\{\frac{1}{2}\qty(\delta_{i3} + \delta_{j3})\,\cC^{ij}_{2L} \\
    &\qquad - \delta_{i3} \delta_{j3}\, \cC_{2R}^{ij} \bigg\} + \beta_Q \gamma_{_H} \delta_{ij} \;, \\
    \mu\frac{d\mathcal{C}^{ij}_{2R}}{d\mu} &= \frac{3\alpha_s C_F }{\pi }\, \tilde{\mathcal{C}}_1 \,\delta_{ij} + \frac{\alpha_t}{16\pi}\qty{\qty( \delta_{i3} + \delta_{j3})\,\cC^{ij}_{2R}  - 2 \delta_{i3}\,\delta_{j3}\, \cC^{ij}_{2L}} +  \frac{\alpha_w}{16\pi}\xi_t \bigg\{\frac{3}{2}\qty(\delta_{i3} + \delta_{j3})\,\cC^{ij}_{2R} \\
    & \qquad - 3\delta_{i3}\delta_{j3}\,\cC^{ij}_{2L} \bigg\} + \beta_u \gamma_{_H} \delta_{ij} \, \;,\\
    \mu\frac{d\mathcal{C}^{ij}_{3L}}{d\mu} & = - \frac{3\alpha_s C_F}{\pi}\, \tilde{\mathcal{C}}_1 \delta_{ij}+ \frac{\alpha_w}{16\pi}\xi_t\qty( \sum_{k} V^*_{3i}V_{3k} \,\cC^{kj}_{3L} + \sum_l V^*_{3l}V_{3j}\, \cC^{il}_{3L} - 2V^*_{3i}V_{3j}\,\cC^
    {33}_{2R})  + \beta_Q \gamma_{_H} \delta_{ij} \\ 
    & \qquad + \frac{3\alpha_s\alpha_w}{8\pi^2}\, C_F V^*_{32}V_{33} \,\xi_t \delta_{i2} \delta_{j3} \, \cC_1 \,, \\
    \mu\frac{d\mathcal{C}^{ij}_{3R}}{d\mu} &= \frac{3\alpha_s C_F}{\pi} \, \tilde{\mathcal{C}}_1 \delta_{ij}  + \beta_d \gamma_{_H}\delta_{ij}\,,\\
    \mu\frac{d\mathcal{C}^{ij}_{eL}}{d\mu} & =  \beta_L \gamma_{_H}\delta_{ij}\,,  \quad \mu\frac{d\mathcal{C}^{ij}_{eR}}{d\mu}  =  \beta_e \gamma_{_H}\delta_{ij}\,.
\end{aligned}
\label{eq:runningWCs}
\end{equation}
}

We now discuss various aspects of these RGEs and clarify certain new notations
\begin{itemize}[leftmargin=*]
    \item  In the first term of the RGEs for axion-quark couplings, we have defined
    \begin{equation}\label{eq:C1tilde}
    \tilde{\mathcal{C}}_1 \equiv \mathcal{C}_1 + \frac{\alpha_s}{8\pi} \sum_i \qty( \mathcal{C}_{2R}^{ii} + \mathcal{C}_{3R}^{ii} - \mathcal{C}_{2L}^{ii} - \mathcal{C}_{3L}^{ii}) \;.
    \end{equation}
    This extra contribution from all diagonal quark couplings comes from the following type of two-loop diagrams~\cite{Bauer:2020jbp}

\begin{figure}[H]
\begin{center}
\resizebox{5cm}{!}{%
\begin{tikzpicture}
    \begin{feynman}
    \vertex (a) {\(a\)};
    \vertex [right=of a] (b) ;
    \vertex [above right=of b] (c) ;
    \vertex [below right=of b] (d) ;
    \vertex [right=of c] (e) ;
    \vertex [right=of d] (f) ;
    \vertex [right=of e] (g) ;
    \vertex [right=of f] (h) ;
    \diagram*{
        (a) -- [scalar](b) -- [fermion](c) -- [fermion] (d) -- [fermion] (b); (e) -- [gluon](c) ; (d) -- [gluon] (f) ; (e) -- [fermion] (f) ; (g) -- [fermion] (e) ; (f) -- [fermion] (h) ;
    };
    \end{feynman}
\end{tikzpicture}
}
\end{center}
\caption{Two-loop diagrams generating axion-quark couplings.}
\label{fig:anomalytwoloop}
\end{figure}

    From Eq.~\eqref{eq:C1tilde} it is clear that this effect is mostly relevant for scenarios where $\mathcal{C}_1 \sim \alpha_s/8\pi$ and all other axion-quark couplings are of $\mathcal{O}(1)$ at the UV scale. Then both terms in $\tilde{\mathcal{C}}_1$ contribute equally, and the two-loop effect cannot be ignored.

    \item In the last terms of the RGEs for axion-fermion couplings, we have defined $\beta_F \equiv \mathcal Y_F/\mathcal Y_H$, the ratio of the hypercharges of fermion and Higgs field, and
\begin{equation}
    \gamma_{_H} = \frac{3\alpha_t}{2\pi} \qty(\cC^{33}_{2L} - \cC^{33}_{2R})\;.
\end{equation}
This term represents the RG contributions from the redundant operator $\mathcal{O}_H$ (see Appendix~\ref{app:RedOp} for the explicit calculation).

 \item The (diagonal) leptonic couplings of the axion $\mathcal{C}_{eL/R}^{ii}$ receive their sole non-zero running contribution from the redundant operator. This well-known result is shown in the last two equations in \eqref{eq:runningWCs}. 

    \item To derive the above expressions, we have made use of the relations $C^{ij}_{2L} = V_{ik}V^*_{jl}C^{kl}_{3L}$, given in Eq.~\eqref{eq:udrelation}, which come from the SM $SU(2)_L$ symmetry in the UV Lagrangian.

    \item The RGEs have been derived with the mass basis fields. We have verified them by converting the flavor basis RGEs in the literature~\cite{Bauer:2020jbp, MartinCamalich:2020dfe, Arteaga:2018cmw} to the mass basis. 

    \item We have also considered: $\xi_i \approx 0$ for $i=d,s,b,u,c$ in deriving the above equations. This also implies: $\sum_{k=d,s,b} V_{ik}^*V_{jk}\xi_k \approx 0$ while $\sum_{k=u,c,t}V_{ki}^* V_{kj}\xi_k \approx V_{ti}^*V_{tj}\xi_t$.
\end{itemize}
For solving these equations, we take the products of the CKM matrix elements up to $\mathcal{O}(\lambda^2)$ in the Wolfenstein parameterization. Higher order terms in $\lambda$ are ignored.

\section{Finite terms}
\label{sec:finite}
The matching coefficient for the $b\to s a$ operator written down at the $M_W$ scale has mainly three components from the theory written down at the $\Lambda$ scale. For clarity, we refer to this theory as the UV theory, acknowledging a minor terminological concession. The first term, denoted by $f_1^{sw}$, arises from the two-loop diagrams shown in Fig.~\ref{fig:8}. Similarly, the axion-diagonal quark operators also contribute to the same, generated by the one-loop process depicted in the leftmost diagram of Fig.~\ref{fig:7}. The final contribution, of course, comes from the general counterterm, described by the operator $b-s-a$, shown in Fig.~\ref{fig:11}.

To compute the matching coefficient for the $b\to s a$ operator, we first have to evaluate the amplitude for the $b \to sa$ process in the UV theory, and we find
\begin{align}
    & \mathcal{M}_{\text{UV}}(b\to sa)  = \sum_{n} \cC_n(\mu) \mel{sa}{\cO_n (\mu)}{b}\;, \nonumber \\
    & = \cC_{4L}(\mu) + \frac{\alpha_w}{16\pi}\sum_{i,j}\Big\{\cC_{2L}^{ij}(\mu)  f_{2L}^{ij;w}(\mu)  + (L\to R)\Big\}  + \frac{\alpha_s}{4\pi}\; \frac{\alpha_w}{16\pi}\,\cC_{1}(\mu)  f_{1}^{sw}(\mu)\;,
\end{align}
where the one-loop matching coefficients are

\begin{equation}
\begin{aligned}
        f_{2L}^{ij;w}(\mu) =  V_{is}^* V_{jb} \bigg[2\log\bigg(\frac{\mu^2}{M_W^2} \bigg)+\frac{2\xi_i^2 (\xi_j-1) \log (\xi_
   i)}{  (\xi_i-1) (\xi_i-\xi_j)}-\frac{2(\xi_i-1) \xi_j^2 \log (\xi_j)}{ (\xi_j-1) (\xi_i-\xi_j)}-1\bigg]
   \end{aligned}
   \label{eq:F2Lw}
   \end{equation}
   \begin{equation}
   \begin{aligned}
   f_{2R}^{ij;w}(\mu) = V_{is}^*V_{jb} \sqrt{\xi_i \xi_j}\bigg[ \log \left(\frac{\mu ^2}{M_W^2}\right)-\frac{\left( \xi_i^2 - 4 \xi_i\right)  \log (\xi_i)}{(\xi_i-1) (\xi_i-\xi_j)}-\frac{(4 \xi_j-\xi_j^2)  \log (\xi_j)}{(\xi_j-1) (\xi_i-\xi_j)}+\frac{1}{2}\bigg]
   \end{aligned}
   \label{eq:F2Rw}
\end{equation}

and the two-loop matching coefficient is obtained as

\footnotesize{
\begin{equation}
\begin{aligned}
& f_{1}^{sw}(\mu)  =  6 C_F \sum_{i}V^\ast_{is}V_{ib} 
 \bigg[ -\xi_i \log^2\qty(\frac{\mu^2}{M_W^2} ) +\log (\frac{\mu^2}{M_W^2} )
   \left(\frac{24 \xi_i -6 \xi_i ^2 (\xi_i +15)}{12 (\xi_i -1) \xi_i }+\frac{2 \xi_i  ((\xi_i -2) \xi_i +4) \log (\xi_i )}{(\xi_i -1)^2}\right)\\
   & +\frac{\left(6 (\xi_i -6) \xi_i ^3-48 \xi_i ^2+24 (\xi_i -2) (\xi_i  (3 \xi_i -1)+1) \log (\xi_i -1)\right) \log (\xi_i )}{12
   (\xi_i -1)^2 \xi_i }+\frac{\pi ^2 ((\xi_i -2) \xi_i  (2 \xi_i -1)-2)}{3 (\xi_i -1)^2 \xi_i }\\
   & -\frac{(\xi_i  (\xi_i  (\xi_i +5)-2)+4) \log ^2(\xi_i )}{(\xi_i -1) \xi_i }+\frac{15 \xi_i  \left(7 \xi_i ^2+\xi_i +2\right)-24 (\xi_i +2) (\xi_i  (2 \xi_i -1)+1)
   \text{Li}_2\left(\frac{\xi_i -1}{\xi_i }\right)}{12 (\xi_i -1) \xi_i }\\
   & -\frac{2 (\xi_i -2) (\xi_i  (3 \xi_i -1)+1) \text{Li}_2\left(\frac{1}{\xi_i }\right)}{(\xi_i -1)^2 \xi_i } + \underbrace{\log (\frac{\mu^2}{m_R^2} ) \left(\xi_i \log \bigg(\frac{\mu^2}{M_W^2} \bigg)+\frac{6 \xi_i ^2 (3 \xi_i +11)-48 \xi_i }{12 (\xi_i -1) \xi_i }-\frac{(\xi_i +2) \xi_i ^2 \log (\xi_i )}{(\xi_i -1)^2}\right)}_{\text{IR-divergent}} \bigg]\;,
\end{aligned}
\label{eq:F1sw}
\end{equation}
}
\normalsize
where the dilogarithm function is denoted as $\text{Li}_2(x)$ in $f_1^{sw}(\mu)$
\begin{equation}
    \text{Li}_2(x) = - \int_0^x \delta t \;\frac{\log(1-t)}{t}\;.
\end{equation}
We observe that there exist $\log^2$ terms in $f_1^{sw}$, which is an artifact of two loops. More importantly, although we were able to remove the pure UV poles and the mixed UV-IR poles, the finite piece $f_1^{sw}$ still has pure IR divergences, shown with the underbrace. As expected, these IR divergences are of the form $\log m_R^2$ and therefore nonanalytic functions of the IR scale. However, both the UV and the EFT theories are local in the IR mass scale. Hence, a consistent matching condition is defined as

\begin{equation}
\text{Matching condition:}\quad \mathcal{M}_{\text{UV}} (\mu) = \mathcal{M}_{\text{EFT}}(\mu)\;,
\end{equation}
should get rid of any IR divergences, and this is what we show in the next section.

\section{Matching at the Electroweak Scale}
\label{sec:Matching}
In this section, we verify whether the IR divergent pieces shown in Eq.~\eqref{eq:F1sw} indeed cancel out once we match the theory to the effective theory written at the $M_W$ scale. Furthermore, the matching condition leads to a change in the two-loop contribution to the Wilson coefficient of the effective operator $b\to sa$. After integrating out the degrees of freedom (dof) heavier than the $M_W$ scale, the effective Lagrangian is expressed as
\begin{equation}\label{eq:LeffmuW}
\mathcal{L}_{\text{EFT}} = \mathcal{L'}_{\text{\tiny SM}} + \frac{1}{2} \left(\partial_\mu a\right)^2 -\frac{m_a^2}{2} a^2 + \sum_{n} \cC'_n \cO'_n\;,
\end{equation}
where $\mathcal{L'}_{\text{\tiny SM}}$ denotes the low energy SM Lagrangian. Operators are explicitly defined as:

\begin{equation}
\cO'_1  = \mathcal{O}_1\;,\quad \mathcal{O}'^{ij}_{2L/R} = \cO^{ij}_{2L/R} \;, \quad \cO'^{ij}_{3L/R} = \cO^{ij}_{3L/R} \;, \quad \cO'_{4L/R} = \mathcal{O}_{4L/R}\;, \quad \cO'_{5L}  = \bar{s} i\sl D P_L b\;,
\label{eq:mwoperators}
\end{equation}

with the covariant derivative $D_\mu = \partial_\mu - ig_s G_\mu^a T^a$, and where $i,j = \{d,s,b\}$ for down-type quarks and $i,j = \{u,c\}$ for up-type quarks, since the top quark is no longer a degree of freedom. The problem now is to compute the Wilson coefficients $\mathcal{C}^\prime_j$'s in the EFT.

\color{black}
\subsection{Framework for Calculating Wilson Coefficients in the EFT}
We define the amplitudes for the initial to final state $(i\to f)$ transition in the UV and the effective theory as
\begin{equation}
    \cM_{\text{UV}}(\mu) = \sum_n \cC_n(\mu) \expval{\cO_n (\mu)}, \quad  \cM_{\text{EFT}}(\mu) = \sum_n \cC'_n(\mu) \expval{\cO'_n (\mu)}\;.
    \label{eq:WEFT}
\end{equation}
where the shorthand notation describes $\expval{\cO_n (\mu)} = \mel{f}{\cO_n (\mu)}{i}$. The Wilson coefficients in the EFT, $\mathcal{C}^\prime$'s can be determined by matching at the $\mu_w=M_W$ scale, that is, $\mathcal{M}_{\text{UV}}\left(\mu_w\right)=\mathcal{M}_{\text{EFT}}\left(\mu_w\right)$. We can now expand Eq.~\eqref{eq:WEFT} order by order in $\alpha$, leading to 

\begin{equation}
\begin{aligned}
    \cC'_n(\mu) & = \cC'^{\;\text{tree}}_n(\mu) + \frac{\alpha_s}{4\pi}\cC'^{s}_n(\mu) + \frac{\alpha_w}{16\pi}\cC'^{w}_n(\mu) + \frac{\alpha_s}{4\pi}\frac{\alpha_w}{16\pi}\cC'^{sw}_n(\mu)\;, \\
    \expval{\cO'_n (\mu)} & = \expval{\cO'_n (\mu)}^{\;\text{tree}} + \frac{\alpha_s}{4\pi}\expval{\cO'_n (\mu)}^{s} + \frac{\alpha_w}{16\pi}\expval{\cO'_n (\mu)}^{w} + \frac{\alpha_s}{4\pi}\frac{\alpha_w}{16\pi}\expval{\cO'_n (\mu)}^{sw}\;,\\
    \cM_\text{UV}(\mu) & = \cM^{\;\text{tree}}_\text{UV}(\mu) + \frac{\alpha_s}{4\pi} \cM^{s}_\text{UV}(\mu) + \frac{\alpha_w}{16\pi} \cM^{w}_\text{UV}(\mu) + \frac{\alpha_s}{4\pi}\frac{\alpha_w}{16\pi}\cM^{sw}_\text{UV}(\mu)\;.
\end{aligned}
\end{equation}

The superscript `{\it tree}' implies the tree-level contribution of the corresponding operators. We reiterate that the terms proportional to $\alpha_s^2$ contribute only as corrections to the RGE equations. In contrast, at the $\mathcal{O}(\alpha_s\alpha_w)$ order, new operators are generated, which is relevant for our phenomenological study. By applying the previously mentioned matching condition, we find the tree level as well as $\mathcal{O}(\alpha_s, \alpha_w)$ order components of the amplitude as
\begin{align}
     \cM^{\;\text{tree}}_\text{UV}(\mu_w) & =  \cC'^{\;\text{tree}}_n(\mu_w) \expval{\cO'_n (\mu_w)}^{\;\text{tree}}\;,
     \label{eq:treeLevelMatchingCondition}\\
     \cM^{s, w}_\text{UV}(\mu_w) & = \cC'^{\;\text{tree}}_n(\mu_w) \expval{\cO'_n (\mu_w)}^{s, w} + \cC'^{s, w}_n(\mu_w) \expval{\cO'_n (\mu_w)}^{\;\text{tree}}\;.
     \label{eq:oneLoopMatchingCondition}
 \end{align}
 In a similar way, $\cM^{sw}_\text{UV}(\mu_w)$ has four contributions at the order $\order{\alpha_s \alpha_w}$. Plugging the relevant initial and final states into the matrix elements one can compute the $\cC'_i$'s by solving these equations. 

\subsection{Evaluation of $\cC'_{4L}$}
For $b\to sa$ production, we only require the coefficient $\cC'_{4L}(\mu_w)$. To determine this, the following components are required: $\cM_\text{UV}(b \to sa)$, $\cM_\text{UV}(b \to s)$, $\mel{sa}{\cO'_n}{b}$, and $\mel{s}{\cO'_n}{b}$ matrix elements. Collecting terms in order by order again, we find
\begin{enumerate}[leftmargin=*]
\item {\bf Tree level Contribution:}

The tree-level matching condition can be evaluated straightforwardly. For example, in the UV theory, the tree-level matrix element is given by
\begin{equation}
    \cM^{\;\text{tree}}_\text{UV}(b \to sa) = \cC_{4L} (\mu_w) \mel{sa}{\cO_{4L}(\mu_w)}{b}^{\;\text{tree}}\;,
\end{equation}
Notice that both the UV theory and EFT have the same operator $\cO_{4L}$. Therefore,  upon using the matching condition given in Eq.~\eqref{eq:treeLevelMatchingCondition}, we find 
\begin{equation}
    \cC'^{\;\text{tree}}_{4L}(\mu_w) = \cC_{4L} (\mu_w)\;.
\end{equation}

\item {\bf Contribution at $\order{\alpha_s}$ and $\order{\alpha_w}$:}

At one-loop level, i.e., at the $\mathcal{O}(\alpha_s)$ and $\mathcal{O}(\alpha_w)$ order, the matrix elements for $b \to sa$ in the UV theory are

\begin{equation}
\begin{aligned}
    \frac{\alpha_w}{16\pi} \cM^{w}_\text{UV}(b \to sa) & = \sum_{i,j}\mathcal{C}_{2L}^{ij} (\mu_w) \mel{sa}{\cO_{2L}^{ij}(\mu_w)}{b}^{w} + \left(L\to R\right) \;,\\
    & = \frac{\alpha_w}{16\pi} \sum_{ij}\qty[\mathcal{C}_{2L}^{ij}(\mu_w) f_{2L}^{ij;w}(\mu_w) + \left(L\to R\right)] \mel{sa}{\cO_{4L}(\mu_w)}{b}^{\;\text{tree}}\;.
\end{aligned}
\end{equation}

Here, the sum is over all the three up-type quark flavors. The finite pieces $f_{2L}^{ij;w}$ and $f_{2R}^{ij;w}$ have already been computed by the insertion of $\cO_{2L}^{ij}$ and $\cO_{2R}^{ij}$ operators in the first diagram of Fig. \ref{fig:7} and by adding the $b \to sa$ counter term. 

\color{black}
\begin{figure}[H]
\centering
\begin{tikzpicture}
    \begin{feynman}
    \vertex (a1) {b};
    \vertex [right=1cm of a1] (b1);
    \vertex [right=0.8cm of b1] (c1);
    \vertex [right=0.8cm of c1] (d1);
    \vertex [right=1cm of d1] (e1) {s};
    \diagram* {
        (a1) -- [fermion ] (b1) -- [fermion] (d1) -- [fermion] (e1), (d1) -- [boson, half right] (b1)
    };
    \end{feynman}
\end{tikzpicture}
\caption{ Diagram for the finite term $f_{bs}^w$.}
\label{fig:12}
\end{figure}
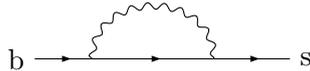
Similarly, at $\mathcal{O}(\alpha_w)$ order, we generate the $b\to s$ transition as shown in Fig.~\ref{fig:12}, and the corresponding amplitude is
\begin{equation}
 \frac{\alpha_w}{16\pi} \cM^{w}_\text{UV}(b \to s)  = \frac{\alpha_w}{16\pi} f_{bs}^{w}(\mu_w) \mel{s}{\cO'_{5L}(\mu_w)}{b}^{\;\text{tree}}\;.
\end{equation}
Although this does not generate axions directly, insertions of such diagrams would indeed generate $b\to sa$. We remind the reader that we have already used insertions of such operators ($1/\epsilon$ pieces) to cancel the UV divergence. Therefore, to match and cancel the purely IR divergent contributions, we only require the finite term from Fig.~\ref{fig:12}, given by
\begin{align}
    f_{bs}^{w}(\mu) & = \sum_{i}V^\ast_{is}V_{ib}\qty[\xi_i  \log \qty(\frac{\mu ^2}{M_W^2}) + \frac{3 \xi_i  (\xi_i +1)}{2 (\xi_i -1)}-\frac{\xi_i ^2 (\xi_i +2) \log \xi_i}{(\xi_i -1)^2}]\;.
\end{align}

In the EFT, $W$-boson ceases to be a dynamic degree of freedom which implies the matrix element $\mel{sa}{\cO'_n}{b}^{w} = 0$ and no flavor changing neutral current (FCNC) is possible at the tree level, rendering the Wilson coefficient  $\cC'^{\text{tree}}_{5L} = 0$. Therefore the one-loop matching condition from Eq.~\eqref{eq:oneLoopMatchingCondition} gives
\begin{align}
    \cC'^{w}_{4L}(\mu_w) & = \sum_{i,j} \cC_{2L}^{ij}(\mu_w) f_{2L}^{ij;w}(\mu_w) +\left(L\to R\right)\;,\quad \text{and}\quad \quad\cC'^{w}_{5L}(\mu_w)  = f_{bs}^{w}(\mu_w)\;.
    \label{eq:WilsonCoefficient_oneLoopContribution}
\end{align}
We note in passing that $\order{\alpha_s}$ and $\order{\alpha_{\text{e}}}$ corrections are the same in the UV and IR theory and therefore would cancel after matching, which implies, $\cC'^{s,e}_{4L}(\mu_w) = 0$.

\item {\bf Contribution at $\order{\alpha_s\alpha_w}$ order:}

Finally, for the two-loop matrix element in UV, the theory for $b \to sa$, the only contribution comes from the $\cO_1$ operator, which is

\begin{equation}
\begin{aligned}
    \frac{\alpha_s}{4\pi}\frac{\alpha_w}{16\pi}\cM^{sw}_\text{UV}(b \to sa) & =  \cC_{1} (\mu_w) \mel{sa}{\cO_{1}(\mu_w)}{b}^{sw}\;,\\
    & = \frac{\alpha_s}{4\pi}\frac{\alpha_w}{16\pi} \cC_1(\mu_w) f_1^{sw}(\mu_w) \mel{sa}{\cO_{4L}(\mu_w)}{b}^{\;\text{tree}}\;,
\end{aligned}
\label{eq:AmpUVsw}
\end{equation}

where the finite term $f_1^{(sw)}$ is evaluated by adding all the two-loop diagrams, its corresponding one-loop counter terms, and the overall counter terms, as illustrated in section \ref{sec:ADMsw}. On the other hand, all the non-zero contributions from the EFT gives

\begin{equation}
\begin{aligned}
    \cM^{sw}_\text{EFT}(b \to sa) & =  \cC'^{w}_{5L} (\mu_w) \mel{sa}{\cO'_{5L}(\mu_w)}{b}^{s} + \cC'^{sw}_{4L}\mel{sa}{\cO'_{4L}(\mu_w)}{b}^{\text{tree}}\;, \\
    & = \qty(\cC'^{w}_{5L} (\mu_w) f_{5L}^{s}(\mu_w) + \cC'^{sw}_{4L} ) \mel{sa}{\cO_{4L}(\mu_w)}{b}^{\text{tree}}.
\end{aligned}
\label{eq:AmpEFTsw}
\end{equation}

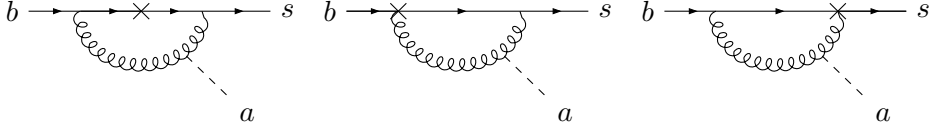
\begin{figure}[H]
\centering
\begin{tikzpicture}
    \begin{feynman}
    \vertex (a) {\(b\)};
    \vertex [right=0.9cm of a] (b);
    \vertex [right=0.8cm of b] (c);
    \vertex [right=0.8cm of c] (e);
    \vertex [right=0.9cm of e] (f) {\(s\)};
    \vertex at ($(b)!0.5!(e)!1.05!135:(b)$) (g);
    \vertex [below right=0.8cm of g] (h) {\(a\)};
    \diagram* {
            (a) -- [fermion] (b) -- [fermion] (c) -- [fermion] (e) -- [fermion] (f), (b) -- [insertion=1.0] (c), (b) -- [gluon, half right, looseness = 1.5] (e), (g) -- [scalar] (h)
        };
        \end{feynman}
\end{tikzpicture} 
\begin{tikzpicture}
    \begin{feynman}
    \vertex (a) {\(b\)};
    \vertex [right=0.9cm of a] (b);
    \vertex [right= 1.6cm of b] (e);
    \vertex [right=0.9cm of e] (f) {\(s\)};
    \vertex at ($(b)!0.5!(e)!1.05!135:(b)$) (g);
    \vertex [below right=0.8cm of g] (h) {\(a\)};
    \diagram* {
            (a) -- [fermion] (b) -- [fermion] (e) -- [fermion] (f), (a) -- [insertion=1.0] (b), (b) -- [gluon, half right] (e), (g) -- [scalar] (h)
        };
        \end{feynman}
\end{tikzpicture}
\begin{tikzpicture}
    \begin{feynman}
    \vertex (a) {\(b\)};
    \vertex [right=0.9cm of a] (b);
    \vertex [right=1.6cm of b] (e);
    \vertex [right=0.9cm of e] (f) {\(s\)};
    \vertex at ($(b)!0.5!(e)!1.05!135:(b)$) (g);
    \vertex [below right=0.8cm of g] (h) {\(a\)};
    \diagram* {
            (a) -- [fermion] (b) -- [fermion] (e) -- [fermion] (f), (e) -- [insertion=0.0] (f), (b) -- [gluon, half right] (e), (g) -- [scalar] (h)
        };
        \end{feynman}
\end{tikzpicture}
\caption{Diagrams for $f_{5L}^{s}$. The cross denotes the $\cO'_{5L}$ operator insertion.}
\label{fig:13}
\end{figure}

\end{enumerate}
where Fig.~\ref{fig:13} gives the following amplitude
\begin{align}
    f_{5L}^{s}(\mu) &= 6 C_F\; \cC_1(\mu) \qty[\log (\frac{\mu^2}{m_R^2} )+\frac{5}{2}]\;.
\end{align}
Finally, using the matching condition at $\order{\alpha_s\alpha_w}$ order, $\cM^{sw}_\text{UV} 
 = \cM^{sw}_\text{EFT}$, we obtain
\begin{equation}
    \cC'^{sw}_{4L}(\mu_w) = \cC_1(\mu_w) f_1^{sw}(\mu_w) - f_{bs}^{w}(\mu_w) f_{5L}^{s}(\mu_w)\;.
    \label{eq:WilsonCoefficient_twoLoopContribution}
\end{equation}
We already saw that in the two-loop finite term $f_1^{(sw)}(\mu)$ has IR divergence pieces $\log(\mu^2/m_R^2)$. However, the EFT at the scale $M_W$ has the same IR structure as in the second term of the equation \eqref{eq:WilsonCoefficient_twoLoopContribution}. Therefore, the final Wilson coefficient contribution from the two-loop diagrams has no IR divergence because it cancels through the matching. Therefore, the total contribution to the $\cC'_{4L}(\mu_w)$ is

\begin{equation}
\boxed{
    \cC'_{4L}(\mu_w) = \cC_{4L}(\mu_w) + \frac{\alpha_w}{16\pi}\;\cC'^{w}_{4L}(\mu_w) + \frac{\alpha_s}{4\pi}\frac{\alpha_w}{16\pi}\;\cC'^{sw}_{4L}(\mu_w).\;
}
    \label{eq:C4LW}
\end{equation}
Plugging all the finite terms in Eq.~\eqref{eq:WilsonCoefficient_oneLoopContribution} and Eq.~\eqref{eq:WilsonCoefficient_twoLoopContribution} respectively, we get

\small
\begin{equation}
\begin{aligned}
    &\cC'^{w}_{4L}(\mu_w)  = \sum_{i,j}\Big\{\cC_{2L}^{ij}(\mu)  f_{2L}^{ij;w}(\mu)  + (L\to R)\Big\}\;,\\
    &\mathcal C'^{sw}_{4L}(\mu_w)  = 6 C_F \sum_{i} V^\ast_{is}V_{ib}\; \cC_{1}(\mu_w)\Bigg[\log (\frac{\mu^2}{M_W^2} ) \left(\frac{2 \xi_i  ((\xi_i -2) \xi_i +4) \log (\xi_i )}{(\xi_i -1)^2}-\frac{(\xi_i +2) (3 \xi_i -1)}{\xi_i -1}\right)\\
    & +\frac{\left(6 (3 \xi_i +2) \xi_i ^3-24 \xi_i ^2+12 (\xi_i -2) (\xi_i  (3 \xi_i -1)+1) \log (\xi_i -1)\right)
   \log (\xi_i )}{6 (\xi_i -1)^2 \xi_i }\\
   &-\frac{2 (\xi_i -2) (\xi_i  (3 \xi_i -1)+1) \text{Li}_2\left(\frac{1}{\xi_i
   }\right)}{(\xi_i -1)^2 \xi_i }-\frac{(\xi_i  (\xi_i  (\xi_i +5)-2)+4) \log ^2(\xi_i )}{(\xi_i -1) \xi_i } \\
   & +\frac{\pi ^2 ((\xi_i -2) \xi_i  (2 \xi_i -1)-2)}{3 (\xi_i -1)^2 \xi_i } +\frac{15 \xi_i  (\xi_i  (2 \xi_i -1)+1)-12 (\xi_i +2) (\xi_i  (2 \xi_i -1)+1) \text{Li}_2\left(\frac{\xi_i -1}{\xi_i }\right)}{6 (\xi_i -1) \xi_i }\Bigg]\;.
\end{aligned}
\end{equation}
\normalsize

Below $M_W$, running effects due to $W, Z$ bosons and the top quark are absent as these heavy particles are integrated. Consequently, axion-fermion diagonal Wilson coefficients run only through the $\mathcal{O}(\alpha_s)$ terms given in Eq.~\eqref{eq:runningWCs}. The off-diagonal axion-fermion operators do not run below $M_W$ therefore, in particular, $\mathcal{C}'_{4L}(m_B) = \mathcal{C}'_{4L}(\mu_w)$ where $m_B$ is the mass of $B$-meson. On the other hand, all right-handed off-diagonal axion-fermion couplings do not run at all. Thus, for the $b\to sa$ right-handed Wilson coefficient $\mathcal{C}'_{4R}(m_B) = \mathcal{C}_{4R}(\Lambda_{\text{UV}})$. The left and right-chiral operators $\mathcal{O}'_{4L}, \,\mathcal{O}'_{4R}$ can be converted to vector and axial-vector operators with corresponding Wilson coefficients given by
\begin{equation}
   \mathcal{C}'_{4V}(m_B) \equiv \frac{\mathcal{C}'_{4R}(m_B) + \mathcal{C}'_{4L}(m_B)}{2}\;, \quad   \mathcal{C}'_{4A}(m_B) \equiv \frac{\mathcal{C}'_{4R}(m_B) - \mathcal{C}'_{4L}(m_B)}{2}\;.
\end{equation}
The vector current part doesn't contribute to $S$-matrix elements since it is conserved below electroweak scale~\cite{Bauer:2020jbp}. Thus, only the axial-vector part needs to be considered for computing the decay width of $B\to Ka$.

\begin{figure}[t]
    \centering
        \includegraphics[width=6in]{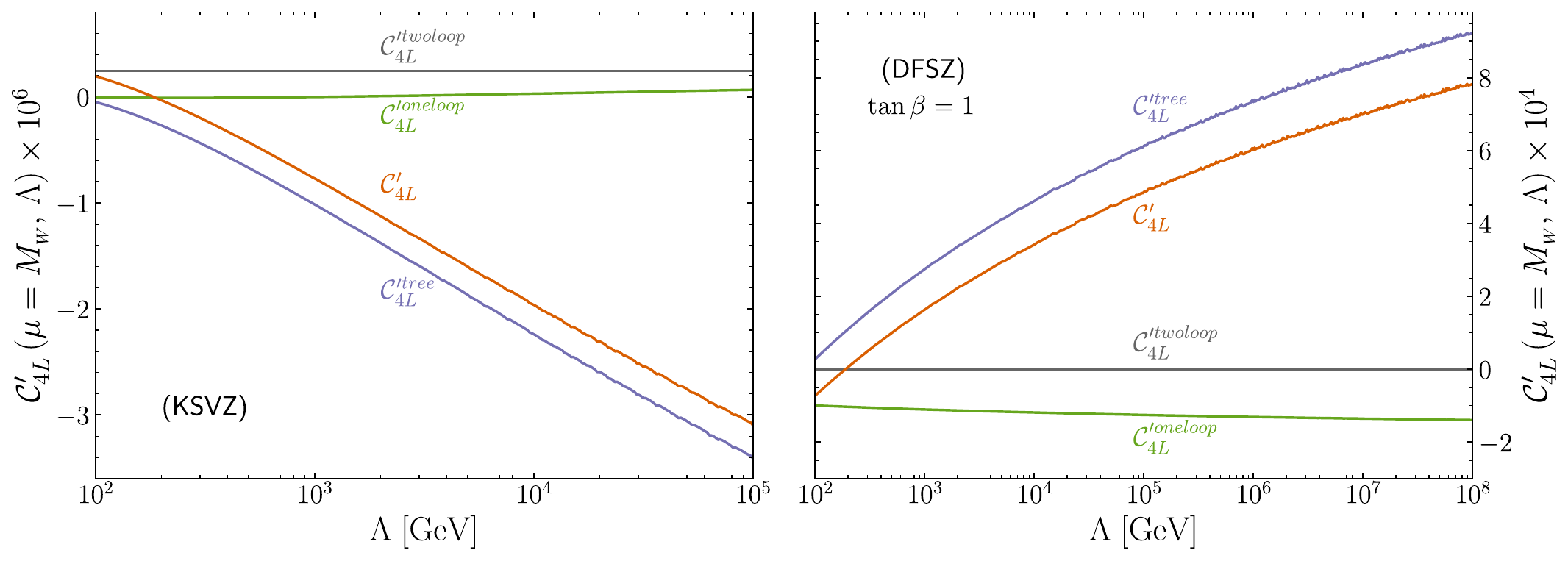}
    \caption{The $\cC'_{4L}(M_{W})$ for the various values of UV scale ($\Lambda$) in the KSVZ model (left) and DFSZ model (right). Violet, green, grey, and orange lines correspond to the tree, one loop, two loop, and total contribution to the $C'_{4L}(M_W)$, Eq.~\eqref{eq:C4LW}, respectively. } 
    \label{fig:C4LW}
\end{figure}

In Fig.~\ref{fig:C4LW}, we show the relative contributions to the Wilson coefficient $\mathcal{C}'_{4L}(M_W)$ as in Eq.~\eqref{eq:C4LW}. On the left-hand side, we present the contributions in the KSVZ-like model for various values of the UV scale. On the other hand, the figure on the right depicts them as a function of $\Lambda$ for DFSZ-like scenarios with $\tan \beta = 1$ (see Eq.~\eqref{eq:tanbeta}). For example, the solid violet line represents the tree-level contribution. Whereas, green, grey, and orange curves depict the one-loop, two-loop, and the total contribution, respectively. This concludes our discussion of the theoretical framework. Before presenting our results, we will first review the existing experimental searches relevant to our framework.

\section{Axion Decays and Branching Fractions}
\label{sec:AxionDecays}

Before discussing the limits on the axion parameter space, it is necessary to first discuss the meson-level decay width for the $B\to K a$ process. As vector current is conserved below the weak scale, only the axial-vector part contributes. We find the decay width to be

\begin{equation}
    \Gamma_{B\to Ka} = |\mathcal{C}'_{4A}(m_B)|^2 \frac{m_B^3}{64\pi f_a^2 }\bigg(1-\frac{m_K^2}{m_B^2} \bigg)^2 \lambda_{Ka}(m_a) f_0(m_a)^2\;,
\end{equation}

where $\lambda_{K_a}$ is the usual K{\"a}ll{\'e}n function, defined as
\begin{equation}
    \lambda_{Ka}(m_a) = \bigg[\bigg(1 - \frac{(m_K + m_a)^2}{m_B^2} \bigg) \bigg(1 - \frac{(m_K - m_a)^2}{m_B^2} \bigg) \bigg]^{1/2} \;,
\end{equation}
and the form factor from lightcone QCD sum rules is given by
\begin{equation}
    f_0(m_a) = \frac{0.330}{1 - [m_a (\text{GeV})]^2/37.5}\;.
\end{equation}
We note in passing that there exists an uncertainty of 10\% in the light-cone form factor.

\begin{figure}[t]
    \centering
    \includegraphics[width=6in]{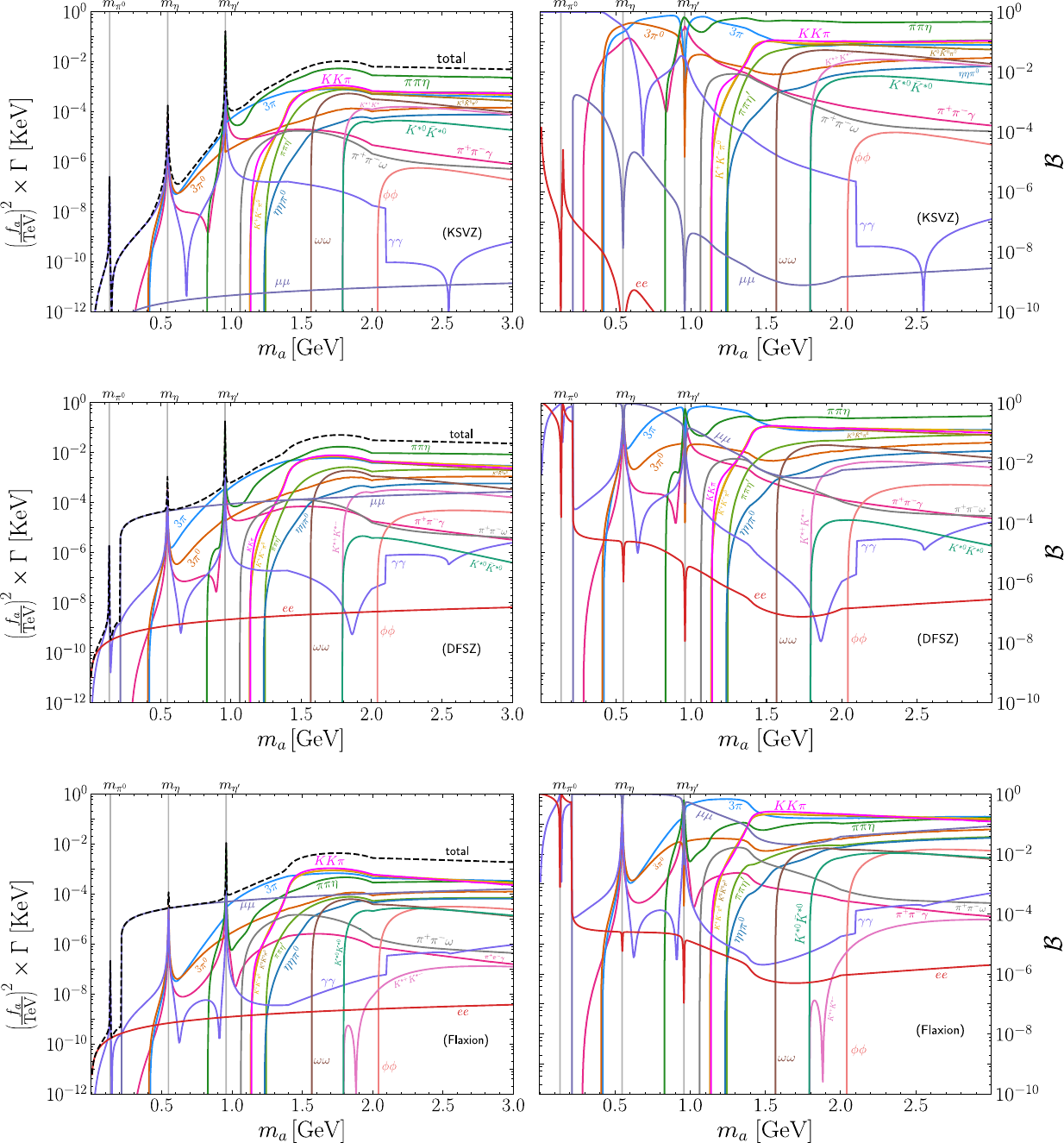}
    \caption{Axion decay widths (left) and branching fractions (right) in KSVZ, DFSZ, and Flaxion models. The total decay width is the sum of these modes, represented by the black dashed curve. $3\pi$ denotes the channel $(\pi^+\pi^-\pi^0)$. Furthermore, the $\pi\pi \eta, \pi \pi \eta'$ and $KK\pi$ labels denote the combined channels $(\pi^0\pi^0 \eta +  \pi^+\pi^-\eta) \;,  (\pi^0\pi^0 \eta' +  \pi^+\pi^-\eta') \;,$ and $ (K^+ \bar{K}^0 \pi^- + K^- K^0 \pi^+)$ respectively.}

\label{fig:DW_BF}
\end{figure}

The produced axion would also decay to various modes. Therefore, in this section, we provide axion decay widths and branching fractions for the KSVZ, DFSZ, and Flaxion models. We use these results in the next section to draw constraints on the space of axion parameters, i.e., the $m_a-f_a$ plane. We take only the most dominant channels such as $a\to\pi\pi\eta$, $KK\pi$, $3\pi$, $\phi\phi, \gamma\gamma$ and $\mu\mu$. However, other modes are also presented in Fig.~\ref{fig:DW_BF} for completeness. To derive the branching fractions, we follow the analysis of~\cite{Cheng:2021kjg} and modify it according to our framework. Detailed calculations are provided in the Appendix \ref{Axion_DW} where ChPT and vector meson dominance are used for matching with the QCD-axion Lagrangian~\eqref{eq:Leffmuchi} at the $\mu_\chi \approx 1-2 \,\unit{GeV}$ scale. Note that the parameters of the QCD-axion Lagrangian are RG evolved with appropriate boundary conditions for different axion-UV models. For the decay widths, we choose the UV scale at $1\,\unit{TeV}$. Since the branching fractions do not vary much with different choices of the UV scales, we use this to give the bounds on the axion parameter space. This framework is then subsequently used to determine the relevant axion decay amplitudes. 

As can be seen in Fig.~\ref{fig:DW_BF}, the decay width and branching fractions for the $a \to \gamma \gamma$ channel differ around $m_a = 2.1\,\unit{GeV}$ for different UV frameworks. This is an artifact of switching off the vector meson-dominance (VMD), whereas the perturbative QCD contribution becomes relevant for $m_a>2.1$ GeV. However, in this region, the differences among the different UV scenarios are mainly attributed to the RG evolved values $\cC_{qq} (\mu_\chi)$. For the other modes, the only contribution comes from ChPT and VMD. This changes only the decay widths and branching fractions quantitatively, but not qualitatively. The peaks around $m_{\pi^0}$, $m_\eta$, and $m_\eta'$ appear due to the divergent nature of the mixing with the axion field; see Eq. \eqref{eq:mixing}. 

\section{Experimental Searches and  Projections}
\label{sec:Exp}

For KSVZ-type scenarios, our analysis focuses mainly on the hadronic interactions of axions, suggesting that the majority of decay channels for the ALP will be hadronic. For other cases, $a\to \mu\mu$ also becomes relevant. Additionally, because of interactions between axions and pions, the axion may also decay into two photons. The branching fractions for these decay modes were evaluated in a data-driven approach in Ref.~\cite{Aloni:2018vki}.

Axion searches can generally be categorized into two types: prompt and displaced. For larger coupling constants or smaller values of $f_a$, the axions decay at the primary vertex, resulting in prompt signals. In contrast, for larger values of $f_a$, the axion lifetime is extended, leading to displaced vertex signatures. The SM background for such processes is relatively small, providing an additional opportunity to probe the axion or ALP parameter space. If the axion has a decay length longer than the dimensions of the detector, it would produce missing-energy signals. However, this scenario implies an even smaller coupling for the axion compared to the displaced topology, resulting in a reduced production cross-section. We first discuss the inclusive $b\to s a$ process followed by tabulating different prompt decays from axions.

\subsection{Inclusive $b\to s a$ process}
We first draw constraints from the $b\to s a$ inclusive process. According to the Particle Data Group, any non-standard decay mode of the $B$-meson should exhibit a branching fraction $<11\%$~\cite{ParticleDataGroup:2020ssz}. This inclusive process is devoid of any hadronic uncertainties and, therefore, extremely robust. The relevant constraints are illustrated by a yellow region in the accompanying plots. For exclusive processes, we also considered the branching fractions of the axions into various final states~\cite{Aloni:2018vki}.

\subsection{Present Constraints}\label{subsec:prompt}
For the present bounds, we mainly consider axions with shorter lifetimes that give rise to prompt decay modes. We note down all the relevant possibilities below.
\begin{enumerate}[leftmargin=*]
\item{$B\to K a \left(a\to\gamma\gamma\right):$} For this particular final state, we follow the strategy from~\cite{Bertholet:2021hjl} where results from BABAR~\cite{BaBar:2021ich} on the branching fraction $\text{Br}\left(B\to K a\right)\times \text{Br}\left(a\to\gamma\gamma\right)$, in the region $0.175~\text{GeV}<m_a<4.78~\text{GeV}$ were revised for heavy QCD axions. Note that, in Ref.~\cite{BaBar:2021ich}, bounds are expressed for different axion lifetimes such as $c\tau_{\text{BBR}}(m_a,f_a)=0$, 0.1, 1 and 10 cm. We took a conservative approach and excluded any values of $m_a$, $f_a$, and the corresponding $c\tau_a$ if it is already excluded by the BABAR analysis satisfying $c\tau_{\text{BBR}}\geq c\tau_a$. We show this exclusion in the {\it slateblue} shaded region of our plots, i.e., Figs.~\ref{fig:KSVZ}, \ref{fig:DFSZ} and \ref{fig:Flaxion}. 

\item{$B\to K a \left(a\to \mu\mu\right):$} Given the sensitivity of muons at Belle and LHCb experiments, a more prominent channel to look for ALPs is via its decay to muons. In the recent past, LHCb experiment has performed both prompt and long-lived searches for scalars and hidden sector bosons, in the channel $B^+\to K^+\chi (\mu^+\mu^-)$\cite{LHCb:2016awg} and $B^0\to K^{0*}\mu^+\mu^-$~\cite{LHCb:2015nkv} respectively. Since, we confine ourselves only with prompt decay modes of the axion, therefore, we only considered the limits for lifetime $\tau=0.1$ ps. The experimental analysis obtained an upper bound on the branching fraction in the ballpark of $10^{-9}$ (see Fig. 3b-S of this \href{https://lhcbproject.web.cern.ch/lhcbproject/Publications/LHCbProjectPublic/LHCb-PAPER-2016-052.html}{link}). As KSVZ framework is devoid of any leptonic couplings at the UV, the bounds obtained from $B\to K a(\mu^+\mu^-)$ are rather conservative. Whereas, for DFSZ and flavorful axion like scenarios, the di-muon channel gives the most promising limits as shown in {\it teal}.

\item $B\to K a \left(a\to \pi\pi \eta\right):$ Ref.~\cite{BaBar:2008rth} reported the branching fractions of $B^+\to \eta_X K^+$, where the $\eta_X$ subsequently decays to $\pi\pi\eta (\pi^0\pi^0\eta+\pi^+\pi^-\eta)$. The identification of $B$-meson candidates was achieved through several kinematic cuts, for example, $1.2~\text{GeV}<m_{\pi\pi\eta}<1.5$~GeV, etc. This invariant mass region is strategically chosen to include a broad spectrum of states below the charm production threshold. From the final reported result, the following constraint~\cite{Aloni:2018vki} was derived $\text{Br}\left(B^{\pm}\to K^{\pm}a\right)\times \text{Br}\left(a\to \pi\pi\eta\right)<2\times 10^{-6}$. Instead of assuming this branching fraction, we follow the strategy of Ref.~\cite{Chakraborty:2021wda}. The experimental collaboration~\cite{BaBar:2008rth} reported events from $B^+\to \eta\pi\pi K^+$ data in a bin size of 6.7 MeV. Motivated by this, we assumed the axion to be present at the boundary of adjacent bins. This is followed by the assumption that the number of events generated by the decay of the axions should be less than the sum of central values in addition to a $2\sigma$ uncertainty of these two bins. As a result, we correct for any spilling-over effects due to smearing. As shown in Figs.~\ref{fig:KSVZ},~\ref{fig:DFSZ} and \ref{fig:Flaxion}, this region provides the strongest bound shown in {\it green} to rule out significant parts of the parameter space in our plots.

\item{$B\to K a \left(a\to K K \pi\right):$} Similar to the previous channel, Ref.~\cite{BaBar:2008rth} also studied the following decay channel $\eta_X\to K K\pi$ with the cut on the invariant mass of the $K\pi$ pair $0.85~\text{GeV}<m_{K\pi}<0.95~\text{GeV}$. This result roughly translates to $\text{Br}\left(B^{\pm}\to K^{\pm}a\right)\times \text{Br}\left(a\to K^{\pm}K_S\pi^{\mp}\right)<1\times 10^{-7}$~\cite{Aloni:2018vki}. This particular final state has also been looked at by LHCb using 3fb$^{-1}$ data~\cite{LHCb:2016utz}. However, their sensitivity remains well below the BaBar experiment. For our purpose, we use the same methodology as discussed for the search $a\to \pi\pi\eta$. Constraints drawn from this search are shown in {\it pink} in our plots.

\item{$B\to K a \left(a\to \phi\phi\right):$} BaBar experiment studied the branching fraction and $CP$ asymmetry for $B\to K\phi\phi$ decay~\cite{BaBar:2011vod} and found to be around $4.5\times 10^{-6}$. Ref.~\cite{Aloni:2018vki} assumed the entire decay process is due to axions and estimated $\text{Br}\left(B\to K a\right)\times \text{Br}\left(a\to\phi\phi\right)<6\times 10^{-6}$. For this particular channel, we assumed the axion to be present in the center of each bin as~\cite{BaBar:2011vod} provides their results in a bin of width 125 MeV. Because of this large bin width, the spilling-over effect due to Gaussian smearing is expected to be well within each of these bins. Finally, we assume that the signal generated from the axion decay is less than the central value augmented with a $2\sigma$ uncertainty. Using this formalism, we exclude the {\it orange} region in our plots.

\item{$B\to K a \left(a\to 3\pi\right):$} The constraint for this particular channel has been derived using the Belle analysis~\cite{Belle:2013nby}, applicable only in the tiny mass window 0.73 GeV -- 0.83 GeV. Recasting their analysis results in a branching fraction $\text{Br}\left[B\to K a\left(a\to\pi^+\pi^-\pi^0\right)\right]<4.9\times 10^{-6}$, derived from $\text{Br}\left(\omega\to3\pi\right)=89\%$ and $\text{Br}\left(B^0\to K^0\omega\right)<5.5\times 10^{-6}$. We use this result to exclude regions shown in {\it blue}.
\end{enumerate}
A crucial point to note is that for searches where the axion decays to $\eta\pi\pi$, $KK\pi$, $\phi\phi$ and $3\pi$, the bounds were roughly stronger by $10\%$ by performing a bin-by-bin analysis rather than using a fixed branching fraction~\cite{Chakraborty:2021wda}.

\subsection{Future Projections}
For future projections, we only use the following prompt signatures 
\subsubsection{$a\to \pi\pi\eta$ and $a\to 3\pi$:}
For prompt signatures, we take the strongest limits coming from $a\to \pi\pi\eta$. As suggested in~\cite{Chakraborty:2021wda}, we extrapolated the background provided in~\cite{BaBar:2008rth} and scaled it with the expected luminosity of Belle II with $5\times 10^{10}$ $B\bar{B}$ pairs. After computing the standard deviation, we assume that the signal from the axion decay should be below the significance $2\sigma$. Similarly, we also follow the same procedure for the $a\to\pi^+\pi^-\pi^0$ channel. In both cases, the experimental resolution of the axion mass is estimated to be $\delta m_a\sim 13.4$ MeV and 40 MeV, respectively, where $\delta m_a\sim \delta m_{\eta\pi\pi}\,m_a/m_{\eta^\prime}$.

\subsubsection{$a\to\gamma\gamma$ and $a\to \mu\mu$}
For projection of the $a\to\gamma\gamma$ channel, we relax the condition of $c\tau_a<10$ cm, as this constraint is too conservative given the size of the Belle II calorimeter. Thereafter, we follow the strategy from~\cite{Bertholet:2021hjl}, where the photon is assumed to be produced promptly at the primary vertex for all lifetime values of the axion. However, the displaced photon analysis is incorporated by assuming a downward smearing on the axion mass, given by
\begin{equation}
    m_{\gamma\gamma}\simeq m_a \left(1-r/S\right)\;, \quad \quad r= \frac{c\tau_a p_a}{m_a}\;,
\end{equation}
where $r$ is the flight distance of the axion, $p_a\sim 2.5$ GeV, is a typical axion momentum. $S\sim 120$ cm, which is roughly the distance between the interaction point and the face of the calorimeter. This translates to a $m_{\gamma\gamma}$ distribution of the form
\begin{equation}
    \frac{dN}{dm_{\gamma\gamma}} = \frac{S}{p_a c\tau_a}\; \text{exp}\left[\frac{S}{p_a c\tau_a}\left(m_{\gamma\gamma}-m_a\right)\right]\Theta\left(m_a-m_{\gamma\gamma}\right)\;,
\end{equation}
We take into account the detector resolution by convoluting this distribution with a Gaussian of width $\sigma_{m_{\gamma\gamma}}\simeq 0.02 m_{\gamma\gamma}$~\cite{BelleIIRes,BaBar:2011bxy}. The expected signal ($N_S$) and background ($N_B$) events are calculated following ref.~\cite{Bertholet:2021hjl} and finally, the projection in the $m_a-f_a$ plane is shown by pink dashed lines when $N_S/\sqrt{N_B}>2$. 

Similarly, for the muon final states, we took a projected luminosity of $300\ \unit{fb}^{-1}$ in light of the LHCb experiment~\cite{LHCb:2021glh}. 

It is important to mention that the signatures of the displaced vertex are also important, as was shown in~\cite{Bertholet:2021hjl}. However, a dedicated study for different UV frameworks considering all channels is required, and we defer this to future work.

\section{Results and Discussions}
\label{sec:results}
In this section, we present our results for different UV scenarios, such as KSVZ, DFSZ, and flavorful axions. The primary difference is the UV boundary condition for the relevant operators defined in Eq.~\eqref{eq:operators}.

Firstly, we define an effective ALP Lagrangian at the UV scale that is manifestly symmetric under the SM gauge group and contains, in general, non-diagonal complex Yukawa interactions (flavor basis)
\begin{equation}
    \mathcal{L}_{\text{UV}}^{\text{flavor}} \supset \mathcal{L}_{\text{\tiny SM}} + \frac{1}{2} \left(\partial_\mu a\right)^2 -\frac{m_a^2}{2} a^2 + \frac{\partial
    _\mu a}{f_a} \sum_F \mathcal{C}_F^{ij} \bar{\psi}_F^i \gamma^\mu \psi_F^j + \mathcal{C}_H \frac{\partial^\mu a}{f_a}(H^\dagger i\overleftrightarrow D_\mu H)  \;. \label{eq:lagflavor}
\end{equation}
Index $F$ runs over SM fermions: $F = \{Q_L, u_R, d_R, L_L, e_R\}$. Firstly, we assume that the redundant operator has been removed at the tree-level as discussed in Appendix~\ref{app:RedOp}. Now, the Yukawa matrices are diagonalized after the following chiral rotations of the fermion fields
\begin{equation}
    u_L \to U_u u_L \;, \quad d_L \to U_d d_L \;, \quad u_R \to W_u u_R \;, \quad d_R \to W_d d_R\;,
\end{equation}
where $U_{u/d}, W_{u/d}$ are unitary matrices. We then arrive at the Lagrangian given in Eq.\eqref{eq:lag}. Thus, the Wilson coefficients in the mass basis are related to those in the flavor basis by
\begin{equation}\label{eq:WCsflavormass}
\begin{aligned}
    &\mathcal{C}_{2L}^{ij} = (U_u^\dagger)^{ik} \mathcal{C}_Q^{kl} U_u^{lj} \;, \quad \mathcal{C}_{3L}^{ij} = (U_d^\dagger)^{ik} \mathcal{C}_Q^{kl} U_d^{lj} \;,\\
    &\mathcal{C}_{2R}^{ij} = (W_u^\dagger)^{ik} \mathcal{C}_{u_R}^{kl} W_u^{lj} \;, \quad\mathcal{C}_{3R}^{ij} = (W_d^\dagger)^{ik} \mathcal{C}_{d_R}^{kl} W_d^{lj} \;.
\end{aligned}   
\end{equation}
In the flavor basis, the coefficients $\mathcal{C}_Q^{ij}$ were the couplings of axion to both up and down-type left-handed quarks due to $SU(2)_L$ symmetry (doublet structure). Whereas in the mass basis, the couplings are $\mathcal{C}_{2L}^{ij}$ and $\mathcal{C}_{3L}^{ij}$. Of course, the number of independent coefficients cannot increase after a basis change. This is ensured by the relation Eq.~\eqref{eq:udrelation} which readily follows from the first two equations in Eq.~\eqref{eq:WCsflavormass} since $V \equiv U_u^\dagger U_d$ is identified as the CKM matrix. Our approach can now be summarized

\begin{itemize}
    \item For models having generic, non-diagonal Yukawas, we first obtain the UV values of the Wilson coefficients in flavor basis by matching with Eq.~\eqref{eq:lagflavor}. Then we use Eq.~\eqref{eq:WCsflavormass} to get the Wilson coefficients in the mass basis.

    \item In some models, e.g. DFSZ-like, the allowed Yukawas can only be diagonal. Consequently, we match directly with the mass basis Lagrangian of Eq.~\eqref{eq:lag}. It must be ensured that Eq.~\eqref{eq:udrelation} is satisfied if the UV model possesses $SU(2)_L$ symmetry.
\end{itemize}

\subsection{KSVZ-Like Model}
The KSVZ model~\cite{Kim:1979if,Shifman:1979if} is described in terms of the UV Lagrangian with a heavy non-SM fermion $Q$ transforming in the fundamental representation of $SU(3)_c$, a singlet representation of $SU(2)_L$ and neutral under hypercharge $U(1)_Y$. 
\begin{equation}
    \mathcal{L}_{\text{KSVZ}} = |\partial_\mu \Phi|^2 + \bar{Q} i \slashed D Q - (y_Q \bar{Q}_L Q_R \Phi + \text{h.c.} ) - V(\Phi)\;.
    \label{eq:KSVZ_Lagrangian}
\end{equation}
\begin{figure}[h]
    \centering
    \includegraphics[width=6in]{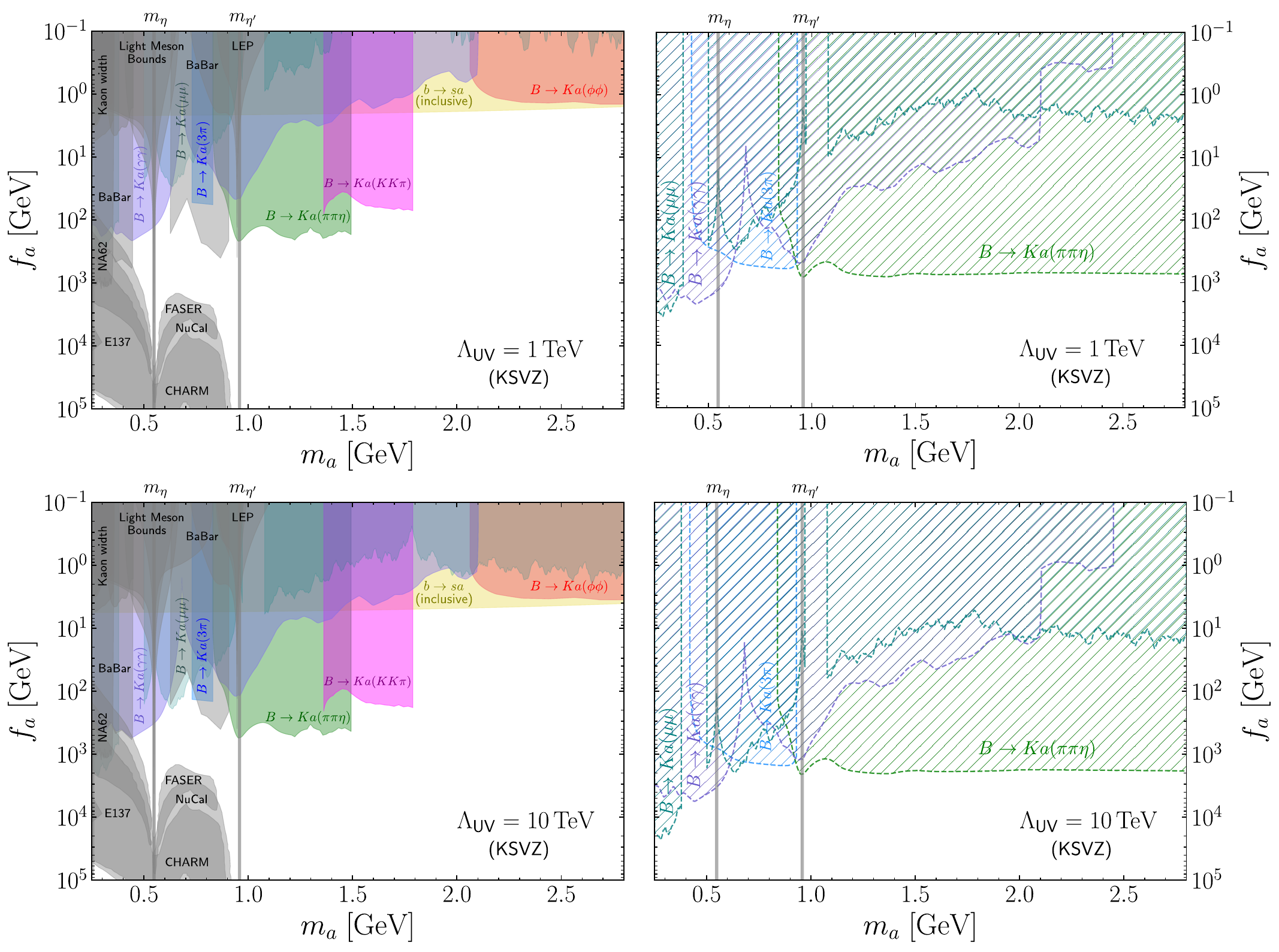}
    \caption{Left panel: We provide constraints on the axion parameter space by solid colored regions by considering the $\Lambda_{\text{UV}}=1$ and 10 TeV respectively. Noticeable constraints come from axion production from $B$-meson and its subsequent prompt decays to $\gamma\gamma$~\cite{BaBar:2021ich} ({\it slate-blue}), $\pi\pi\eta$~\cite{BaBar:2008rth,LHCb:2016utz} ({\it green}), $KK\pi$~\cite{BaBar:2008rth} ({\it pink}), $\phi\phi$~\cite{BaBar:2011vod} ({\it orange}),  $3\pi$ ({\it blue})~\cite{BaBar:2011vod} and $\mu\mu$ ({\it teal})~\cite{LHCb:2015nkv, LHCb:2016awg}. The grey shaded region collectively rules out parameter space from different experiments and bounds such as light meson decays~\cite{Aloni:2018vki,Aloni:2019ruo,Gori:2020xvq,Bauer:2021wjo,OPAL:2002vhf,Knapen:2016moh}, CHARM \cite{CHARM:1985anb}, E137~\cite{Bjorken:1988as}, NuCal~\cite{Blumlein:1990ay, Blumlein:1991xh}, NA62~\cite{NA62:2020pwi, NA62:2021zjw, NA62:2023olg}, BaBar~\cite{BaBar:2021ich}, FASER~\cite{FASER:2024bbl}, and total kaon decay width~\cite{Goudzovski:2022vbt}
    (also see~\cite{Chakraborty:2021wda,Bertholet:2021hjl} for a similar study). Right panel: The future projections from Belle with 3 ab$^{-1}$ and LHCb with 300 fb$^{-1}$ data are shown in dashed lines using prompt decays of the ALP.} 
    \label{fig:KSVZ}
\end{figure}
Eq.~\eqref{eq:KSVZ_Lagrangian} has a global symmetry $U(1)_{PQ}$. After spontaneous symmetry breaking, the scalar field $\Phi$ acquires a vacuum expectation value $f_a/\sqrt{2}$, and the corresponding Goldstone mode is identified as the axion. To obtain the interactions of the axion, one can perform the following chiral transformation on the heavy quark fields
\begin{equation}
    Q \to e^{-i\gamma_5 a/2f_a} Q\;.
\end{equation}
Under such a transformation, the measure of the action is not invariant and it picks up the QCD anomaly term
\begin{equation}
    \mathcal{L}_{\text{KSVZ}} \supset -\frac{\alpha_s}{8\pi f_a} a G \tilde{G}\;.
    \label{eq:KSVZ}
\end{equation}

Comparing Eq.~\eqref{eq:lagflavor} and Eq.~\eqref{eq:KSVZ} gives us the matching coefficient for our operators defined in Eq.~\eqref{eq:operators} at the $\Lambda_{\text{UV}}$ scale. Therefore, we get
\begin{equation}
    \mathcal{C}_1 = - \frac{\alpha_s(\Lambda_{\text{UV}})}{8\pi}\;, \quad \mathcal{C}^{ij}_{2L/R} = \mathcal{C}^{ij}_{3L/R} = 0\;, \mathcal{C}_{eL/R}^{ij} = 0 \;,  \mathcal{C}_H = 0\;.
    \label{eq:initial_conditions_KSVZ}
\end{equation}

Note that axion-fermion couplings will be generated via loops. In \cite{Chakraborty:2021fkp}, such loop-induced UV values were considered for the quark operators, and the sensitivity to UV values was explicitly demonstrated in the exclusion plots. We have verified the same. However, we present our results assuming that only the $a-G-\widetilde{G}$ operator is present in the UV. For such a scenario, the shaded regions in Fig.~\ref{fig:KSVZ} show the present bounds on $f_a$ coming from several final-state searches. The dashed lines represent the projected limits on $f_a$ using the $3\ \text{ab}^{-1}$ data, which is also the projected luminosity of Belle II. Notice that in the literature, sometimes bounds are provided by assuming $\Lambda=f_a$ or $4\pi f_a$. We have checked that, for the former case, we do not get any bounds at all. However, for the latter as well as for projections, we can indeed exclude certain parts of the ALP parameter space, as illustrated in the Appendix~\ref{appn:KSVZ_bound}. The primary reason for obtaining a bound in this case is that we treat the UV scale and $f_a$ independently. As a result, large logarithms from RG running contribute to the Wilson coefficients. However, because of the choice $\Lambda=f_a$, such large logs are absent. Furthermore, the choice $\Lambda=4\pi f_a$ is motivated by a strongly coupled UV theory, which defeats the purpose of a weakly coupled UV framework to solve the strong-CP problem. Nevertheless, we remain agnostic about the UV framework and use only the UV values of the Wilson coefficient motivated by KSVZ. Finally, from Fig.~\ref{fig:KSVZ}, we conclude that the roughly 100 GeV and 1 TeV values of $f_a$ can already be excluded from the present intensity frontier data. This is similar to the limits obtained in~\cite{Chakraborty:2021wda}. 

\subsection{DFSZ-Like Model}\label{sub:DFSZ}
The DFSZ model~\cite{Dine:1982ah, Dine:1981rt} uses a complex scalar field $\Phi$ to decouple the $PQ$ symmetry-breaking scale from the electroweak scale. However, unlike KSVZ, it implements the QCD anomaly as in the Weinberg-Wilczek model through SM quarks charged under $PQ$ symmetry. Thus, two Higgs doublets $H_u$ and $H_d$ are introduced, having Yukawa couplings with SM quarks and leptons.

The $PQ$-symmetry is spontaneously broken when $\Phi$ gets a vev at the scale $v_\Phi$. The exact physical axion field is defined by a linear combination of Goldstone modes of $\Phi, H_u, H_d$. However, since $v_\Phi >> v_u, v_d$ we can approximate the Goldstone mode of $\Phi$ as the full axion for RG analysis above the $M_W$ scale.  Following the analysis of~\cite{DiLuzio:2023tqe},
the couplings of axion are generated by making an axion-dependent redefinition 
\begin{equation}
    \psi_F \to e^{-i\mathcal{X}_F a/f_a}\psi_F \;,
\end{equation}

\begin{figure}[h]
    \centering
    \includegraphics[width=6in]{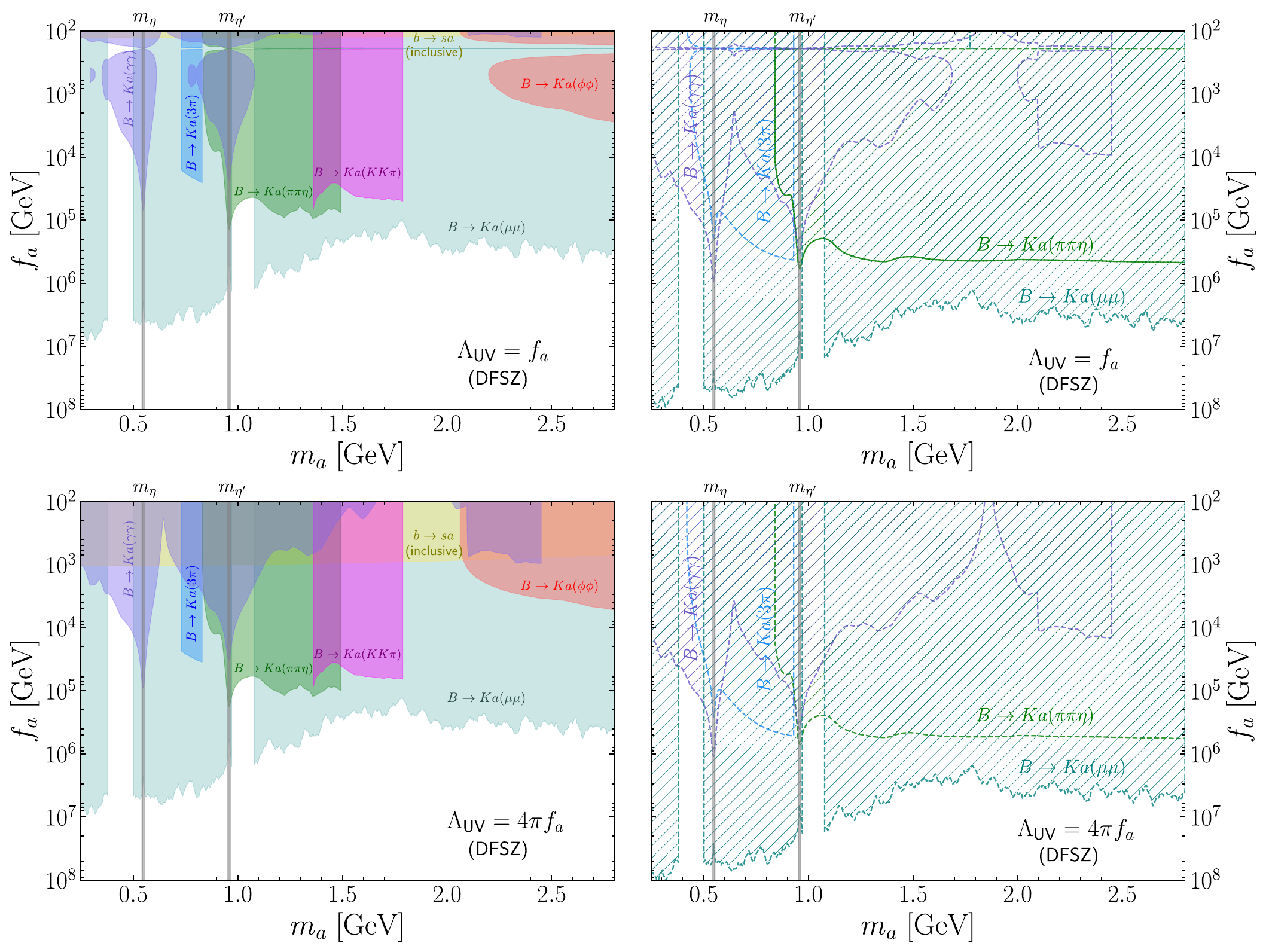}
    \caption{Left panel: Limits on the axion parameter space corresponding to the DFSZ model having initial conditions Eq.~\eqref{eq:initial_conditions_DFSZ} with $\tan \beta = 1$. Right panel: The future projections from Belle, shown in dashed lines using both prompt decay signatures. The shaded regions and dashed lines are to be interpreted similarly to the bounds obtained in the KSVZ model Fig.~\ref{fig:KSVZ}. The grey experimental bounds shown in Fig.~\ref{fig:KSVZ} are all based on the axion-gluon dominated scenario, i.e. KSVZ-like, so we do not keep them here. For different UV models such as DFSZ, quark couplings of the axions are equally dominant and those bounds need to be revisited.}
    \label{fig:DFSZ}
\end{figure}

which leads to the axion couplings $\mathcal{C}_F = \mathcal{X}_F$. We take the charges corresponding to the DFSZ-I scenario: $\mathcal{X}_{Q_L} = \mathcal{X}_{L_L} = 0, \mathcal{X}_{u_R} = -\cos^2\beta, \mathcal{X}_{d_R} = -\sin^2\beta$ and we consider UV values after matching with the Lagrangian having only a single SM-like Higgs doublet~\cite{DiLuzio:2023tqe}. This translates to the following values for the mass basis Wilson coefficients at $\Lambda_{\text{UV}}$

\begin{equation}\label{eq:initial_conditions_DFSZ}
\begin{aligned}
        &\mathcal{C}_1 = \frac{3\alpha_s(\Lambda_{\text{UV}})}{8\pi}\,,\;  \mathcal{C}^{ij}_{2L} = \mathcal{C}^{ij}_{3L} = 0 \,,\; \mathcal{C}^{ij}_{2R} = -\cos^2\beta \;\delta_{ij}\,,\; \mathcal{C}^{ij}_{3R} = -\sin^2\beta \;\delta_{ij}\,\;, \\
        &\mathcal{C}_{eL}^{ij} = 0 \;, \mathcal{C}_{eR}^{ij}= -\sin^2\beta\  \delta_{ij} \;,  \mathcal{C}_H = 0 \;.
\end{aligned}
\end{equation}

\color{black}

The unitarity bound on the scattering amplitude of fermions translates to the following bound on $\beta$ as~\cite{ParticleDataGroup:2024cfk}
 \begin{equation}\label{eq:tanbeta}
     \tan\beta \in [0.28, 140]\;.
 \end{equation}

For a typical value of $\tan\beta=1$, we show the constraints on the $m_a-f_a$ plane for DFSZ-like scenarios in the left panel of Fig.~\ref{fig:DFSZ}. The projections are provided on the right panel with 3 ab$^{-1}$ data from Belle and $300\ \unit{fb}^{-1}$ from LHCb. Unlike the KSVZ framework, here the flavor violation is generated at the one-loop level, which translates to much stronger limits on $f_a$. Notice an island-like region for the top left plot shown in Fig.~\ref{fig:DFSZ}. This is an artifact of the variation of $\mathcal{C}_{4L}^\prime (M_W)$ with $f_a$, where the Wilson coefficient itself varies from negative to positive values. This is due to the interplay between different signs in the loop-induced and tree-level Wilson coefficients. Moreover, the expression for the branching fraction explicitly depends on $1/f_a^2$. The cumulative features enable us to investigate higher values of $f_a$ as depicted in the shaded island region.

\subsection{Flaxion Model}
In this model \cite{Ema:2016ops}, the axion emerges as the Goldstone mode of the BSM complex scalar field $\phi$ (called \textit{flavon}). The model is constructed economically. The flavon field also addresses the flavor structure of the SM, via the Froggatt-Nielsen mechanism \cite{Froggatt:1978nt}. So, the global Abelian flavor symmetry: $U(1)_F$  that generates the Yukawa couplings also acts as the PQ symmetry. 

After the flavon gets a vev, hierarchical Yukawa couplings are generated. We make field-dependent redefinitions proportional to $U(1)_F$ charges
\begin{equation}
    \psi_F \to e^{i q_{_F} a/\sqrt{2} v_\phi} \psi_F
\end{equation}
that generate axion couplings to SM fermion multiplets. The charges $q_{_F}$ are given as
\begin{equation}\label{eq:flaxion_charges}
\begin{pmatrix}
    q_{Q_1} & q_{Q_2} & q_{Q_3}\\
    q_u & q_c & q_t\\
    q_d & q_s & q_b
\end{pmatrix} = \begin{pmatrix}
    3 & 2 & 0\\
    -5 & -1 & 0\\
    -4 & -3 & -3
\end{pmatrix}\;.
\end{equation}
These diagonal, flavor non-universal couplings are in the flavor basis i.e., Yukawas are not diagonalized. To get the mass basis Wilson coefficients, we employ Eq.~\eqref{eq:WCsflavormass} and use the unitary transformation matrices of~\cite{Ema:2016ops}. We note that to reproduce the signs of the CKM elements correctly, we had to incorporate signs in the transformation matrices also. Thus, for completeness, we provide them here in our notation
\small
\begin{equation}
\begin{aligned}
       & U_u = \begin{pmatrix}
        1 & -\epsilon^{q_{Q_1} - q_{Q_2}} & \epsilon^{q_{Q_1} - q_{Q_3}} \\
        \epsilon^{q_{Q_1} - q_{Q_2}} & 1 & -\epsilon^{q_{Q_2} - q_{Q_3}} \\
        \epsilon^{q_{Q_1} - q_{Q_3}} & \epsilon^{q_{Q_2} - q_{Q_3}} & 1 
    \end{pmatrix} \;, U_d = \begin{pmatrix}
        1 & \epsilon^{q_{Q_1} - q_{Q_2}} & \epsilon^{q_{Q_1} - q_{Q_3}} \\
        - \epsilon^{q_{Q_1} - q_{Q_2}} & 1 & \epsilon^{q_{Q_2} - q_{Q_3}} \\
        \epsilon^{q_{Q_1} - q_{Q_3}} & -\epsilon^{q_{Q_2} - q_{Q_3}} & 1 
    \end{pmatrix} \;, \\
   & W_u = \begin{pmatrix}
        1 & \epsilon^{q_c -q_u} & \epsilon^{q_t -q_u}\\
        \epsilon^{q_c -q_u} & 1 & \epsilon^{q_t -q_c}\\
        \epsilon^{q_t -q_u} & \epsilon^{q_t -q_c} & 1
    \end{pmatrix} \;, W_d = \begin{pmatrix}
        1 & \epsilon^{q_s -q_d} & \epsilon^{q_b -q_d}\\
        \epsilon^{q_s -q_d} & 1 & \epsilon^{q_b -q_s}\\
        \epsilon^{q_b -q_d} & \epsilon^{q_b -q_s} & 1
    \end{pmatrix} \;.
\end{aligned}
\end{equation}
\normalsize
Since $ \sqrt{2} v_\phi = f_a N_{\text{DW}}$, we find the following UV values ($\epsilon \sim 0.2, N_{\text{DW}} = 26$) 
\small
\begin{equation}
\begin{aligned}
      &\mathcal{C}_1 = - \frac{\alpha_s(\Lambda_{\text{UV}})}{8\pi}\;, \mathcal{C}_H = 0\;,\\
      & \mathcal{C}_{2L} = \frac{1}{N_{\text{DW}}}\begin{pmatrix}
          -3.08 & 0.20 & -0.008 \\
 0.2 & -2.12 & 0.0848 \\
 -0.008 & 0.0848 & -0.00339 \\
      \end{pmatrix} \;, \mathcal{C}_{3L} = \frac{1}{N_{\text{DW}}}\begin{pmatrix}
           -3.08 & -0.20 & -0.008 \\
 -0.2 & -2.12 & -0.0848 \\
 -0.008 & -0.0848 & -0.00339 \\
      \end{pmatrix} \;, \\
      & \mathcal{C}_{2R} = \frac{1}{N_{\text{DW}}}\begin{pmatrix}
          5. & 0.0096 & 0.00192 \\
 0.0096 & 1.00 & 0.2 \\
 0.00192 & 0.20 & 0.040 \\
      \end{pmatrix} \;, \mathcal{C}_{3R} = \frac{1}{N_{\text{DW}}}\begin{pmatrix}
          4.24 & 2.0 & 2.0 \\
 2.0 & 6.16 & 6.16 \\
 2.0 & 6.16 & 6.16 \\
      \end{pmatrix} \;, \\ 
      & \mathcal{C}_{eL} = \frac{1}{N_{\text{DW}}} \begin{pmatrix}
           -1. & -0.2 & -0.2 \\
 -0.2 & -0.04 & -0.04 \\
 -0.2 & -0.04 & -0.04 \\
      \end{pmatrix} \;, \mathcal{C}_{eR} = \frac{1}{N_{\text{DW}}} \begin{pmatrix}
          8.00 & 0.024 & 0.00192 \\
 0.024 & 5.005 & 0.080 \\
 0.00192 & 0.080  & 3.01 \\ 
      \end{pmatrix}\;.
\end{aligned}
\label{eq:initial_conditions_flaxion}
\end{equation}
\normalsize

\color{black}
\begin{figure}[t]
    \centering
    \includegraphics[width=6in]{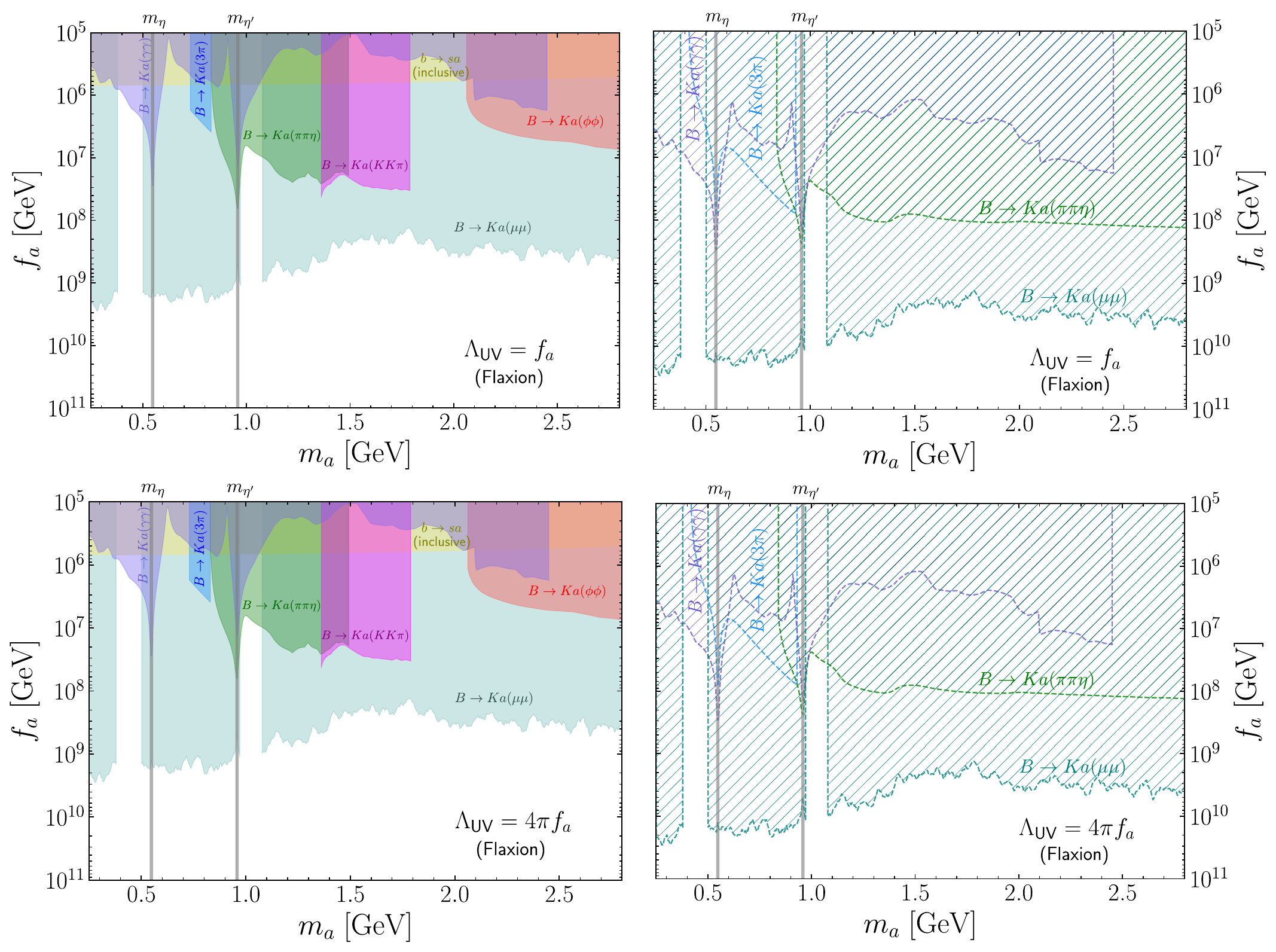}
    \caption{Limits on axion parameter space corresponding to the Flaxion model with the initial conditions of Eq.~\eqref{eq:initial_conditions_flaxion}. As already discussed in the previous plots, the colors and the dashed curves represent the present and future projection bounds on the axion parameter space. Furthermore, we do not show the grey experimental bounds for the same reasons as discussed in Fig.~\ref{fig:DFSZ}}
    \label{fig:Flaxion}
\end{figure}

Finally, we present our results in the left panel and projections in the right panel of Fig.~\ref{fig:Flaxion}. The input parameter values used for Figs.~\ref{fig:KSVZ}, \ref{fig:DFSZ} and \ref{fig:Flaxion} are given in Appendix~\ref{app:input}. We note in passing that non-zero $\mathcal{C}_{4L}$ and $\mathcal{C}_{4R}$ at UV are possible within such flavorful axion models as discussed in~\cite{Berezhiani:1990wn, Berezhiani:1989fp} and evident in the case of Flaxion from Eq.~\eqref{eq:initial_conditions_flaxion}.

For both DFSZ and Flaxion models, we present our bounds and projections choosing $\Lambda=f_a, 4\pi f_a$. We noticed that the bounds on $f_a$ overshoot the UV scale, if we take a similar approach to the KSVZ scenario. This completely invalidates the effective framework. As a result, all our bounds and projections for these two frameworks are given following the choice as mentioned before. We conclude from Figs.~\ref{fig:DFSZ} and~\ref{fig:Flaxion}, that $f_a\sim 10^7$ GeV and $10^9$ GeV can be ruled out using the present experimental data. For a Flaxion-like framework, the limits hardly change while changing the UV scale because the dominant contribution comes from the large tree-level Wilson coefficient, which are unchanged by $\mathcal{O}(\alpha_w)$  effects due to RGEs.

\section{Conclusion}
\label{sec:Conclusion}

Heavy ALPs, with masses in the range of $\mathcal O(100)~\text{MeV}<m_a<\mathcal O(1)~\text{GeV}$, are well motivated because not only they can address the strong-$CP$ problem but also ameliorate the axion quality problem. Such a region is also technically challenging, due to the difficulties in the rigorous perturbative computation of the $b\to s$ transition rate. Moreover, such an axion can also play an important role in explaining the matter-antimatter asymmetry of the universe. Searches for such an axion have gained much attention in the recent past, mostly from the perspective of novel signatures at intensity frontier experiments. In this work, we provide a comprehensive study of such an ALP and provide present bounds and future projections based on experiments such as Belle and BaBar, etc. Our focus mainly lies on hadronic production and decay modes of the axion, along with the unavoidable $a\to \mu\mu$ and $a\to\gamma\gamma$ channels.

We start with the KSVZ-like model where the axion only interacts with SM gluon fields, inducing flavor-changing neutral current processes at two-loop. Such diagrams usually have UV, UV-IR mix, and purely IR divergences. We show that appropriate counterterms, viz. axion-quark diagonal and off-diagonal operators, eliminate both UV and UV-IR mix divergences. These counterterms are simply new operators at the UV, which renormalize the theory. Although all axion-quark off-diagonal operators are generated at the UV, our main focus is on operators that give rise to $b\to s$ transitions. We compute the anomalous dimension matrix at $\mathcal{O}(\alpha_s)$, $\mathcal{O}(\alpha_w)$, $\mathcal{O}(\alpha_t)$, and $\mathcal{O}(\alpha_s\alpha_w)$ (this two-loop contribution is dominant in heavy QCD axion scenarios). The $\mathcal{O}(\alpha_s^2)$ contributions, important for precision studies, only provide corrections to existing terms and do not generate any operators via operator mixing, so we ignore them. The Wilson coefficients are evolved from the UV scale to the electroweak scale $M_W$ by the corresponding RGEs. Afterwards, we show that the pure IR divergences from two loops cancel upon matching with the effective theory below $M_W$. Throughout, we work in naive dimensional regularization with the $\overline{\text{MS}}$ scheme. Finally, we consider only prompt vertex searches of the ALP and provide bounds and future projections on the axion parameter space based on existing data and the 3 ab$^{-1}$ and $300\ \unit{fb}^{-1}$ projected luminosity from Belle and LHCb, respectively. We analyze different hadronic decay modes of the axion such as $a\to \pi\pi\eta$, $a\to K K\pi$ etc as well as $a\to \gamma\gamma$ and $a\to\mu\mu$. For the KSVZ-like model and $\Lambda_{\text{UV}} = 1$ and $10\ \unit{TeV}$, we observe that $f_a$ values of 100 GeV and 1 TeV can be ruled out, respectively, using the present data set from the Belle and BaBar experiments. In addition, larger areas of the parameter space up to $f_a \sim 10$ TeV can also be probed with the projected luminosity of Belle II. Note that for all KSVZ bounds and projections, the axion only talks to SM gluon fields in the UV, and the initial values of all other Wilson coefficients are set to zero. We have checked that the limits are both quantitatively and qualitatively more or less robust if one introduces loop-induced values of the UV coefficients of other operators. 

We show similar exclusions and projections for DFSZ and Flaxion frameworks as well. The UV values of the operators are motivated by ~\cite{Dine:1982ah, Dine:1981rt, Ema:2016ops}. Because of the presence of direct coupling between the axion and SM fermion fields, one obtains stronger bounds for both these cases. It is important to note that the operator $b-s-a$ runs only at $\mathcal{O}(\alpha_w)$. Therefore, we notice negligible sensitivity to the UV scale in the Flaxion model. To conclude, for the DSFZ and Flaxion frameworks, the values of $f_a$ that can be probed with the full luminosity of the Belle II and LHCb data are $10^8$ GeV and $10^{10}$ GeV, respectively. In particular, for the DFSZ-like framework, certain islands have already been ruled out with $f_a \sim 10$ TeV. This peculiar feature is an artifact of the variation of the Wilson coefficient $\mathcal{C}_{4L}^\prime$ as a function of $f_a$, as already discussed in the text.  

\acknowledgments
\vspace{-0.2cm} 
We thank Joydeep Chakrabortty, Manfred Kraus, Vazha Loladze, Takemichi Okui, Ennio Salvioni and Kohsaku Tobioka for discussions and valuable comments on the draft. SC thanks the Science and Engineering Research Board, Government of India (Grant No. SRG/2023/001162), and the IIT-Kanpur initiation grant (PHY/2022220) for financial support. SC also acknowledges the hospitality at ICTS, Bengaluru, during the `School for Advanced Topics in Particle Physics' and TIFR,  Mumbai, where part of the work was done.

\newpage
\appendix
\section{Interaction vertices}
\label{app:FD}
The following vertices arise from the terms $a-q_i - q_j$ and $a-G-\widetilde{G}$ of the UV scale Lagrangian.

\begin{figure}[H]
\begin{tikzpicture}
    \begin{feynman}
    \vertex (a);
    \vertex [below left=of a] (b) {\(\psi_{L,R}\)};
    \vertex [above left=of a] (c) {\(\bar\psi_{L,R}\)};
    \vertex [right=of a] (d) {\(a\)};
    \diagram*{
        (b) -- [fermion, momentum' = {[arrow shorten=0.25] \(p_2\)}](a) -- [fermion, rmomentum' = {[arrow shorten=0.25] \(p_1\)}](c); (d) -- [scalar, rmomentum = {[arrow shorten=0.25] \(l\)}](a)
    };
    \vertex [right=of d] {\( \qquad : \dfrac{\mathcal{C}_{{\psi}_{_{L,R}}}}{f_a} (-\slashed l P_{L,R})\)};
    \end{feynman}
\end{tikzpicture}\\

\begin{tikzpicture}
    \begin{feynman}
    \vertex (a);
    \vertex [below left=of a] (b) {\(\nu ; b \)};
    \vertex [above left=of a] (c) {\(\mu ;a\)};
    \vertex [right=of a] (d) {\(a\)};
    \diagram*{
        (b) -- [gluon, momentum' = {[arrow shorten=0.25] \(p_2\)}](a); (c) -- [gluon, momentum = {[arrow shorten=0.25] \(p_1\)}](a); (d) -- [scalar, rmomentum = {[arrow shorten=0.25] \(l\)}](a)
    };
    \vertex [right=2cm of a] {\(: \dfrac{4i \mathcal{C}_1}{f_a} \delta^{ab}\epsilon^{\mu\nu\rho\sigma} {p_{1\rho}} {p_{2 \sigma}}\)};
    \end{feynman}
\end{tikzpicture}
%\end{figure}

%\begin{figure}[H]
\begin{tikzpicture}
    \begin{feynman}
    \vertex (a);
    \vertex [below left=of a] (b) {\(\rho ;c\)};
    \vertex [above left=of a] (c) {\(\nu ;b\)};
    \vertex [below right=of a] (d) {\(a\)};
    \vertex [above right=of a] (e) {\(\mu ;a\)};
    \diagram*{
        (b) -- [gluon, momentum = {[arrow shorten=0.25] \(p_3\)}](a); (c) -- [gluon, momentum = {[arrow shorten=0.25] \(p_2\)}](a); (e) -- [gluon, momentum = {[arrow shorten=0.25] \(p_1\)}](a); (d) -- [scalar, rmomentum = {[arrow shorten=0.25] \(l\)}](a)
    };
    \vertex [right=2cm of a] {\(: -\dfrac{4\mathcal{C}_1}{f_a} g_s f^{abc}\epsilon^{\mu\nu\rho\sigma}(p_1+p_2+p_3)_\sigma\)};
    \end{feynman}
\end{tikzpicture} 
\caption{The vertex factors of relevant axion couplings with SM quarks and gluons.}
\label{fig:vertices}
\end{figure}

\section{Redundant Operator $\mathcal{O}_H$}\label{app:RedOp}

There is a redundant gauge-invariant operator $\mathcal{O}_H$ possible at dimension-$5$ coupling axion with the SM Higgs doublet $H$. It is convenient here to work in the flavor basis of the SM. The relevant pieces of the Lagrangian at the UV scale are then
\begin{equation}\label{eq:lagred}
    \mathcal{L}^{\text{flavor}}_{\text{UV}} \supset \sum_{F}\mathcal{C}_F^{ij} \frac{\partial_\mu a}{f_a} \bar{\psi}_F^i \gamma^\mu \psi_F^j  + \mathcal{C}_H \underbrace{\frac{\partial^\mu a}{f_a} (H^{\dagger}i \overleftrightarrow{D}_\mu H )}_{\mathcal{O}_H: \text{ redundant}}\;.
\end{equation}
Here $F = \{Q_L, L_L, u_R, d_R, e_R\}$ are the SM chiral fermion multiplets. We use $i,j$ as generation indices. The $\mathcal{O}_H$ operator is redundant because it can be removed from the Lagrangian by the following field-dependent redefinitions~\cite{Bauer:2020jbp, MartinCamalich:2020dfe} 
\begin{equation}\label{eq:redef}
    H \to e^{ic\mathcal{Y}_H a/f_a} H \;, \quad \mathcal{\psi}_F^i \to e^{-ic \mathcal{Y}_F a/f_a }\psi_F^i \;.
\end{equation}
Here $c$ is the redefinition parameter and $\mathcal{Y}_F, \mathcal{Y}_H$ are the respective $U(1)_Y$ hypercharges. This redefinition generates dimension-$5$ axion couplings from the renormalizable part of the SM Lagrangian, shifting the Wilson Coefficients in Eq.~\eqref{eq:lagred} as
\begin{equation}
    \mathcal{C}_H \to \mathcal{C}_H' =  \mathcal{C}_H - c\mathcal{Y}_H  \;, \quad \mathcal{C}_F^{ij} \to (\mathcal{C}_F')^{ij} =  \mathcal{C}_F^{ij} + c \mathcal{Y}_F \delta_{ij} \;.
\end{equation}
Using hypercharge in the redefinition naturally ensures that the SM Yukawa interactions remain unchanged and anomalous axion couplings with gauge fields are not generated. Fixing $c = \mathcal{C}_H/\mathcal{Y}_H$ we get $\mathcal{C}'_H = 0$, i.e., $\mathcal{O}_H$ is removed from the tree-level Lagrangian. Converting the resulting Lagrangian to mass basis gives us the form in Eq.~\eqref{eq:lag}. 

However, removing the redundant operator at the tree-level is insufficient as it can be generated at one-loop, Fig.~\ref{fig:aHH}, affecting the renormalization group evolution. To take this into account, we follow the standard treatment in the literature~\cite{Bauer:2020jbp, MartinCamalich:2020dfe, Arteaga:2018cmw, Jenkins:2013zja}.

\begin{figure}[H]
\begin{center}
\begin{tikzpicture}
    \begin{feynman}
    \vertex (a) {\(a\)};
    \vertex [right=of a] (b) ;
    \vertex [above right=of b] (c) ;
    \vertex [below right=of b] (d) ;
    \vertex [right=of c] (e) {\(H\)};
    \vertex [right=of d] (f) {\(H\)};
    \diagram*{
        (a) -- [scalar](b) -- [fermion](c) -- [fermion] (d) -- [fermion] (b); (c) -- [scalar](e) ; (d) -- [scalar] (f) ;
    };
    \end{feynman}
\end{tikzpicture}
\end{center}
\caption{One-loop operator insertion diagram generating the redundant operator $\mathcal{O}_H$. All SM fermions run in the triangle loop.}
\label{fig:aHH}
\end{figure}

 In the renormalization of any given fermionic bare operator $\mathcal{O}^{ij\ (0)}_F$ we add the contribution from the redundant operator as 
\begin{equation}
\mathcal{O}_{F}^{ij\ (0)} = \mathcal{Z}(F,F')_{ij;kl}\ \mathcal{O}_{F'}^{kl} + \mathcal Z^{\text{red}}(F)_{ij;} \ \mathcal{O}_H \;,
\end{equation} 
$F'$ runs over SM fermion multiplets and $i,j,k,l$ are flavor indices as before. Repeated indices are assumed to be summed. The redundant operator can be written as $\mathcal{O}_H = \mathcal O_{\text{EOM}} + \beta_{F'} \mathcal{O}^{kk}_{F'} $. By the equation of motion for fermions and Higgs field, the $\mathcal O_{\text{EOM}}$ operator vanishes. Therefore, we find 
\begin{equation}
\mathcal{O}_{F}^{ij\ (0)} =  \qty[ \mathcal{Z}(F,F')_{ij;kl} + \mathcal{Z}^{\text{red}} (F)_{ij;} \,\beta_{F'}\, \delta_{kl}]\, \mathcal{O}_{F'}^{kl} \;,
\end{equation}   
where $\beta_F = \mathcal Y_F/ \mathcal Y_H$. To calculate $\mathcal Z^{\text{red}}(F)_{ij;}$ we insert the axion-fermion operators in the $H \to aH$ matrix element (see Fig.~\ref{fig:aHH}) and obtain
\begin{equation}
\begin{aligned}
        \sum_{F, i,j} \mathcal{Z}^{\text{red}}(F)_{ij;}  &=  \frac{1}{16\pi^2} \text{Tr}\bigg[N_c\qty( Y_d^\dagger \mathcal{C}_{Q_L} Y_d - Y_u^\dagger \mathcal{C}_{Q_L} Y_u 
        + Y_u \, \mathcal{C}_{u_R} Y_u^\dagger - Y_d \, \mathcal{C}_{d_R} Y_d^\dagger) \\
        & \quad  + \, Y_e^\dagger \mathcal{C}_{L_L} Y_e - Y_e \, \mathcal{C}_{e_R} Y_e^\dagger \bigg] \frac{1}{\epsilon}\;.
\end{aligned}
\end{equation}
This gives the following contribution to the ADM 
\begin{equation}
    \gamma_{_H} = -2 \sum_{F, i,j} \mathcal{Z}_{(1)}^{\text{red}}(F)_{ij;}  =  \frac{3\alpha_t}{2\pi} \qty(\cC^{33}_{2L} - \cC^{33}_{2R})\;,
\end{equation}
where $\mathcal{Z}_{(1)}$ is defined in Eq.~\eqref{eq:Renormalization_matrices_expansion}. We have ignored all other Yukawa couplings except for the top quark. This directly enters the RGEs for diagonal axion-fermion Wilson coefficients, as shown in Eq.~\eqref{eq:runningWCs}.

\section{Example of a two-loop amplitude}\label{app:two_loop}
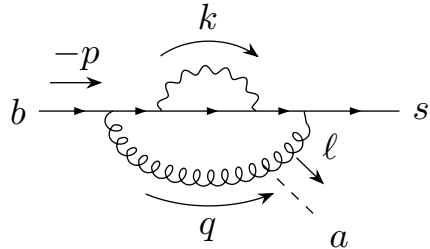
\begin{figure}[H]
\begin{center}
\resizebox{6cm}{!}{%
\begin{tikzpicture}
    \begin{feynman}
    \vertex (a) {\(b\)};
    \vertex [right=1cm of a] (b);
    \vertex [right=0.5cm of b] (c);
    \vertex [right=1cm of c] (d);
    \vertex [right=0.5cm of d] (e);
    \vertex [right=1cm of e] (f) {\(s\)};
    \vertex at ($(b)!0.5!(e)!0.90!135:(b)$) (g);
    \vertex [below right=0.8cm of g] (h) {\(a\)};
%    \vertex []
    \diagram* {
            (a) -- [fermion , momentum = { \(-p\)}] (b) -- [fermion] (c) -- [fermion] (d) -- [fermion] (e) -- [fermion] (f), (c) -- [boson, half left,  momentum = {[arrow shorten=0.35] \(k\)}] (d), (b) -- [gluon, half right, looseness = 1.2, momentum' = {[arrow shorten=0.4] \(q\)}] (e), (g) -- [scalar, momentum' = {[arrow shorten=0.25] \(l\)}] (h)
        };
        \end{feynman}
\end{tikzpicture}
}
\end{center}
\caption{Illustrative example of a 1PI two-loop  diagram contributing to $b\to sa$.}
\label{fig:appendixB1}
\end{figure}

In this appendix, we work out an explicit calculation illustrating the detailed steps we have employed to evaluate the two-loop 1PI diagrams in our analysis. We take the example of Fig.~\ref{fig:appendixB1} as an illustration.

The amplitude for the $b\to s a$ process initiated by the $a-G-\widetilde{G}$ coupling is
\begin{multline}
        i \Sigma_2^W  = \mu^{8-2d} \sum_i \int  \frac{d^dk}{(2\pi)^d} \frac{d^dq}{(2\pi)^d} \; \Bigl\{i g_s \gamma^\mu T^a_{ik}\Bigl\} \qty[\frac{i (- \sl p - \sl q)}{(p+q)^2}] \qty{\frac{i g}{\sqrt{2}}V_{is}^\ast \gamma^\nu P_L}  \\
        \times \qty[\frac{i (- \sl p - \sl k - \sl q + m_u)}{(p+k+q)^2 - m_i^2}] \qty{\frac{i g}{\sqrt{2}}V_{ib} \gamma^\rho P_L} \qty[\frac{i (- \sl p - \sl q)}{(p+q)^2}] \qty{i g_s \gamma^\sigma T^b_{kj}} \\
        \times \qty[\frac{-i g_{\rho\nu}}{k^2 -M_W^2} ]\qty[\frac{-i g_{\mu\alpha} \,\delta_{ac}}{(q-l)^2 - m_R^2}] \qty{\frac{4 i \mathcal C_1}{f_a} \epsilon^{\alpha\beta\kappa\omega} (- q +l)_\kappa\, q_\omega\, \delta_{cd}} \qty[\frac{-i g_{\beta\sigma} \, \delta_{db}}{q^2 - m_R^2}],
\end{multline}
where the sum is over all the up-type quark flavors. We have taken the external light quark states to be massless, i.e., $m_s, m_b \to 0$. This also induces an IR divergence. To properly obtain the UV poles, we have used a fictitious mass parameter $m_R$ for the gluon that acts as an IR regulator. 
\color{black}
We now expand the external momenta about $p,l = 0$, where $l$ is the axion momentum. We only retain the leading-order term(s) in the numerator since we work in the approximation of vanishing external momenta. For the above integral, this gives
\begin{multline}
    i \Sigma_2^W  \approx 2i \,g^2 g_s^2 \,C_F \delta_{ij} \qty(\frac{C_1}{f_a}) \mu^{8-2d} \int  \frac{d^dk}{(2\pi)^d} \frac{d^dq}{(2\pi)^d} \;\sum_i V_{is}^\ast V_{ib} \\ \times \frac{\gamma_\alpha (-\sl q + m_R)\gamma^\nu P_L (- \sl k - \sl q + m_u) \gamma_\nu P_L (-\sl q + m_R) \gamma_\beta}{\qty(k^2 - M_W^2)\,\qty(q^2 - m_R^2)^2 \,  (q^2)^2 \, \qty[(k+q)^2 - m_i^2]} \, \epsilon^{\alpha\beta\rho\sigma} l_\rho q_\sigma \;.
\end{multline}

\vspace*{0.5cm}
\noindent \textbf{Feynman parameterization:}  The denominator can now be clubbed via Feynman parameterization, and we define
\begin{equation}
\begin{aligned}
        \mathcal{T}(a,b,c) & \equiv \mu^{8-2d} \int  \frac{d^dk}{(2\pi)^d} \frac{d^dq}{(2\pi)^d} \;\frac{1}{\qty[k^2 - M^2(x)]^a \, \qty[q^2 - M_R^2(y)]^b \, \qty[(k+q)^2 - m_i^2]^c} \\
        & = \mu^{8-2d} \int  \frac{d^dk}{(2\pi)^d} \frac{d^dq}{(2\pi)^d} \; \frac{1}{\mathcal{P}_1^a\, \mathcal{P}_2^b \,\mathcal{P}_3^c}\;.
\end{aligned}
\label{eq:appBpropagators}
\end{equation}
Here $x,y$ are the Feynman parameters. We have $M^2(x) = x\,m_i^2 + (1-x)M_W^2$ for the first, third, and fourth two-loop diagrams (Fig.~\ref{fig:7}) and $M^2(x) = M_W^2$ for the second diagram (this example). Similarly, $M_R^2(y) = m_R^2$ for the first diagram and $M_R^2(y) = m_R^2(1-y)$ for other diagrams. We now provide the details for the tensor reduction. 

\vspace*{0.5cm}
\noindent \textbf{Tensor reduction:} Once the Dirac algebra of the numerator is done in $d$-dimensions, the tensorial quantities are expressed in terms of dot products via the following relations:

\begin{equation}
\begin{aligned}
& T_{40}(q_\mu q_\nu q_\alpha q_\beta) = \frac{(q.q)^2}{d(d+2)}(\eta_{\mu\nu}\eta_{\alpha\beta} + \eta_{\mu\alpha}\eta_{\nu\beta} + \eta_{\mu \beta}\eta_{\nu\alpha} )\;, \\
& T_{04}(k_\mu k_\nu k_\alpha k_\beta) = \frac{(k.k)^2}{d(d+2)}(\eta_{\mu\nu}\eta_{\alpha\beta} + \eta_{\mu\alpha}\eta_{\nu\beta} + \eta_{\mu \beta}\eta_{\nu\alpha} )\;,  \\
& T_{31}(q_\mu q_\nu q_\alpha k_\beta) = \frac{(q.q)(q.k)}{d(d+2)}(\eta_{\mu\nu}\eta_{\alpha\beta} + \eta_{\mu\alpha}\eta_{\nu\beta} + \eta_{\mu \beta}\eta_{\nu\alpha} )\;, \\
& T_{13}(k_\mu k_\nu k_\alpha q_\beta)  = \frac{(k.k)(k.q)}{d(d+2)}(\eta_{\mu\nu}\eta_{\alpha\beta} + \eta_{\mu\alpha}\eta_{\nu\beta} + \eta_{\mu \beta}\eta_{\nu\alpha} )\;, \\ 
& T_{22}(q_\mu q_\nu k_\alpha k_\beta) = A\eta_{\mu\nu}\eta_{\alpha\beta} + B\eta_{\mu\alpha}\eta_{\nu\beta} + C \eta_{\mu\beta}\eta_{\nu\alpha}\;, \\
& T_{30}(q_\mu q_\nu q_\alpha) = 0\;, \quad  T_{03}(k_\mu k_\nu k_\alpha) = 0\;,  \\
& T_{02}(k_\alpha k_\beta)  =  \frac{k.k}{d}\eta_{\alpha\beta}\;,\quad T_{20}(q_\mu q_\nu)  = \frac{q.q}{d}\eta_{\mu\nu}\;,\quad T_{11}(q_\mu k_\alpha) = \frac{q.k}{d}\eta_{\mu\alpha}\;.
\end{aligned}
\end{equation}
The coefficients $A, B$ and $C$ can be obtained in a straightforward way by taking appropriate dot products with $\eta^{\star\star}$, where the $\star$'s denote appropriate Lorentz indices.
\begin{equation}
     A = \frac{(1+d)k^2q^2 - 2(q.k)^2}{d(d-1)(d+2)} \;, \quad\quad B = C = \frac{d(q.k)^2 - q^2k^2}{d(d-1)(d+2)}\;. 
\end{equation}
Then, using \eqref{eq:appBpropagators}, the following substitutions are made in the numerator
\begin{equation}
    k.k = \mathcal{P}_1 + M^2(y),\quad q.q = \mathcal{P}_2 + m_R^2,\quad k.q = \frac{1}{2}(\mathcal{P}_3 - \mathcal{P}_2 - \mathcal{P}_1 + m_q^2 -M^2(y) - m_R^2)\;.
\end{equation}
This converts every term in the two-loop integral to the form $\mathcal{T}(a,b,c)$ (multiplied by mass scales and numerical factors). Next, we use \texttt{KIRA}~\cite{Maierhofer:2017gsa} for expressing these integrals on the basis of the four master integrals $\mathcal{T}(1,1,1)$, $\mathcal{T}(1,0,1)$, $\mathcal{T}(1,1,0)$, $\mathcal{T}(0,1,1)$~\cite{Davydychev:1992mt, Adams:2015gva}.
%here we added the citations.

\vspace*{0.5cm}
\noindent\textbf{Final evaluation}: We now substitute the explicit expressions of these four master integrals. The result is expanded about $m_R = 0$. The leading order term in the series contains a logarithm of $m_R$ and all higher order terms drop out in the limit $m_R \to 0$. Finally, integrating over the Feynman parameter $y$ with the appropriate numerator factor, we obtain the result for Fig. \ref{fig:appendixB1} as:

\begin{equation}
    i\Sigma^2_W = \kappa \bigg[-\frac{1}{\epsilon^2} + \frac{1}{\epsilon}\qty{- 2\log \qty(\frac{m_R^2}{M_W^2})  - \frac{1}{2} } + \;\text{F.T.} \bigg](-\slashed l P_L)\; ,
\end{equation}

where $\kappa$ is defined in Eq.~\eqref{eq:kappa}. The result of the corresponding Goldstone diagram has to be added for the complete evaluation of the diagram. As mentioned in the text, the UV and UV-IR divergences go away due to the insertions of appropriate counterterms. However, we still have pure IR divergences, which we have shown explicitly cancel out as a result of matching with the EFT amplitude.

\section{Bounds for KSVZ-Like scenario ($\Lambda=4\pi f_a$)}
\label{appn:KSVZ_bound}
\begin{figure}[H]
    \centering
    \includegraphics[width=6in]{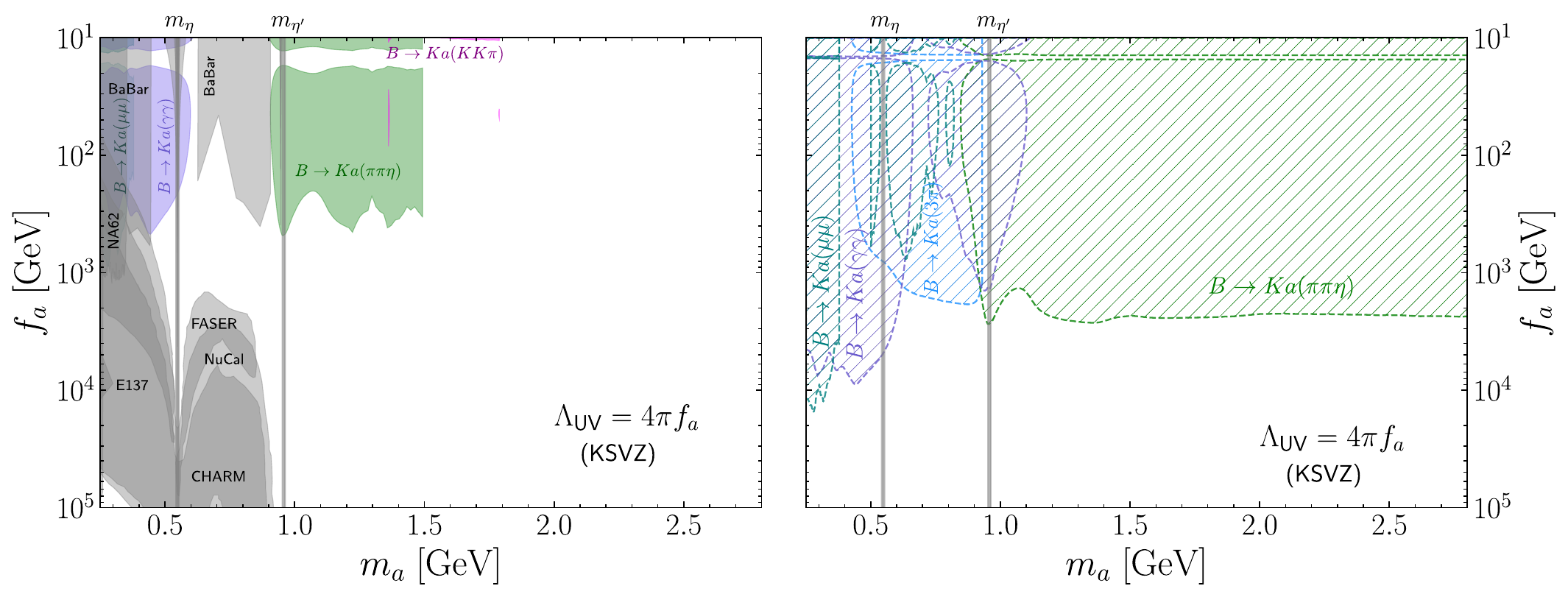}
    \caption{Bounds and projections for the KSVZ-like framework with the choice $\Lambda=4\pi f_a$. The reasons for the island like features have already been discussed in the text, see Sec.~\ref{sub:DFSZ}.} 
    \label{fig:KSVZ2}
\end{figure}

\section{Input Parameters}\label{app:input}

In table~\ref{tab:parameters_values}, we collect the experimental values of the important parameters for numerical estimation from PDG-2024 \cite{ParticleDataGroup:2024cfk}.
\begin{table}[H]
\centering
\begin{tabular}{|c|c||c|c||c|c|}
     \hline
     \textbf{Parameters} & \textbf{Values} & \textbf{Parameters} & \textbf{Values} & \textbf{Parameters} & \textbf{Values} \\
     \hline
     &&&&&\\[-2ex]
     $G_F $ & $ \num{1.1664e-5} \,\unit{GeV}^{-2}$ & $M_W $ & $ 80.369 \,\unit{GeV} $ & $M_Z $ & $ 91.188 \,\unit{GeV}$ \\ 
     $\alpha_s \,(M_Z) $ & $ 0.118$ & $\alpha_w \,(M_W) $ & $ 0.339$ & $\alpha_e \,(M_Z) $ & $ 1/127.9 $ \\ 
     $\sin^2(\theta_W) $ & $ 0.231$ & $m_B $ & $ 5.28 \,\unit{GeV}$ & $m_K $ & $ 0.498 \,\unit{GeV}$ \\
     $m_d \,(2\,\unit{GeV}) $ & $ 0.007 \,\unit{GeV}$ & $m_s \,(2\,\unit{GeV}) $ & $ 0.12 \,\unit{GeV}$ & $m_b \,(m_b) $ & $ 4.3 \,\unit{GeV}$ \\ 
     $m_u \,(2\,\unit{GeV}) $ & $ 0.003 \,\unit{GeV}$ & $m_c \,(m_c) $ & $ 1.2 \,\unit{GeV}$ & $m_t \,(m_t) $ & $ 162.69 \,\unit{GeV}$ \\ 
     $m_t^{pole} $ & $ 172.5 \,\unit{GeV}$ & $y_t \,(m_t^{pole}) $ & $ 1.150$ & $v $ & $ 246 \,\unit{GeV}$\\
     $V_{ud} $ & $ 0.974$ & $V_{us} $ & $ 0.224$ & $V_{ub} $ & $  0.004$ \\
     $V_{cd} $ & $ -0.221$ & $V_{cs} $ & $ 0.975$ & $V_{cb} $ & $  0.0411$ \\
     $V_{td} $ & $ 0.009$ & $V_{ts} $ & $ -0.042$ & $V_{tb} $ & $  1.010$ \\[1ex]
     \hline
\end{tabular}
\caption{We present the values of the parameters used in our work.}
\label{tab:parameters_values}
\end{table}

\section{Axion decay width} \label{Axion_DW}

%\allowdisplaybreaks
The ALP Lagrangian, described in Eq.~\eqref{eq:LeffmuW} can be further evolved down to $\mu_\chi = 1-2 \,\unit{GeV}$, where the only relevant degrees of freedom are the three light quarks, i.e. $q = \{u,d,s\}$ and the leptons $\ell = \{e, \mu\}$. 
\begin{equation}
\begin{aligned}\label{eq:Leffmuchi}
    {\cal L}_{\rm eff}(\mu_\chi) & = \frac12 \qty( \partial_\mu a)^2 - \frac{m_{a}^2}{2}\,a^2
    + \cC_{gg}\,\frac{\alpha_s}{4\pi}\,\frac{a}{f_a}\,G_{\mu\nu}^b\,\tilde G^{\mu\nu,b}
    + \cC_{\gamma\gamma}\,\frac{\alpha}{4\pi}\,\frac{a}{f_a}\,F_{\mu\nu}\,\tilde F^{\mu\nu} \\
    &\quad + \bar q(i\sl D - \bm{m}_q)q \,
    + \,\frac{\partial_\mu a}{f_a} \,\bar q \,\bm{\cC}_{qq}\gamma^\mu \gamma_5 q \,
    + \,\frac{\partial_\mu a}{f_a} \,\bar \ell \,\bm{\cC}_{\ell\ell}\gamma^\mu \gamma_5 \ell  \,,
\end{aligned}
\end{equation}
where $\bm{m}_q = \text{diag}\,(m_u,m_d,m_s)$. For generality, we keep the term of the axion-photon interaction, but consider $\cC_{\gamma\gamma} = 0$ and $\bm{\mathcal{C}}_{qq} = \text{diag}(\mathcal{C}_{uu}, \mathcal{C}_{dd}, \mathcal{C}_{ss})$ for our analysis. Comparing with Eq.~\eqref{eq:lag}, $\mathcal{C}_1 = C_{gg}\ (\alpha_s/4\pi)$. To facilitate matching with the Chiral Lagrangian, we expressed the chiral axion-fermion currents in the form of vector and axial-vector currents. The vector current part doesn't contribute to $S$-matrix elements since it is conserved below electroweak scale~\cite{Bauer:2020jbp}. Thus, only the axial-vector part needs to be considered for low-energy phenomenology.

It is customary and convenient to rotate away the axion-gluon interaction by performing a chiral rotation on the quark fields given by
\begin{equation}
    q \to \exp(-i \bm{\kappa}_q \cC_{gg} \frac{a}{f_a} \gamma_5) q\;.
\end{equation}
Imposing the condition $\Tr(\bm{\kappa_q}) = 1$, where $\bm{\kappa}_q$ is a real $3\times 3$ matrix, leads to the effective Lagrangian
\begin{equation}\label{Leffmuchi_chiral}
\begin{aligned}
        {\cal L}_{\rm eff}(\mu_\chi) &= \frac12 \qty( \partial_\mu a)^2 - \frac{m_{a}^2}{2}\,a^2 + \hat\cC_{\gamma\gamma}\,\frac{\alpha}{4\pi}\,\frac{a}{f_a}\,F_{\mu\nu}\,\tilde F^{\mu\nu} +  \bar q\,i\sl Dq - (\bar q_{_R}\bm{\hat m}_q^\dagger\,q_{_L} + \bar q_{_L}\bm{\hat m}_q\,q_{_R}) \\
    & \quad + \frac{\partial_\mu a}{f_a} \,\bar q \,\bm{\hat{\cC}}_{qq}\gamma^\mu \gamma_5 q \, + \,\frac{\partial_\mu a}{f_a} \,\bar \ell \,\bm{\cC}_{\ell\ell}\gamma^\mu \gamma_5 \ell ,
\end{aligned}
\end{equation}
where,

\begin{equation}
\begin{aligned}
    \bm{\hat{m}}_q & =  e^{-i \bm{\kappa}_q \cC_{gg} a /f_a} \,\bm{m}_q \, e^{-i \bm{\kappa}_q\cC_{gg} a /f_a}\,, \\
    \bm{\hat{\cC}}_{qq} & = \bm{\cC}_{qq} +  \bm{\kappa}_q  \cC_{gg}\,,\\
    \hat{\cC}_{\gamma\gamma} & = \cC_{\gamma\gamma} - 2 N_c \,\cC_{gg} \sum_q Q_q^2{\kappa}_q\,.
\end{aligned}
\end{equation}

The major advantage of such a chiral rotation is that the theory is now completely written in terms of the quark fields. As a result, it can be matched with an effective theory of mesons and baryons, such as the chiral Lagrangian. However, the interaction between the axion and vector mesons can be included in a phenomenological way through the vector meson dominance interactions. So, the Lagrangian is given by~\cite{Georgi:1986df}
\begin{equation}
\begin{aligned}
    {\cal L}_{\text{ChPT}} 
    &= \frac12\,\partial^\mu a\,\partial_\mu a - \frac{m_{a}^2}{2}\,a^2
     + \frac{f_\pi^2}{8}\,\Tr\big[ D^\mu\bm{\Sigma}\,D_\mu\bm{\Sigma}^\dagger \big] 
     + \frac{f_\pi^2}{4}\,B_0\,\Tr\big[ \bm{\Sigma}\,\hat{\bm{m}}_q^\dagger
     + \hat{\bm{m}}_q\,\bm{\Sigma}^\dagger \big] \\
    & \quad + \frac{if_\pi^2}{4}\,\frac{\partial^\mu a}{f_a}\,
     \Tr\big[ \hat{\bm{\cC}}_{qq} (\bm{\Sigma}\,D_\mu\bm{\Sigma}^\dagger 
     - \bm{\Sigma}^\dagger D_\mu\bm{\Sigma}) \big]
     + \hat \cC_{\gamma\gamma}\,\frac{\alpha}{4\pi}\,\frac{a}{f_a}\,F_{\mu\nu}\,\tilde F^{\mu\nu} + \cL_{\text{VMD}}\,,
 \end{aligned}
\end{equation}
with,
\begin{equation}\label{P_matrix}
    \bm\Sigma = \exp(2i\bm{P}/f_\pi), \qquad \bm{P} = \mqty(\frac{\pi^0}{\sqrt{2}} + \frac{\eta}{\sqrt{3}} + \frac{\eta'}{\sqrt{6}} & \pi^+ & K^+ & \\ \pi^- & - \frac{\pi^0}{\sqrt{2}} + \frac{\eta}{\sqrt{3}} + \frac{\eta'}{\sqrt{6}} & K^0 \\ K^- & \overline{K}^{\,0} & - \frac{\eta}{\sqrt{3}} + \frac{2\eta'}{\sqrt{6}})\;.
\end{equation}
We consider $f_\pi$ or the pion decay constant to be $\sim 130 \,\unit{MeV}$. The chiral condensate parameter that captures the isospin-breaking effects is $B_0 \approx m_\pi^2/(m_u + m_d)$, the covariant derivative also has the usual form $D_\mu\bm{\Sigma}=\partial_\mu\bm{\Sigma}-ieA_\mu [\bm{Q},\bm{\Sigma}]$, where $\bm Q = \text{diag}(Q_u,Q_d,Q_s)$. The usual choice of $\bm{\kappa}_q = \bm{m}_q^{-1}/\Tr(\bm{m}_q^{-1})$ eliminates the mass mixing between axion and $\pi^0$, $\eta_8$ mesons. Furthermore, the physical $\eta,\eta'$ are related to the octet and singlet fields by \cite{ParticleDataGroup:2024cfk}
\begin{equation}
\mqty(\eta \\ \eta') = \mqty(\cos\theta & - \sin\theta \\ \sin\theta & \cos\theta)  \mqty(\eta_8 \\ \eta_0), \quad \sin\theta \approx - \frac{1}{3}\,,\quad \cos\theta \approx \frac{2\sqrt{2}}{3}\,.
\end{equation}
Finally, the vector meson dominance Lagrangian \cite{Fujiwara:1984mp} is given by
\begin{equation}
 \begin{aligned} \label{eq:VMD}
    \mathcal{L}_{\text{VMD}} =&\, \frac{g_
    {VVP}}{2}  \Tr \big( \bm{P} \bm{V}_{\mu\nu} \bm{\widetilde{V}}^{\mu\nu} \big) +  \frac{i \,N_c e}{6\pi^2 f_\pi^3} \epsilon^{\mu\nu\rho\sigma} A_\mu \Tr( \bm{Q} \partial_\nu \bm{P} \partial_\rho \bm{P} \partial_\sigma \bm{P}) \\ 
    & \quad +\, f_\pi^2 \qty[\Tr(\frac{g}{\sqrt 2} \bm{V}_\mu - e A_\mu \bm{Q} - \frac{i}{2 f_\pi^2} [\bm{P}, \partial_\mu \bm{P} ])]^2 \,,
\end{aligned} 
\end{equation}
where the vector meson matrix is
\begin{equation} \label{V_matrix}
    \bm{V} =  \mqty( \frac{\rho_0 + \omega}{\sqrt{2}} & \rho^+  & K^{\ast +} \\ \rho^- & \frac{-\rho_0 + \omega}{\sqrt{2}} & K^{\ast 0} \\ K^{\ast -} & \overline{K}^{\ast 0} & \phi ) \,, \;\;\text{with}\;\; g_{VVP} = - \frac{N_c g^2}{8\pi^2 f_\pi}\;, \; g = \frac{m_\rho}{f_\pi}\approx 6.0\;.
\end{equation}

Diagonalization of the kinetic and mass term at the $\order{f_\pi/f_a}$ in the ChPT Lagrangian leads to a shift in the axion and meson fields by \cite{Cheng:2021kjg,Aloni:2018vki, DallaValleGarcia:2023xhh}
\begin{equation}\label{eq:mixing}
    a \to a - \frac{f_\pi}{\sqrt 2 f_a} \sum_{P} \frac{m_P^2}{m_a^2} \Tr(\bm{aP}) P\,, \; P \to P + \frac{f_\pi}{\sqrt 2 f_a} \Tr(\bm{aP})\, a \,, \; \Tr(\bm{aP}) = \frac{m_a^2 K_{aP}}{m_P^2 - m_a^2}\,,
\end{equation}
where $P = \pi^0, \eta, \eta'$, and $K_{aP}$ is defined as
\begin{equation}
    K_{a\pi^0} = \hat\cC_{uu} - \hat\cC_{dd}\,, \quad K_{a \eta} = \sqrt{\frac{2}{3}} (\hat\cC_{uu} + \hat\cC_{dd} - \hat\cC_{ss})\,,\quad K_{a\eta'} = \frac{1}{\sqrt{3}} (\hat\cC_{uu} + \hat\cC_{dd} + 2 \,\hat\cC_{ss}).
\end{equation}
The effects of isospin breaking due to $m_u \neq m_d$ are neglected. Following the notation of~\cite{Aloni:2018vki}, we assign the axion field in $U(3)$ representation as
\begin{equation}
    \bm a = \sum_P \Tr(\bm{aP})\bm P\,,
\end{equation}
where from (\ref{P_matrix}), we see
\begin{equation}
    \bm{\pi^0} = \frac{1}{\sqrt 2}\; \text{diag}(1,-1,0), \quad \bm\eta = \frac1{\sqrt3}\; \text{diag}(1,1,-1), \quad \bm{\eta'} = \frac1{\sqrt{6}}\; \text{diag}(1,1,2).
\end{equation}

We use the above definitions in the decay width expressions given by
\begin{equation} \label{eq:DW}
\begin{aligned}
    \Gamma_{a \to \gamma\gamma} & = \frac{\alpha^2 m_{a}^3}{(4\pi)^3 f_a^2} \left| C_\gamma^{\chi \text{rot}} + C_\gamma^{\rm VMD} + C_\gamma^{{\rm pQCD},uds} + C_\gamma^{{\rm pQCD},cbt}  \,+\, C_\gamma^{\text {lepton}}\right|^2 \,, \\
     \Gamma_{a\to \phi\phi} & =  \frac{N_c^2}{2048\pi^5}  \frac{m_a^3}{f_a^2} \left| g^2 \Tr(\bm{a\phi\phi}) \mathcal{F}(m_a) \right|^2 \qty( 1 - \frac{4m_\phi^2}{m_a^2} )^{3/2}\,, \\
    \Gamma_{a \to 3-\text{body}} & = \frac{1}{32(2\pi)^3 S m_a^3} \int dm_{12}^2 \,dm_{23}^2 \,\overline{|\mathcal{M}_{a \to 3-\text{body}}|^2}\;, \\
    \Gamma_{a \to \ell^+ \ell^-} & = \frac{\mathcal{C}_{\ell \ell}^2}{2\pi f_a^2} m_a m_\ell^2 \bigg(1 - \frac{4m_\ell^2}{m_a^2} \bigg)^{1/2} \;,
\end{aligned}
\end{equation}
where $S$ is the symmetry factor and $m_{ij}^2=(p_i+p_j)^2$, with $p_{i,j}$ denoting the momenta of the final-state particles. The Chiral rotation contribution in $\hat{C}_{\gamma \gamma}$ gives
\begin{equation}
    C_{\gamma}^{\chi rot}  = - 2 N_c \,\cC_{gg} \sum_{q= u,d,s} Q_q^2{\kappa}_{q } \,\Theta (m_{\eta'} - m_a)\;,
\end{equation}
The expressions for other Wilson coefficients in $\Gamma_{a \to \gamma \gamma}$ and amplitude expressions for three body decay modes are explicitly given in Ref.~\cite{Cheng:2021kjg}. The different inputs presented in~\cite{Cheng:2021kjg,Aloni:2018vki} can be used in the above expressions with the following replacements
\begin{equation}
f_\pi \to  f_\pi/\sqrt{2}\;, \qquad \langle \bm{aP} \rangle \to \Tr(\bm{aP})\;,\qquad \langle \bm{aVV} \rangle \to  \frac{1}{\sqrt{2}} \Tr(\bm{aVV})\;,
\end{equation}
which converts their definitions (LHS) to ours (RHS). Finally, the meson masses, decay widths, and relevant meson couplings are taken from~\cite{ParticleDataGroup:2024cfk, Fariborz:1999gr, BaBar:2015kii}.

\bibliographystyle{JHEP}
\bibliography{references}

\end{document}